\documentclass[12pt]{article} 
\def\square{\kern1pt\vbox{\hrule height 1.2pt\hbox{\vrule width 1.2pt\hskip 3pt
   \vbox{\vskip 6pt}\hskip 3pt\vrule width 0.6pt}\hrule height 0.6pt}\kern1pt}
\def\ts{\textstyle}
\def\ss{\scriptstyle}
\def\sss{\scriptscriptstyle}
\def\gsim{\mathrel{\raise.3ex\hbox{$>$\kern-.75em\lower1ex\hbox{$\sim$}}}} 
\def\lsim{\mathrel{\raise.3ex\hbox{$<$\kern-.75em\lower1ex\hbox{$\sim$}}}} 
\def\eq#1{eq.$\,$(\ref{#1})} 
\def\eqs#1{eqs.$\,$(\ref{#1})} 
\def\Eq#1{Eq.$\,$(\ref{#1})} 
\def\Eqs#1{Eqs.$\,$(\ref{#1})} 
\def\fig#1{fig.$\,$(\ref{#1})} 
 
\def\Fig#1{Fig.$\,$(\ref{#1})} 
 
\def\e{{\ts\varepsilon}}

\def\eps{\epsilon}

\def\Mt{\widetilde{M}}

\def\tbfd{{\bf\tilde\Delta}}
\def\td{{\tilde\Delta}}

\def\tbfs{{\bf\tilde S}}
\def\st{{\tilde S}}

\def\V{{\bf V}}

\def\ra{\rangle}
\def\la{\langle} 
\def\lra{\leftrightarrow}
\def\d{\partial} 
\def\ph{\varphi} 
\def\Lam{\Lambda} 
\def\lam{\lambda}

\def\C{{\cal C}}
\def\D{{\cal D}}
\def\H{{\cal H}}
\def\L{{\cal L}}

\def\P{{\cal P}}
\def\T{{\cal T}}
\def\tr{{\rm Tr}}

\def\p{{\bf p}} 
\def\x{{\bf x}} 
\def\y{{\bf y}} 
\def\k{{\bf k}}
\def\ph{\varphi}

\def\chid{\chi^\dagger}
\def\ad{a^\dagger}
\def\bd{b^\dagger}
\def\dd{d^\dagger}

\def\dpt{\widetilde{dp}}
\def\dtx{d^{3\!}x}
\def\dty{d^{3\!}y}
\def\dtk{d^{3\!}k}
\def\dtp{d^{3\!}p}

\def\dfx{d^{4\!}x}
\def\dfy{d^{4\!}y}
\def\dfz{d^{4\!}z}
\def\dfw{d^{4\!}w}
\def\dfk{d^{4\!}k}
\def\dfp{d^{4\!}p}
\def\dfq{d^{4\!}q}
\def\dfl{d^{4\!}\ell}

\def\ddq{d^{d\!}q}
\def\ddl{d^{d\!}\ell}

\def\half{{\textstyle{1\over2}}}
\def\w{\omega}
\def\mut{{\tilde\mu}}
\def\psl{{\rlap{\kern1pt /}{p}}}
\def\pslp{{\rlap{\kern1pt /}{p}}\kern1pt{}'}
\def\asl{{\rlap{/}{a}}}
\def\Asl{{\rlap{\kern2.25pt /}{A}}}
\def\bsl{{\rlap{/}{b}}}
\def\csl{{\rlap{/}{c}}}
\def\ddsl{{\rlap{/}{d}}}
\def\Dsl{{\rlap{\kern2.25pt /}{D}}}
\def\Asl{{\rlap{\kern2.25pt /}{A}}}
\def\dsl{{\rlap{\kern0.5pt /}{\partial}}}
\def\dslx{{\rlap{\kern0.5pt /}{\partial_x}}}
\def\dsly{{\rlap{\kern0.5pt /}{\partial_y}}}
\def\ksl{{\rlap{\kern0.5pt /}{k}}}
\def\kslp{{\rlap{\kern0.5pt /}{k}}{}'}
\def\lsl{{\rlap{\kern-0.5pt /}{\ell}}}
\def\qsl{{\rlap{/}{q}}}
\def\zsl{{\rlap{/}{z}}}

\def\eslp{{\rlap{/}{\varepsilon}}\kern1pt{}'}
\def\rsl{{\rlap{/}{r}}}
\def\ssl{{\rlap{/}{s}}}
\def\cd{\!\cdot\!}
\def\g{\gamma}
\def\ubar{\overline{u}}
\def\ubarp{\overline{u}\kern1pt{}'}
\def\vbar{\overline{v}}
\def\vbarp{\overline{v}\kern1pt{}'}
\def\Psit{\Psi^{\rm\sss T}}
\def\Psibar{\overline{\Psi}}
\def\Psibart{\Psibar^{\kern1pt\lower1pt\hbox{$\rm\sss T$}}}

\def\psid{\psi^\dagger}
\def\psit{\psi^{\rm\sss T}}
\def\etabar{\overline{\eta}}

\def\chibar{\bar{\chi}}
\def\adot{{\dot a}}
\def\bdot{{\dot b}}
\def\ccdot{{\dot c}}

\def\edot{{\dot e}}
\def\LK{Lehmann-K\"all\'en\ }

\usepackage{epsfig}
\usepackage{amsbsy}
\begin{document}

\rightline{$\phantom{X}$}

\vskip0.5in

\begin{center}
\large
Quantum Field Theory
\vskip0.25in
Part II: Spin One Half
\vskip0.5in
Mark Srednicki
\end{center}
\vskip0.1in
\begin{center}
Department of Physics

University of California

Santa Barbara, CA 93106
\vskip0.1in
mark@physics.ucsb.edu
\end{center}

\vskip0.5in

\baselineskip=16pt

\noindent
This is a draft version of Part II of a three-part 
textbook on quantum field theory. 

\vskip0.2in

\vfill\eject

\begin{center}
\large{Part II: Spin One Half}
\end{center}

33) Representations of the Lorentz Group (2)

34) Left- and Right-Handed Spinor Fields (3, 33)

35) Manipulating Spinor Indices (34)

36) Lagrangians for Spinor Fields (4, 22, 35)

37) Canonical Quantization of Spinor Fields I (36)

38) Spinor Technology (37)

39) Canonical Quantization of Spinor Fields II (38)

40) Parity, Time Reversal, and Charge Conjugation (39)

41) LSZ Reduction for Spin-One-Half Particles (39)

42) The Free Fermion Propagator (39)

43) The Path Integral for Fermion Fields (9, 42)

44) Formal Development of Fermionic Path Integrals (43)

45) The Feynman Rules for Dirac Fields and Yukawa Theory (10, 13, 41, 43)

46) Spin Sums (45)

47) Gamma Matrix Technology (36)

48) Spin-Averaged Cross Sections in Yukawa Theory (46, 47)

49) The Feynman Rules for Majorana Fields (45)

50) Massless Spin-One-Half Particles and Spinor Helicity (48)

51) Loop Corrections in Yukawa Theory (19, 40, 48)

52) Beta Functions in Yukawa Theory (27, 51)

53) Functional Determinants (44, 45)

\vfill\eject

\noindent Quantum Field Theory  \hfill   Mark Srednicki

\vskip0.5in

\begin{center}
\large{33: Representations of the Lorentz Group}
\end{center}
\begin{center}
Prerequisite: 2
\end{center}

\vskip0.5in

In section 2, we saw that we could define a unitary operator $U(\Lam)$ that
implemented a Lorentz transformation on a scalar field $\ph(x)$ via
\begin{equation}
U(\Lam)^{-1} \ph(x) U(\Lam) = \ph(\Lam^{-1}x)\;.
\label{uphu32}
\end{equation}
As shown in section 2, 
this implies that the derivative of the field transforms as
\begin{equation}
U(\Lam)^{-1} \d^\mu\ph(x) U(\Lam) 
= \Lam^\mu{}_\rho\bar\d^\rho\ph(\Lam^{-1}x)\;,
\label{udphu32}
\end{equation}
where the bar on he derivative means 
that it is with respect to the argument $\bar x=\Lam^{-1}x$.

\Eq{udphu32} suggests that we could define a {\it vector field\/}
$A^\mu(x)$ which would transform as
\begin{equation}
U(\Lam)^{-1} A^\rho(x) U(\Lam) 
= \Lam^\mu{}_\rho A^\rho(\Lam^{-1}x)\;,
\label{udau}
\end{equation}
or a {\it tensor field\/} $B^{\mu\nu}(x)$ which would transform as
\begin{equation}
U(\Lam)^{-1} B^{\mu\nu}(x) U(\Lam) 
= \Lam^\mu{}_\rho \Lam^\nu{}_\sigma 
B^{\rho\sigma}(\Lam^{-1}x)\;.
\label{udbu}
\end{equation}
Note that if $B^{\mu\nu}$ is either symmetric, 
$B^{\mu\nu}(x)=B^{\nu\mu}(x)$, or antisymmetric,
$B^{\mu\nu}(x)=-B^{\nu\mu}(x)$, then this symmetry property is preserved
by the Lorentz transformation.  Also, if we take the trace to get
$T(x)\equiv g_{\mu\nu}B^{\mu\nu}(x)$, 
then, using 
$g_{\mu\nu}\Lam^\mu{}_\rho \Lam^\nu{}_\sigma = g_{\rho\sigma}$,
we find that $T(x)$ transforms like a scalar field,
\begin{equation}
U(\Lam)^{-1} T(x) U(\Lam) = T(\Lam^{-1}x)\;.
\label{upsu}
\end{equation}
Thus, given a tensor field $B^{\mu\nu}(x)$ with no particular symmetry, 
we can write
\begin{equation}
B^{\mu\nu}(x) = A^{\mu\nu}(x) + S^{\mu\nu}(x) + {\ts{1\over4}}g^{\mu\nu}T(x)\;,
\label{bast}
\end{equation}
where $A^{\mu\nu}$ is antisymmetric 
($A^{\mu\nu}=-A^{\nu\mu}$) and
$S^{\mu\nu}$ is symmetric ($S^{\mu\nu}=S^{\nu\mu}$) and traceless 
($g_{\mu\nu}S^{\mu\nu}=0$).
The key point is that the fields $A^{\mu\nu}$, $S^{\mu\nu}$, and $T$ 
do not mix with each other under Lorentz transformations.

Is it possible to further break apart these fields into still smaller sets that
do not mix under Lorentz transformations?  How do we make this decomposition
into {\it irreducible representations\/} of the Lorentz group for a field
carrying $n$ vector indices?  Are there any other kinds of indices we
could consistently assign to a field?  If so, how do these behave under
a Lorentz transformation?

The answers to these questions are to be found in the theory of 
{\it group representations}.  Let us see how this works for
the Lorentz group in four spacetime dimensions.

For an infinitesimal transformation
$\Lam^\mu{}_\nu = \delta^\mu{}_\nu + \delta\w^\mu{}_\nu$,
we can write
\begin{equation}
U(1{+}\delta\w)=I + {\ts{i\over2}}\delta\w_{\mu\nu}M^{\mu\nu}\;,
\label{udw32}
\end{equation}
where $M^{\mu\nu}=-M^{\nu\mu}$ is a set of hermitian operators,
the {\it generators of the Lorentz group}.  
As shown in section 2, these obey the commutation relations
\begin{equation}
[M^{\mu\nu},M^{\rho\sigma}]=i\Bigl(g^{\mu\rho}M^{\nu\sigma}
                            -(\mu{\leftrightarrow}\nu)\Bigr)
                            -(\rho{\leftrightarrow}\sigma)\;.
\label{comm32}
\end{equation}
We can identify the components of the angular momentum operator $\vec J$ 
as $J_i\equiv\half\e_{ijk}M^{jk}$ 
and the components of the boost operator $\vec K$ as $K_i\equiv M^{i0}$.
We then find from \eq{comm32} that
\begin{eqnarray}
{[}J_i,J_j{]} &=& +i\e_{ijk}J_k \;,
\label{commjj} \\
{[}J_i,K_j{]} &=& +i\e_{ijk}K_k \;,
\label{commjk} \\
{[}K_i,K_j{]} &=& -i\e_{ijk}J_k \;.
\label{commkk}
\end{eqnarray}
We would now like to find all the {\it representations\/} of 
eqs.$\,$(\ref{commjj}--\ref{commkk}).
A representation is a set of finite-dimensional matrices with the
same commutation relations.  For example, if we restrict our attention
to \eq{commjj} alone, we know (from standard results in the
quantum mechanics of angular momentum) that we can find three
$(2j{+}1)\times(2j{+}1)$ hermitian matrices 
${\cal J}_1$, ${\cal J}_2$, and ${\cal J}_3$
that obey \eq{commjj}, and
that the eigenvalues of (say) ${\cal J}_3$
are $-j,-j{+}1,\ldots,+j$,
where $j$ has the possible values $0,\half,1,\ldots\,$.
We further know that these matrices constitute all of the
inequivalent, irreducible representations of SO(3),
the rotation group in three dimensions.
({\it Inequivalent\/} means not related by 
a unitary transformation; 
{\it irreducible\/} means cannot be made block-diagonal by
a unitary transformation.)
We would like to extend these conclusions to encompass
the full set of eqs.$\,$(\ref{commjj}--\ref{commkk}).

In order to do so, it is helpful to
define some nonhermitian operators whose physical significance
is obscure, but which simplify the commutation relations.  These are
\begin{eqnarray}
N_i &\equiv& \half(J_i-iK_i)\;,
\label{njk} \\
N^\dagger_i &\equiv& \half(J_i+iK_i)\;.
\label{nnd}
\end{eqnarray}
In terms of $N_i$ and $N^\dagger_i$, 
eqs.$\,$(\ref{commjj}--\ref{commkk}) become
\begin{eqnarray}
{[}N_i,N_j{]} &=& i\varepsilon_{ijk}N_k \;,
\label{ncomm} \\
{[}N^\dagger_i,N^\dagger_j{]} &=& i\varepsilon_{ijk}N^\dagger_k \;,
\label{ndcomm} \\
{[}N_i,N^\dagger_j{]} &=& 0 \;.
\label{nndcomm} 
\end{eqnarray}
We recognize these as the commutation relations of two independent SO(3) 
groups [or, equivalently, SU(2); see section 32].
Thus the Lorentz group in four dimensions is equivalent to
SO(3)$\times$SO(3).
And, as just discussed, we are already familiar with
the representation theory of SO(3).
We therefore conclude that the representations of the
Lorentz group in four spacetime dimensions are specified by two
numbers $n$ and $n'$, each a nonnegative integer or half-integer.

This turns out to be correct, but there is a complication.
To derive the usual representation theory of SO(3), as is done
in any text on quantum mechanics, we need to use the fact that
the components $J_i$ of the angular momentum operator are hermitian.
The components $N_i$ of \eq{nnd}, on the other hand, are not.
This means that we have to redo the usual derivation of the 
representations of SO(3), and see what changes.

As we have already noted, the final result is the naive one, that the
representations of the Lorentz group in four dimensions are the same as
the representations of SO(3)$\times$SO(3).  Those uninterested in the
(annoyingly complicated) details can skip ahead all the way ahead
to the last four paragraphs of this section.

We begin by noting that ${\vec N}^2$ commutes with $N_i$;
this is easily derived from \eq{ncomm}. 
Similarly,  ${\vec N}^{\dagger 2}$ commutes with $N^\dagger_i$.
\Eq{nndcomm} then implies that 
${\vec N}^2$, $N_3$, ${\vec N}^{\dagger 2}$, and $N^\dagger_3$ 
are all mutually commuting.
Therefore, we can define a set of simultaneous eigenkets
$|n,m;n',m'\rangle$, where the eigenvalues of 
$\vec N^2$, $N_3$, $\vec N^{\dagger 2}$, and $N^\dagger_3$ 
are $f(n)$, $m$, $f(n')$, and $m$, respectively.
[Later we will see that $n$ and $n'$ must be nonnegative
integers or half-integers, and that $f(n)=n(n{+}1)$, as expected.]
We also define a set of bra states $\la n,m;n',m'|$
that, {\it by definition}, obey
\begin{equation}
\la n_2,m_2;n_2',m_2'|n_1,m_1;n_1',m_1'\ra =
\delta_{n_{\ss 2}n_{\ss 1}}
\delta_{m_{\ss 2}m_{\ss 1}}
\delta_{n'_{\ss 2}n'_{\ss 1}}
\delta_{m'_{\ss 2}m'_{\ss 1}}
\equiv\Delta_{21} 
\label{braket}
\end{equation}
and
\begin{equation}
\sum | n,m;n',m'\ra\la n,m;n',m'| = 1 \;.
\label{complete}
\end{equation}
In \eq{complete}, 
the sum is over all allowed values of $n$, $m$, $n'$, and $m'$;
our goal is to determine these allowed values.

From the discussion so far, we can conclude that
\begin{eqnarray}
\la n_2,m_2;n'_2,m'_2|\,\vec N^2\,|n_1,m_1;n'_1,m'_1\ra 
&=& f(n_1)\,\Delta_{21} \;,
\label{n2} \\
\la n_2,m_2;n'_2,m'_2|\,N_3\,| n_1,m_1;n'_1,m'_1\ra 
&=& \phantom{f(n_1)}\llap{$\smash{m_1}$}\,\Delta_{21} \;,
\label{n3} \\
\la n_2,m_2;n'_2,m'_2|\,\vec N^{\dagger 2}\,|n_1,m_1;n'_1,m'_1\ra 
&=& f(n'_1)\,\Delta_{21} \;,
\label{nd2} \\
\la n_2,m_2;n'_2,m'_2|\,N_3\,| n_1,m_1;n'_1,m'_1\ra 
&=& \phantom{f(n_1)}\llap{$\smash{m'_1}$}\,\Delta_{21} \;.
\label{nd3}
\end{eqnarray}
Note that we have not yet made any assumptions about the properties of the
states under hermitian conjugation.  From \eqs{ncomm} and (\ref{ndcomm}),
we see that hermitian conjugation exchanges the two SO(3) groups.
Therefore, we must have
\begin{eqnarray}
|n,m;n',m'\ra^\dagger &=& \la n',m';n,m| \;,
\label{herm1} \\
\la n,m;n',m'|^\dagger &=& | n',m';n,m\ra \;,
\label{herm2}
\end{eqnarray}
up to a possible phase factor that turns out to be irrelevant.
Compare the ordering of the labels in 
\eqs{herm1} and (\ref{herm2}) with those in \eqs{braket} and (\ref{complete}); 
a state $|n,m;n',m'\ra$ has zero inner product with its own hermitian conjugate if 
$n\ne n'$ or $m\ne m'$.

Next, take the hermitian conjugates of \eqs{n2} and (\ref{n3}), using
\eqs{herm1} and (\ref{herm2}).  We get
\begin{eqnarray}
\la n'_1,m'_1;n_1,m_1|\,\vec N^{\dagger 2}\,|n'_2,m'_2;n_2,m_2\ra 
&=& [f(n_1)]^*\,\Delta_{21} \;,
\label{n2conj} \\
\la n'_1,m'_1;n_1,m_1|\,N_3^\dagger\,|n'_2,m'_2;n_2,m_2\ra 
&=& \phantom{[f(n_1)]^*}\llap{$\smash{m^*_1}$}\,\Delta_{21} \;,
\label{n3conj} 
\end{eqnarray}
Comparing \eqs{n2conj} and (\ref{n3conj}) with \eqs{nd2} and (\ref{nd3}), 
we find that the allowed values of $f(n)$ and $m$ are real.

We now define the raising and lowering operators
\begin{eqnarray}
N_\pm &\equiv& N_1\pm iN_2 \;,
\label{npm} \\
(N^\dagger)_\pm &\equiv& N^\dagger_1\pm iN^\dagger_2 \;;
\label{ndpm}
\end{eqnarray}
note that
\begin{equation}
(N_\pm)^\dagger=(N^\dagger)_\mp \;.
\label{npmdndmp}
\end{equation}
The commutation relations (\ref{ncomm}) become
\begin{eqnarray}
{[}N_3,N_\pm{]} &=& \pm N_\pm \;,
\label{n3npm} \\
{[}N_+,N_-{]} &=& 2N_3 \;,
\label{npnm}
\end{eqnarray}
plus the equivalent with $N\to N^\dagger$. 
By inserting a complete set of states into \eq{n3npm}, 
and mimicking the usual procedure in quantum mechanics,
it is possible to show that
\begin{eqnarray}
\la n_2,m_2{+}1;n'_2,m'_2|\,N_+\,|n_1,m_1;n'_1,m'_1\ra 
&=& \lambda_+(n_1,m_1)\,\Delta_{21} \;,
\label{lamnp} \\
\la n_1,m_1;n'_1,m'_1|\,N_-\,|n_2,m_2{+}1;n'_2,m'_2\ra 
&=& \lambda_-(n_1,m_1)\,\Delta_{21} \;,
\label{lamnm}
\end{eqnarray}
where $\lambda_+(n,m)$ and $\lambda_-(n,m)$ are functions to be determined.
By inserting a complete set of states into \eq{npnm}, and using 
\eqs{lamnp} and (\ref{lamnm}), we can show that
\begin{equation}
 \lambda_+(n,m{-}1)\lambda_-(n,m{-}1) -\lambda_+(n,m)\lambda_-(n,m)=2m \;.
\label{lamdiff}
\end{equation}
The solution of this recursion relation is
\begin{equation}
\lambda_+(n,m)\lambda_-(n,m)=C(n)-m(m{+}1) \;,
\label{recurs}
\end{equation}
where $C(n)$ is an arbitrary function of $n$.

Next we need the {\it parity operator\/} $P$, introduced in section 23.
From the discussion there, we can conclude that
\begin{equation}
P^{-1}M^{\mu\nu}P = {\cal P}^\mu{}_\rho {\cal P}^\nu{}_\sigma 
M^{\rho\sigma} \;,
\label{pmmunup}
\end{equation}
where 
\begin{equation}
\P^\mu{}_\nu  = \pmatrix{ +1 & & & \cr
                                         & -1 & & \cr
                                         & & -1 & \cr
                                         & & & -1 \cr}.
\label{p32}
\end{equation}
\Eq{pmmunup} implies
\begin{eqnarray}
P^{-1}J_i P &=& +J_i\;,
\label{pjp} \\
P^{-1}K_i P &=& -K_i\;,
\label{pkp}
\end{eqnarray}
or, equivalently,
\begin{eqnarray}
P^{-1}N_i P &=& N^\dagger_i\;,
\label{pnp} \\
P^{-1}N^\dagger_i P &=& N_i\;.
\label{pndp}
\end{eqnarray}
Since $P$ exchanges $N_i$ and $N^\dagger_i$, it must be that
\begin{eqnarray}
P|n,m;n',m'\ra &=&  |n',m';n,m\ra \;,
\label{pnmnm} \\
P^{-1}|n,m;n',m'\ra &=&  |n',m';n,m\ra \;,
\label{pinmnm}
\end{eqnarray}
up to a possible phase factor that turns out to be irrelevant.  
Taking the hermitian conjugate of \eqs{pnmnm} and (\ref{pinmnm}),
we get
\begin{eqnarray}
\la n',m';n,m|P^{-1} &=&  \la n,m;n',m'| \;,
\label{pnmnmherm} \\
\la n',m';n,m|P^{\phantom{-1}} &=& \la n,m;n',m'| \;,
\label{pinmnmherm}
\end{eqnarray}
where we have used the fact that $P$ is unitarity: $P^\dagger = P^{-1}$.

Now we can take \eq{lamnp} and insert $PP^{-1}$ on either side of
$N_+$ to get
\begin{eqnarray}
\lambda_+(n_1,m_1)\,\Delta_{21}
&=& \la n_2,m_2{+}1;n'_2,m'_2|\,PP^{-1}N_+ PP^{-1}\,|n_1,m_1;n'_1,m'_1\ra 
\nonumber \\
&=& \la n'_2,m'_2;n_2,m_2{+}1|\,P^{-1}N_+ P\,|n'_1,m'_1;n_1,m_1\ra
\nonumber \\
&=& \la n'_2,m'_2;n_2,m_2{+}1|\,(N^\dagger)_+ \,|n'_1,m'_1;n_1,m_1\ra \;.
\label{lamnpd}
\end{eqnarray}
In the second line, we used \eqs{pinmnm} and (\ref{pinmnmherm}).
In the third, we used \eq{pnp}.  
Now taking the hermitian conjugate of \eq{lamnpd},
and using \eqs{herm1}, (\ref{herm2}), and (\ref{npmdndmp}), we find
\begin{equation}
\la n_1,m_1;n'_1,m'_1|\,N_-\,|n_2,m_2{+}1;n'_2,m'_2\ra
= [\lambda_+(n_1,m_1)]^*\,\Delta_{21} \;.
\label{lamnpconj}
\end{equation}
Comparing \eq{lamnpconj} with \eq{lamnm},
we see that
\begin{equation}
\lambda_-(n,m)=[\lambda_+(n,m)]^* \;.
\label{lamlam}
\end{equation}
This is the final ingredient.  Putting \eq{lamlam} into \eq{recurs}, we get
\begin{equation}
|\lambda_+(n,m)|^2=C(n)-m(m{+}1) \;.
\label{lamsq}
\end{equation}
From here, everything can be done by mimicking the usual
procedure in the quantum mechanics of angular momentum.  
We see that the left-hand side of \eq{lamsq} is real and nonnegative,
while the right-hand side becomes negative for sufficiently large $|m|$.
This is not a problem if there are two values of $m$, differing by
an integer, for which
$\lambda_+(n,m)$ is zero.  From this we can deduce that the allowed values
of $m$ are real integers or half-integers, and that if we choose
$C(n)=n(n{+}1)$,
then $n$ is an integer or half-integer such that the allowed values of $m$
are $-n,\,-n\!+\!1,\,\ldots,\,+n$.  We can also show that $f(n)=C(n)=n(n{+}1)$.
Thus the representations of the Lorentz group in four dimensions are just
the same as those of SO(3)$\times$SO(3).

We will label these representations as $(2n{+}1,2n'{+}1)$; the number
of components of a representation is then $(2n{+}1)(2n'{+}1)$.
Different components within a representation can also 
be labeled by their angular momentum
representations.  To do this, we first note that, from
\eqs{njk} and (\ref{nnd}), we have
$\vec J = \vec N +{\vec N}^\dagger$.
Thus, deducing the allowed values of $j$ given $n$ and $n'$
becomes a standard problem in the addition of angular momenta.  
The general result is that the allowed values of $j$ are
$|n{-}n'|, |n{-}n'|{+}1,\ldots,n{+}n'$,
and each of these values appears exactly once.

The four simplest and most often encountered representations are
$(1,1)$, $(2,1)$, $(1,2)$, and $(2,2)$.  These are given special names:
\begin{eqnarray}
(1,1) &=& \hbox{\it Scalar {\rm or} singlet}
\nonumber \\
(2,1) &=& \hbox{\it Left-handed spinor}
\nonumber \\
(1,2) &=& \hbox{\it Right-handed spinor}
\nonumber \\
(2,2) &=& \hbox{\it Vector}
\label{replist}
\end{eqnarray}
It may seem a little surprising that $(2,2)$
is to be identified as the vector representation.
To see that this must be the case, we first note that
the vector representation is irreducible: 
all the components of a four-vector mix with each other under
a general Lorentz transformation.  Secondly, the vector
representation has four components.  Therefore, the only
candidate irreducible representations are 
$(4,1)$, $(1,4)$, and $(2,2)$.
The first two of these contain angular momenta $j={3\over2}$ only,
whereas $(2,2)$ contains $j=0$ and $j=1$.  This is just right for
a four-vector, whose time component is a scalar under spatial
rotations, and whose space components are a three-vector.

In order to gain a better understanding of what it means for $(2,2)$ to be
the vector representation, we must 
first investigate the spinor representations $(1,2)$ and $(2,1)$,
which contain angular momenta $j=\half$ only.

\vskip0.5in

\begin{center}
Problems
\end{center}

\vskip0.25in

33.1) Express $A^{\mu\nu}(x)$, $S^{\mu\nu}(x)$, and $T(x)$ in terms of
$B^{\mu\nu}(x)$.

33.2) Verify that eqs.$\,$(\ref{ncomm}--\ref{nndcomm}) follow from 
eqs.$\,$(\ref{commjj}--\ref{commkk}).

\vfill\eject

\noindent Quantum Field Theory  \hfill   Mark Srednicki

\vskip0.5in

\begin{center}
\large{34: Left- and Right-Handed Spinor Fields}
\end{center}
\begin{center}
Prerequisite: 33
\end{center}

\vskip0.5in

Consider a {\it left-handed spinor field\/} $\psi_a(x)$, 
also known as a {\it left-handed Weyl field}, which is 
in the $(2,1)$ representation of the Lorentz group.
Here the index $a$ is a {\it left-handed spinor index\/} 
that takes on two possible values.
Under a Lorentz transformation, we have
\begin{equation}
U(\Lam)^{-1} \psi_a(x) U(\Lam) = L_a{}^b(\Lam)\psi_b(\Lam^{-1}x)\;,
\label{upsiu}
\end{equation}
where $L_a{}^b(\Lam)$ is a matrix in the $(2,1)$ representation.
These matrices satisfy the group composition rule
\begin{equation}
L_a{}^b(\Lam_1) L_b{}^c(\Lam_2)  = L_a{}^c(\Lam_1\Lam_2) \;.
\label{lll}
\end{equation}
For an infinitesimal transformation
$\Lam^\mu{}_\nu = \delta^\mu{}_\nu + \delta\w^\mu{}_\nu$,
we can write
\begin{equation}
L_a{}^b(1{+}\delta\w) = \delta_a{}^b 
+ {\ts{i\over2}}\delta\w_{\mu\nu}(S_{\rm\sss L}^{\mu\nu})_a{}^b\;,
\label{liw}
\end{equation}
where $(S_{\rm\sss L}^{\mu\nu})_a{}^b=-(S_{\rm\sss L}^{\nu\mu})_a{}^b$
is a set of $2\times2$ matrices that obey the same commutation relations
as the generators $M^{\mu\nu}$, namely
\begin{equation}
[S_{\rm\sss L}^{\mu\nu},S_{\rm\sss L}^{\rho\sigma}]
=i\Bigl(g^{\mu\rho}S_{\rm\sss L}^{\nu\sigma}
                            -(\mu{\leftrightarrow}\nu)\Bigr)
                            -(\rho{\leftrightarrow}\sigma)\;.
\label{commsl}
\end{equation}
\Eq{upsiu} becomes
\begin{equation}
[\psi_a(x),M^{\mu\nu}] = 
-i(x^\mu\d^\nu{-}x^\nu\d^\mu)\psi_a(x) 
+ (S_{\rm\sss L}^{\mu\nu})_a{}^b\psi_b(x)\;,
\label{psim}
\end{equation}
where $U(1{+}\delta\w)=I + {\ts{i\over2}}\delta\w_{\mu\nu}M^{\mu\nu}$.
The first term on the right-hand side of
\eq{psim} would also be present for a scalar field,
and is not the focus of our current interest; we will suppress it
by evaluating the fields at the spacetime origin, $x^\mu=0$.  Recalling
that $M^{ij}=\e^{ijk}J_k$, where $J_k$ is the angular momentum operator, we have
\begin{equation}
\e^{ijk}[\psi_a(0),J_k] = 
(S_{\rm\sss L}^{ij})_a{}^b\psi_b(0)\;.
\label{psij}
\end{equation}
Recall that the $(2,1)$ representation of the Lorentz group includes
angular momentum $j=\half$ only.
For a spin-one-half operator, the standard convention is that 
the matrix on the right-hand side of \eq{psij} is 
$\half\e^{ijk}\sigma_k$, 
where $\sigma_k$ is a Pauli matrix:
\begin{equation}
\sigma_1=\pmatrix{0 & 1\cr 1 &  0\cr}, \quad
\sigma_2=\pmatrix{0 & -i\cr i & 0\cr}, \quad
\sigma_3=\pmatrix{1 & 0\cr 0 & -1\cr} .
\label{pauli0}
\end{equation}
We therefore conclude that 
\begin{equation}
(S_{\rm\sss L}^{ij})_a{}^b = \half\e^{ijk}\sigma_k \;.
\label{sijsigma}
\end{equation}
Thus, for example, setting $i{=}1$ and $j{=}2$ yields
$(S_{\rm\sss L}^{12})_a{}^b = \half\e^{12k}\sigma_k=\half\sigma_3$,
where the subscript $a$ is the row index (and the superscript $b$
is the column index) of the matrix $\half\sigma_3$.
Therefore, 
$(S_{\rm\sss L}^{12})_1{}^1=+\half$,
$(S_{\rm\sss L}^{12})_2{}^2=-\half$, and
$(S_{\rm\sss L}^{12})_1{}^2=(S_{\rm\sss L}^{12})_2{}^1=0$.

Once we have the $(2,1)$ representation matrices for the 
angular momentum operator $J_i$, we can easily get 
them for the boost operator $K_k=M^{k0}$.  
This is because
$J_k=N_k+N_k^\dagger$ and 
$K_k=i(N_k-N_k^\dagger)$,
and, in the $(2,1)$ representation, 
$N_k^\dagger$ is zero.
Therefore, the representation matrices for $K_k$ are simply 
$i$ times those for $J_k$, and so
\begin{equation}
(S_{\rm\sss L}^{k0})_a{}^b = \half i\sigma_k \;.
\label{sk0sigma}
\end{equation}

Now consider taking the hermitian conjugate of the left-handed spinor field
$\psi_a(x)$.  Recall that hermitian conjugation swaps the two SO(3) factors
in the Lorentz group.  Therefore, the hermitian conjugate of a field in
the $(2,1)$ representation should be a field in the $(1,2)$ representation;
such a field is called a {\it right-handed spinor field},
or a {\it right-handed Weyl field}.
We will distinguish the indices of the $(1,2)$ representation 
from those of the $(2,1)$ representation
by putting dots over them.  Thus, we write
\begin{equation}
[\psi_a(x)]^\dagger = \psid_\adot(x) \;.
\label{psidag}
\end{equation}
Under a Lorentz transformation, we have
\begin{equation}
U(\Lam)^{-1} \psid_\adot(x) U(\Lam) 
= R_\adot{}^\bdot(\Lam)\psid_\bdot(\Lam^{-1}x) \;,
\label{upsidagu}
\end{equation}
where $R_\adot{}^\bdot(\Lam)$ is a matrix in the $(1,2)$ representation.
These matrices satisfy the group composition rule
\begin{equation}
R_\adot{}^\bdot(\Lam_1) R_\bdot{}^{\dot c}(\Lam_2)  
= R_\adot{}^{\dot c}(\Lam_1\Lam_2)  \;.
\label{rrr}
\end{equation}
For an infinitesimal transformation
$\Lam^\mu{}_\nu = \delta^\mu{}_\nu + \delta\w^\mu{}_\nu$,
we can write
\begin{equation}
R_\adot{}^\bdot(1{+}\delta\w) = 
\delta_\adot{}^\bdot + {\ts{i\over2}}\delta\w_{\mu\nu}
(S_{\rm\sss R}^{\mu\nu})_\adot{}^\bdot\;,
\label{riw}
\end{equation}
where $(S_{\rm\sss R}^{\mu\nu})_\adot{}^\bdot
=-(S_{\rm\sss R}^{\nu\mu})_\adot{}^\bdot$
is a set of $2\times2$ matrices that obey the same commutation relations
as the generators $M^{\mu\nu}$.  We then have
\begin{equation}
[\psid_\adot(0),M^{\mu\nu}] = 
(S_{\rm\sss R}^{\mu\nu})_\adot{}^\bdot\psid_\bdot(0)\;.
\label{psidm}
\end{equation}
Taking the hermitian conjugate of this equation, we get
\begin{equation}
[M^{\mu\nu},\psi_a(0)] = 
[(S_{\rm\sss R}^{\mu\nu})_\adot{}^\bdot]^*\psi_b(0)\;.
\label{psidmconj}
\end{equation}
Comparing this with \eq{psim}, we see that
\begin{equation}
(S_{\rm\sss R}^{\mu\nu})_\adot{}^\bdot =
-[(S_{\rm\sss L}^{\mu\nu})_a{}^b]^* \;.
\label{srconjsl}
\end{equation}

In the previous section, we examined the Lorentz-transformation
properties of a field carrying two vector indices.  To help us
get better acquainted with the properties of spinor indices,
let us now do the same for a field that carries two $(2,1)$ indices.
Call this field $C_{ab}(x)$.  Under a Lorentz transformation, we have
\begin{equation}
U(\Lam)^{-1} C_{ab}(x) U(\Lam) 
= L_a{}^c(\Lam) L_b{}^d(\Lam)C_{cd}(\Lam^{-1}x)\;.
\label{uphabu}
\end{equation}
The question we wish to address is whether or not 
the four components of $C_{ab}$ can be grouped into smaller sets
that do not mix with each other under Lorentz transformations.

To answer this question, recall from quantum mechanics that 
two spin-one-half particles can be in a state of total spin zero, 
or total spin one.  Furthermore,
the single spin-zero state is the unique {\it antisymmetric\/} 
combination of the two spin-one-half states, 
and the three spin-one states are the three 
{\it symmetric\/} combinations of the two spin-one-half states.
We can write this schematically as 
$2\otimes2=1_{\rm A}\oplus3_{\rm S}$, where we
label the representation of SO(3) by the number of its components,
and the subscripts S and A indicate whether
that representation appears in the symmetric or antisymmetric
combination of the two 2's.
For the Lorentz group, the relevant relation is
$(2,1)\otimes(2,1)=(1,1)_{\rm A}\oplus(3,1)_{\rm S}$.
This implies that we should be able to write
\begin{equation}
C_{ab}(x) = \e_{ab}D(x) + G_{ab}(x) \;,
\label{cdg}
\end{equation}
where $D(x)$ is a scalar field, $\e_{ab}=-\e_{ba}$ is an antisymmetric set
of constants, and $G_{ab}(x)=G_{ba}(x)$. 
The symbol $\e_{ab}$ is uniquely determined by its symmetry properties up to an
overall constant; we will choose $\e_{21}=-\e_{12}=+1$.

Given that $D(x)$ is a Lorentz scalar, \eq{cdg} is consistent with \eq{uphabu}
only if 
\begin{equation}
L_a{}^c(\Lam) L_b{}^d(\Lam) \e_{cd} = \e_{ab}\;.
\label{lle}
\end{equation}
This means that $\e_{ab}$ is an {\it invariant symbol\/} of the Lorentz group:
it does not change under a Lorentz transformation that acts on all of its
indices.
In this way, $\e_{ab}$ is analogous to the metric $g_{\mu\nu}$, which
is also an invariant symbol, since
\begin{equation}
\Lam_\mu{}^\rho\Lam_\nu{}^\sigma g_{\rho\sigma}=g_{\mu\nu} \;.
\label{llge}
\end{equation}
We use $g_{\mu\nu}$ and its inverse $g^{\mu\nu}$ to raise and lower
vector indices, and
we can use $\e_{ab}$ and and its inverse $\e^{ab}$ to raise and lower
left-handed spinor indices.  Here we define $\e^{ab}$ via
\begin{equation}
\e^{12}=\e_{21}=+1\;,\qquad \e^{21}=\e_{12}=-1 \;.
\label{eenorm}
\end{equation}
With this definition, we have
\begin{equation}
\e_{ab}\e^{bc}=\delta_a{}^c \;,
\qquad 
\e^{ab}\e_{bc}=\delta^a{}_c \;.
\label{eed}
\end{equation}
We can then define
\begin{equation}
\psi^a(x) \equiv \e^{ab}\psi_b(x) \;.
\label{psiup}
\end{equation}
We also have (suppressing the spacetime argument of the field)
\begin{equation}
\psi_a=\e_{ab}\psi^b=\e_{ab}\e^{bc}\psi_c=\delta_a{}^c\psi_c \;,
\label{psidown}
\end{equation}
as we would expect.  However,
the antisymmetry of $\e^{ab}$ means that
we must be careful with minus signs; for example,
\eq{psiup} can be written in various ways, such as
\begin{equation}
\psi^a = \e^{ab}\psi_b = -\e^{ba}\psi_b = -\psi_b\e^{ba} = \psi_b\e^{ab} \;.
\label{psis1}
\end{equation}
We must also be careful about signs when we contract indices, since
\begin{equation}
\psi^a\chi_a=\e^{ab}\psi_b\chi_a=-\e^{ba}\psi_b\chi_a=-\psi_b\chi^b \;.
\label{psis3}
\end{equation}
In section 35, we will (mercifully) develop an index-free notation 
that automatically keeps track
of these essential (but annoying) minus signs.

An exactly analogous discussion applies to the second SO(3) factor;
from the group-theoretic relation
$(1,2)\otimes(1,2)=(1,1)_{\rm A}\oplus(1,3)_{\rm S}$,
we can deduce the existence of an invariant symbol
$\varepsilon_{\adot\bdot}=-\varepsilon_{\bdot\adot}$.
We will normalize $\varepsilon^{\adot\bdot}$ according to 
\eq{eenorm}.  Then eqs.$\,$(\ref{eed}--\ref{psis3}) hold
if {\it all\/} the undotted indices are replaced by dotted indices.

Now consider a field carrying one undotted and one dotted index,
$A_{a\adot}(x)$.  Such a field is in the $(2,2)$ representation,
and in section 33 we concluded that the $(2,2)$ representation was
the vector representation.  We would more naturally write a field
in the vector representation as $A^\mu(x)$.  There must, then, be a dictionary
that gives us the components of $A_{a\adot}(x)$ in terms of the components of
$A^\mu(x)$; we can write this as
\begin{equation}
A_{a\adot}(x)=\sigma^\mu_{a\adot}A_\mu(x)\;,
\label{aadotmu}
\end{equation}
where $\sigma^\mu_{a\adot}$ is another invariant symbol.  That such a symbol
must exist can be deduced from the group-theoretic relation
\begin{equation}
(2,1)\otimes(1,2)\otimes(2,2)=(1,1)\oplus\ldots\;.
\label{122122}
\end{equation}
As we will see in section 35, 
it turns out to be consistent with our already established conventions
for $S_{\rm\sss L}^{\mu\nu}$ and $S_{\rm\sss R}^{\mu\nu}$
to choose 
\begin{equation}
\sigma^\mu_{a\adot}=(I,\vec\sigma)\;.
\label{snorm0}
\end{equation}
Thus, for example,
$\sigma^3_{1\dot 1}=+1$,
$\sigma^3_{2\dot 2}=-1$,
$\sigma^3_{1\dot 2}=\sigma^3_{2\dot 1}=0$.

In general, whenever the product of a set of representations
includes the singlet, there is a corresponding invariant symbol.
For example, we can deduce the existence of $g_{\mu\nu}=g_{\nu\mu}$ from 
\begin{equation}
(2,2)\otimes(2,2)=(1,1)_{\rm S} \oplus (1,3)_{\rm A}
                  \oplus (3,1)_{\rm A} \oplus (3,3)_{\rm S} \;.
\label{2222}
\end{equation}
Another invariant symbol, the {\it Levi-Civita symbol}, follows from
\begin{equation}
(2,2)\otimes (2,2)\otimes (2,2)\otimes (2,2) = (1,1)_{\rm A} \oplus \ldots \;,
\label{levic}
\end{equation}
where the subscript A denotes the completely antisymmetric part.  
The Levi-Civita symbol is $\e^{\mu\nu\rho\sigma}$, 
which is antisymmetric on exchange of any pair of its indices,
and is normalized via $\e^{0123}=+1$.
To see that $\e^{\mu\nu\rho\sigma}$ is invariant, we note that 
$\Lam^\mu{}_\alpha
\Lam^\nu{}_\beta
\Lam^\rho{}_\gamma
\Lam^\sigma{}_\delta
\e^{\alpha \beta \gamma \delta}$ is antisymmetric
on exchange of any two of its uncontracted
indices, and therefore must be
proportional to  
$\e^{\mu\nu\rho\sigma}$.  The constant of proportionality works out to be
$\det\Lam$, which is $+1$ for a proper Lorentz transformation.

We are finally ready to answer a question 
we posed at the beginning of section 33.
There we considered a field $B^{\mu\nu}(x)$ carrying two vector indices, 
and we decomposed it as
\begin{equation}
B^{\mu\nu}(x) = A^{\mu\nu}(x) + S^{\mu\nu}(x) + {\ts{1\over4}}g^{\mu\nu}T(x)\;,
\label{bast33}
\end{equation}
where $A^{\mu\nu}$ is antisymmetric 
($A^{\mu\nu}=-A^{\nu\mu}$) and
$S^{\mu\nu}$ is symmetric ($S^{\mu\nu}=S^{\nu\mu}$) and traceless 
($g_{\mu\nu}S^{\mu\nu}=0$).
We asked whether further decomposition into still smaller irreducible
representations was possible.  The answer to this question can be found
in \eq{2222}.
Obviously, $T(x)$ corresponds to $(1,1)$, and $S^{\mu\nu}(x)$ to $(3,3)$.
[Note that a symmetric traceless tensor has three independent diagonal
components, and six independent off-diagonal components, for a total of nine,
the number of components of the $(3,3)$ representation.]
But, according to \eq{2222}, the antisymmetric field $A^{\mu\nu}(x)$ should
correspond to $(3,1)\oplus(1,3)$.  A field in the $(3,1)$ representation 
carries a symmetric pair of left-handed (undotted) spinor indices; its 
hermitian conjugate is 
a field in the $(1,3)$ representation that 
carries a symmetric pair of right-handed (dotted) spinor indices.
We should, then, be able to find a mapping, analogous to \eq{aadotmu},
that gives $A^{\mu\nu}(x)$ in terms of a field $G_{ab}(x)$ and 
its hermitian conjugate $G^\dagger_{\adot\bdot}(x)$.

This mapping is provided by the generator matrices $S_{\rm\sss L}^{\mu\nu}$ and
$S_{\rm\sss R}^{\mu\nu}$.
We first note that the Pauli matrices are traceless, and
so \eqs{sijsigma} and (\ref{sk0sigma}) imply that 
$(S_{\rm\sss L}^{\mu\nu})_a{}^a=0$.  Using \eq{psiup},
we can rewrite this as $\e^{ab}(S_{\rm\sss L}^{\mu\nu})_{ab}=0$.
Since $\e^{ab}$ is antisymmetric,
$(S_{\rm\sss L}^{\mu\nu})_{ab}$ must be symmetric on exchange
of its two spinor indices.  An identical argument shows that
$(S_{\rm\sss R}^{\mu\nu})_{\adot\bdot}$ must be symmetric on exchange
of its two spinor indices.
Furthermore, according to \eqs{sijsigma} and (\ref{sk0sigma}), we have
\begin{equation}
(S_{\rm\sss L}^{10})_a{}^b = -i(S_{\rm\sss L}^{23})_a{}^b \;.
\label{s01s23}
\end{equation}
This can be written covariantly with the Levi-Civita symbol as
\begin{equation}
(S_{\rm\sss L}^{\mu\nu})_a{}^b 
= -{\ts{i\over2}}\e^{\mu\nu\rho\sigma}
  (S_{{\rm\sss L}\,\rho\sigma})_a{}^b \;. 
\label{smnsps}
\end{equation}
Similarly,
\begin{equation}
(S_{\rm\sss R}^{\mu\nu})_\adot{}^\bdot 
= +{\ts{i\over2}}\e^{\mu\nu\rho\sigma}
  (S_{{\rm\sss R}\,\rho\sigma})_\adot{}^\bdot \;.
\label{smnspsdots}
\end{equation}
\Eq{smnspsdots} follows from taking the complex conjugate of
\eq{smnsps} and using \eq{srconjsl}.

Now, given a field $G_{ab}(x)$ in the $(3,1)$ representation,
we can map it into a {\it self-dual antisymmetric tensor\/} $G^{\mu\nu}(x)$ via
\begin{equation}
G^{\mu\nu}(x) \equiv (S_{\rm\sss L}^{\mu\nu})^{ab} G_{ab}(x) \;.
\label{gmngab}
\end{equation}
By {\it self-dual}, we mean that $G^{\mu\nu}(x)$ obeys
\begin{equation}
G^{\mu\nu}(x) = -{\ts{i\over2}}\e^{\mu\nu\rho\sigma}G_{\rho\sigma}(x) \;.
\label{egpig}
\end{equation}
Taking the hermitian conjugate of \eq{gmngab}, and using \eq{srconjsl}, we get
\begin{equation}
G^{\dagger\mu\nu}(x)
= -(S_{\rm\sss R}^{\mu\nu})^{\adot\bdot} G^\dagger_{\adot\bdot}(x) \;,
\label{gmngabdots}
\end{equation}
which is {\it anti-self-dual},
\begin{equation}
G^{\dagger\mu\nu}(x) =
+{\ts{i\over2}}\e^{\mu\nu\rho\sigma}G^\dagger_{\rho\sigma}(x) \;.
\label{egmig}
\end{equation}
Given a hermitian antisymmetric tensor field $A^{\mu\nu}(x)$, 
we can extract its self-dual and anti-self-dual parts via
\begin{eqnarray}
G^{\mu\nu}(x) &=& \half A^{\mu\nu}(x) 
- {\ts{i\over4}}\e^{\mu\nu\rho\sigma}\!A_{\rho\sigma}(x) \;,
\label{gasd} \\
\noalign{\smallskip}
G^{\dagger\mu\nu}(x) &=& \half A^{\mu\nu}(x) 
+ {\ts{i\over4}}\e^{\mu\nu\rho\sigma}\!A_{\rho\sigma}(x) \;.
\label{gasdd} 
\end{eqnarray}
Then we have
\begin{equation}
A^{\mu\nu}(x) = G^{\mu\nu}(x) + G^{\dagger\mu\nu}(x) \;.
\label{aggd}
\end{equation}
The field $G^{\mu\nu}(x)$ is in the $(3,1)$ representation, and
the field $G^{\dagger\mu\nu}(x)$ is in the $(1,3)$ representation;
these do not mix under Lorentz transformations.

\vskip0.5in

\begin{center}
Problems
\end{center}

\vskip0.25in

34.1) Verify that \eq{psim} follows from \eq{upsiu}.

34.2) Verify that \eqs{sijsigma} and (\ref{sk0sigma}) obey \eq{commsl}.

34.3) Consider a field $C^{a\ldots c\,\dot a\ldots\dot c}(x)$, 
with $N$ undotted indices and $M$ dotted indices, 
that is furthermore symmetric on exchange of any pair of undotted indices,
and also symmetric on exchange of any pair of dotted indices.
Show that this field corresponds to a single irreducible representation
$(2n{+}1,2n'{+}1)$ of the Lorentz group, and identify $n$ and $n'$.

\vfill\eject

\noindent Quantum Field Theory  \hfill   Mark Srednicki

\vskip0.5in

\begin{center}
\large{35: Manipulating Spinor Indices}
\end{center}
\begin{center}
Prerequisite: 34
\end{center}

\vskip0.5in

In section 34 we introduced the invariant symbols $\e_{ab}$, $\e^{ab}$, 
$\e_{\adot\bdot}$, and $\e^{\adot\bdot}$, where
\begin{equation}
\e^{12}=\e^{\dot 1\dot 2}=\e_{21}=\e_{\dot 2\dot 1}=+1 \;,
\qquad 
\e^{21}=\e^{\dot 2\dot 1}=\e_{12}=\e_{\dot 1\dot 2}=-1 \;.
\label{eenorm34}
\end{equation}
We use the $\e$ symbols to raise and lower spinor indices,
contracting the second index on the $\e$. 
(If we contract the first index instead, 
then there is an extra minus sign).

Another invariant symbol is 
\begin{equation}
\sigma^\mu_{a\adot}=(I,\vec\sigma) \;,
\label{snorm}
\end{equation}
where $I$ is the $2\times2$ identity matrix, and
\begin{equation}
\sigma_1=\pmatrix{0 & 1\cr 1 &  0\cr}, \quad
\sigma_2=\pmatrix{0 & -i\cr i & 0\cr}, \quad
\sigma_3=\pmatrix{1 & 0\cr 0 & -1\cr} 
\label{pauli}
\end{equation}
are the Pauli matrices.  

Now let's consider some combinations of invariant symbols with some indices
contracted, such as $g_{\mu\nu}\sigma^\mu_{a\adot}\sigma^\nu_{b\bdot}$.  
This object must also be invariant.  Then,
since it carries two undotted and two dotted spinor indices, it must
be proportional to $\e_{ab}\e_{\adot\bdot}$.
Using \eqs{eenorm34} and (\ref{snorm}), 
we can laboriously check this; it turns out to
be correct.   [If it wasn't, then \eq{snorm} would not be a tenable 
choice of numerical values for this symbol.] 
The proportionality constant works out to be minus two:
\begin{equation}
\sigma^\mu_{a\adot}\sigma_{\mu b\bdot} = -2\e_{ab}\e_{\adot\bdot}\;.
\label{ssee}
\end{equation}
Similarly, 
$\e^{ab}\e^{\adot\bdot}\sigma^\mu_{a\adot}\sigma^\nu_{b\bdot}$ 
must be proportional to $g^{\mu\nu}$, 
and the proportionality constant is again minus two:
\begin{equation}
\e^{ab}\e^{\adot\bdot}\sigma^\mu_{a\adot}\sigma^\nu_{b\bdot} =-2g^{\mu\nu} \;.
\label{eessg}
\end{equation}

Next, let's see what we can learn about the generator matrices
$(S_{\rm\sss L}^{\mu\nu})_a{}^b$ and $(S_{\rm\sss R}^{\mu\nu})_\adot{}^\bdot$
from the fact that $\e_{ab}$, $\e_{\adot\bdot}$, and $\sigma^\mu_{a\adot}$
are all invariant symbols.  Begin with
\begin{equation}
\e_{ab}=L(\Lam)_a{}^c L(\Lam)_b{}^d \e_{cd} \;,
\label{elle}
\end{equation}
which expresses the Lorentz invariance of $\e_{ab}$.
For an infinitesimal transformation
$\Lam^\mu{}_\nu = \delta^\mu{}_\nu + \delta\w^\mu{}_\nu$,
we have
\begin{equation}
L_a{}^b(1{+}\delta\w) = \delta_a{}^b 
+ {\ts{i\over2}}\delta\w_{\mu\nu}(S_{\rm\sss L}^{\mu\nu})_a{}^b\;,
\label{liw34}
\end{equation}
and \eq{elle} becomes
\begin{eqnarray}
\e_{ab} &=& \e_{ab} + {\ts{i\over2}}\delta\w_{\mu\nu}
\Bigl[\,(S_{\rm\sss L}^{\mu\nu})_a{}^c\e_{cb} +
 (S_{\rm\sss L}^{\mu\nu})_b{}^d\e_{ad}\,\Bigr] + O(\delta\w^2)
\nonumber \\
\noalign{\medskip}
&=& \e_{ab} + {\ts{i\over2}}\delta\w_{\mu\nu}
\Bigl[\,-(S_{\rm\sss L}^{\mu\nu})_{ab} 
 +(S_{\rm\sss L}^{\mu\nu})_{ba}\,\Bigr] + O(\delta\w^2) \;.
\label{elle2}
\end{eqnarray}
Since \eq{elle2} holds for any choice of $\delta\w_{\mu\nu}$,
it must be that the factor in square brackets vanishes.  Thus we 
conclude that $(S_{\rm\sss L}^{\mu\nu})_{ab}=(S_{\rm\sss L}^{\mu\nu})_{ba}$,
which we had already deduced in section 34 by a different method.
Similarly, starting from the Lorentz invariance of $\e_{\adot\bdot}$,
we can show that
$(S_{\rm\sss R}^{\mu\nu})_{\adot\bdot}=(S_{\rm\sss R}^{\mu\nu})_{\bdot\adot}$.

Next, start from
\begin{equation}
\sigma^\rho_{a\adot}=\Lam^\rho{}_\tau L(\Lam)_a{}^b R(\Lam)_\adot{}^\bdot
\sigma^\tau_{b\bdot} \;,
\label{slamee}
\end{equation}
which expresses the Lorentz invariance of $\sigma^\rho_{a\adot}$.
For an infinitesimal transformation, we have
\begin{eqnarray}
\Lam^\rho{}_\tau &=& \delta^\rho{}_\tau 
+\half\delta\w_{\mu\nu}(g^{\mu\rho}\delta^\nu{}_\tau
                       -g^{\nu\rho}\delta^\mu{}_\tau) \;,
\label{lamrhotau} \\
\noalign{\medskip}
L_a{}^b(1{+}\delta\w) &=& \delta_a{}^b 
+ {\ts{i\over2}}\delta\w_{\mu\nu}(S_{\rm\sss L}^{\mu\nu})_a{}^b\;,
\label{liw342} \\
\noalign{\medskip}
R_\adot{}^\bdot(1{+}\delta\w) &=& 
\delta_\adot{}^\bdot + {\ts{i\over2}}\delta\w_{\mu\nu}
(S_{\rm\sss R}^{\mu\nu})_\adot{}^\bdot\;.
\label{riw34}
\end{eqnarray}
Substituting eqs.$\,$(\ref{lamrhotau}--\ref{riw34}) into \eq{slamee}
and isolating the coefficient of $\delta\w_{\mu\nu}$ yields
\begin{equation}
(g^{\mu\rho}\delta^\nu{}_\tau-g^{\nu\rho}\delta^\mu{}_\tau)\sigma^\tau_{a\adot}
      +i(S_{\rm\sss L}^{\mu\nu})_a{}^b\sigma^\rho_{b\adot}
      +i(S_{\rm\sss R}^{\mu\nu})_\adot{}^\bdot\sigma^\rho_{\smash{a\bdot}} =0\;.
\label{coeffzero}
\end{equation}
Now multiply by $\sigma_{\rho c\ccdot}$ to get
\begin{equation}
 \sigma^\mu_{c\ccdot}\sigma^\nu_{a\adot}
-\sigma^\nu_{c\ccdot}\sigma^\mu_{a\adot}
+i(S_{\rm\sss L}^{\mu\nu})_a{}^b\sigma^\rho_{b\adot}\sigma_{\rho c\ccdot}
+i(S_{\rm\sss R}^{\mu\nu})_\adot{}^\bdot\sigma^\rho_{\smash{a\bdot}}
                                                    \sigma_{\rho c\ccdot}=0\;.
\label{coeffzero2}
\end{equation}
Next use \eq{ssee} in each of the last two terms to get
\begin{equation}
 \sigma^\mu_{c\ccdot}\sigma^\nu_{a\adot}
-\sigma^\nu_{c\ccdot}\sigma^\mu_{a\adot}
+2i(S_{\rm\sss L}^{\mu\nu})_{ac}\e_{\adot\ccdot}
+2i(S_{\rm\sss R}^{\mu\nu})_{\adot\ccdot}\e_{ac} = 0\;.
\label{coeffzero3}
\end{equation}
If we multiply \eq{coeffzero3} by $\e^{\adot\ccdot}$, and remember that
$\e^{\adot\ccdot}(S_{\rm\sss R}^{\mu\nu})_{\adot\ccdot}=0$ and that
$\e^{\adot\ccdot}\e_{\adot\ccdot}=-2$, we get a formula for 
$(S_{\rm\sss L}^{\mu\nu})_{ac}$, namely
\begin{equation}
(S_{\rm\sss L}^{\mu\nu})_{ac} = 
{\ts{i\over 4}}\e^{\adot\ccdot}(\sigma^\mu_{a\adot}\sigma^\nu_{c\ccdot}
                                          -\sigma^\nu_{a\adot}\sigma^\mu_{c\ccdot}) \;.
\label{slmunu}
\end{equation}
Similarly, if we multiply \eq{coeffzero3} by $\e^{ac}$, we get
\begin{equation}
(S_{\rm\sss R}^{\mu\nu})_{\adot\ccdot} = 
{\ts{i\over 4}}\e^{ac}(\sigma^\mu_{a\adot}\sigma^\nu_{c\ccdot}
                              -\sigma^\nu_{a\adot}\sigma^\mu_{c\ccdot}) \;.
\label{srmunu}
\end{equation}
These formulae can be made to look a little nicer if we define
\begin{equation}
\bar\sigma^{\mu\adot a} \equiv \e^{ab}\e^{\adot\bdot}\sigma^\mu_{b\bdot}\;.
\label{sigmabar}
\end{equation}
Numerically, it turns out that
\begin{equation}
\bar\sigma^{\mu\adot a} = (I,-\vec\sigma)\;.
\label{sbnorm}
\end{equation}
Using $\bar\sigma^\mu$, we can write \eqs{slmunu} and (\ref{srmunu}) as
\begin{eqnarray}
(S_{\rm\sss L}^{\mu\nu})_a{}^b &=&
+{\ts{i\over 4}}(\sigma^\mu\bar\sigma^\nu
                -\sigma^\nu\bar\sigma^\mu)_a{}^b \;,
\label{slmunu2} \\
\noalign{\medskip}
(S_{\rm\sss R}^{\mu\nu})^\adot{}_\bdot &=&
-{\ts{i\over 4}}(\bar\sigma^\mu\sigma^\nu
                -\bar\sigma^\nu\sigma^\mu)^\adot{}_\bdot \;.
\label{srmunu2}
\end{eqnarray}
In \eq{srmunu2},
we have suppressed a contracted pair of undotted indices arranged as
${}^c{\,}_c$,
and in \eq{slmunu2},
we have suppressed a contracted pair of dotted indices arranged as
${}_\ccdot{\,}^\ccdot$. 

We will adopt this as a general convention: a missing pair of contracted,
undotted indices is understood to be written as ${}^c{\,}_c$, and 
a missing pair of contracted,
dotted indices is understood to be written as
${}_\ccdot{\,}^\ccdot$.
Thus, if $\chi$ and $\psi$ are two
left-handed Weyl fields, we have
\begin{equation}
\chi\psi = \chi^a\psi_a 
\quad \hbox{and} \quad 
\chi^\dagger\psi^\dagger =\chi^\dagger_\adot\psi^{\dagger\adot} \;.
\label{chipsi}
\end{equation}
We expect Weyl fields to describe spin-one-half particles, and
(by the spin-statistics theorem) these particles must be {\it fermions}.
Therefore the correspoding fields must {\it anticommute}, rather than
commute.  That is, we should have
\begin{equation}
\chi_a(x)\psi_b(y) = -\psi_b(y)\chi_a(x)\;.
\label{mchipsi}
\end{equation}
Thus we can rewrite \eq{chipsi} as 
\begin{equation}
\chi\psi=\chi^a\psi_a=-\psi_a\chi^a=\psi^a\chi_a=\psi\chi\;.
\label{chipsi2}
\end{equation}
The second equality follows from anticommutation of the fields,
and the third from switching 
${}_a{\,}^a$ to ${}^a{\,}_a$ (which introduces an extra minus sign).
\Eq{chipsi2} tells us that $\chi\psi=\psi\chi$,
which is a nice feature of this notation.  Furthermore,
if we take the hermitian conjugate of $\chi\psi$, we get
\begin{equation}
(\chi\psi)^\dagger = (\chi^a\psi_a)^\dagger
= (\psi_a)^\dagger(\chi^a)^\dagger 
= \psi^\dagger_\adot\chi^{\dagger\adot}
=\psi^\dagger\chi^\dagger\;. 
\label{chipsidag}
\end{equation}
That $(\chi\psi)^\dagger = \psi^\dagger\chi^\dagger$
is just what we would expect if we ignored the indices
completely.  Of course, by analogy with \eq{chipsi2},
we also have 
$\psi^\dagger\chi^\dagger=\chi^\dagger\psi^\dagger$.

In order to tell whether a spinor field is left-handed or 
right-handed when its spinor index is suppressed,
we will adopt the convention that a right-handed
field is always written as the hermitian conjugate of a left-handed
field.  Thus, a right-handed field is always written with a dagger, 
and a left-handed field is always written without a dagger.  

Let's try computing the hermitian conjugate of something a little
more complicated:
\begin{equation}
\psi^\dagger \bar\sigma^\mu\chi = 
\psi^\dagger_\adot \,\bar\sigma^{\mu\adot c}\,\chi_c \;.
\label{psidagschi}
\end{equation}
This behaves like a vector field under Lorentz transformations,
\begin{equation}
U(\Lam)^{-1}[\psi^\dagger \bar\sigma^\mu\chi]U(\Lam) = 
\Lam^\mu{}_\nu[\psi^\dagger \bar\sigma^\nu\chi] \;.
\label{upsichiu}
\end{equation}
(To avoid clutter, we suppressed the spacetime argument of the fields;
as usual, it is $x$ on the left-hand side and $\Lam^{-1}x$ on the right.)

The hermitian conjugate of \eq{psidagschi} is
\begin{eqnarray}
[\psi^\dagger \bar\sigma^\mu\chi ]^\dagger
&=& [\psi^\dagger_\adot \,\bar\sigma^{\mu\adot c}\,\chi_c]^\dagger
\nonumber \\
\noalign{\smallskip}
&=& \chi^\dagger_\ccdot\,(\bar\sigma^{\mu a\ccdot})^*\,\psi_a
\nonumber \\
\noalign{\smallskip}
&=& \chi^\dagger_\ccdot \,\bar\sigma^{\mu\ccdot a}\,\psi_a
\nonumber \\
\noalign{\smallskip}
&=& \chi^\dagger\bar\sigma^{\mu}\psi \;.
\label{pscdag}
\end{eqnarray}
In the third line, 
we used the hermiticity of the matrices $\bar\sigma^\mu=(I,-\vec\sigma)$. 

We will get considerably more practice with this notation in the following
sections.

\vskip0.5in

\begin{center}
Problems
\end{center}

\vskip0.25in

35.1) Verify that \eq{sbnorm} follows from \eqs{snorm} and (\ref{sigmabar}).

35.2) Verify that \eq{slmunu2} is consistent with
\eqs{sijsigma} and (\ref{sk0sigma}).
 
35.3) Verify that \eq{srmunu2} is consistent with \eq{srconjsl}.

35.4) Verify \eq{eessg}.  

Hint for all problems:
write everything in ``matrix multiplication'' order, and
note that, numerically, $\e^{ab}=-\e_{ab}=i\sigma_2$.
Then make use of the properties of the Pauli matrices.

\vfill\eject

\noindent Quantum Field Theory  \hfill   Mark Srednicki

\vskip0.5in

\begin{center}
\large{36: Lagrangians for Spinor Fields}
\end{center}
\begin{center}
Prerequisite: 4, 22, 35
\end{center}

\vskip0.5in

Suppose we have a left-handed spinor field $\psi_a$.  We would like to find
a suitable lagrangian for it.  This lagrangian must be Lorentz invariant, 
and it must be hermitian.   We would also like it to be
quadratic in $\psi$ and its hermitian conjugate $\psi_\adot^\dagger$,
because this will lead to a linear equation of motion, 
with plane-wave solutions.
We want plane-wave solutions because these describe free particles, 
the starting point for a theory of interacting particles.

Let us begin with terms with no derivatives.  The only possibility is
$\psi\psi=\psi^a\psi_a=\e^{ab}\psi_b\psi_a$, plus its hermitian conjugate.
Alas, $\psi\psi$ appears to be zero,
since $\psi_b\psi_a=\psi_a\psi_b$, while $\e^{ab}=-\e^{ba}$.

However, from the spin-statistics theorem, we expect that spin-one-half
particles must be {\it fermions}, described by fields that {\it anticommute}.
Therefore, we should have $\psi_b\psi_a=-\psi_a\psi_b$ rather than
$\psi_b\psi_a=+\psi_a\psi_b$.  Then $\psi\psi$ does not vanish, and we can
use it as a term in $\L$.

Next we need a term with derivatives.  
The obvious choice is $\d^\mu\psi\d_\mu\psi$,
plus its hermitian conjugate.  This, however, 
yields a hamiltonian that is unbounded below, which is unacceptable.  
To get a bounded hamiltonian, the kinetic term must involve
both $\psi$ and $\psi^\dagger$.  A candidate is
$i\psi^\dagger\bar\sigma^\mu\d_\mu\psi$.  This not hermitian, but
\begin{eqnarray}
(i\psi^\dagger\bar\sigma^\mu\d_\mu\psi)^\dagger
&=& (i\psi_\adot^\dagger\,\bar\sigma^{\mu\adot c}\d_\mu\psi_c)^\dagger
\nonumber \\
\noalign{\smallskip}
&=& -i\d_\mu\psi_\ccdot^\dagger\,(\bar\sigma^{\mu a\ccdot })^*\psi_a
\nonumber \\
\noalign{\smallskip}
&=& -i\d_\mu\psi_\ccdot^\dagger\,\bar\sigma^{\mu \ccdot a}\psi_a
\nonumber \\
\noalign{\smallskip}
&=& i\psi_\ccdot^\dagger\,\bar\sigma^{\mu \ccdot a}\d_\mu\psi_a
   -i\d_\mu(\psi_\ccdot^\dagger\,\bar\sigma^{\mu \ccdot a}\psi_a).
\nonumber \\
\noalign{\smallskip}
&=& i\psid\bar\sigma^\mu\d_\mu\psi
   -i\d_\mu(\psid\bar\sigma^\mu\psi) \;.
\label{kinpsidag}
\end{eqnarray}
In the third line, we used the hermiticity of the matrices 
$\bar\sigma^\mu=(I,-\vec\sigma)$. 
In the fourth line, we used $A\d B=-(\d A)B+\d(AB)$.
In the last line, the second term is a total divergence,
and vanishes (with suitable boundary conditions on the fields
at infinity) when we integrate it over $\dfx$ to get the action $S$.
Thus $i\psi^\dagger\bar\sigma^\mu\d_\mu\psi$ has the 
hermiticity properties necessary for a term in $\L$.

Our complete lagrangian for $\psi$ is then
\begin{equation}
\L = i\psi^\dagger\bar\sigma^\mu\d_\mu\psi - \half m\psi\psi
     -\half m^*\psid\psid \;,
\label{ellpsi}
\end{equation}
where $m$ is a complex parameter with dimensions of mass.
The phase of $m$ is actually irrelevant: if $m=|m|e^{i\alpha}$,
we can set $\psi=e^{-i\alpha/2}\,\tilde\psi$ in \eq{ellpsi}; 
then we get a lagrangian for $\tilde\psi$ that is identical to
\eq{ellpsi}, but with $m$ replaced by $|m|$.  So we can,
without loss of generality, take $m$ to be real and positive
in the first place, and that is what we will do, setting $m^*=m$
in \eq{ellpsi}.

The equation of motion for $\psi$ is then
\begin{equation}
0=-{\delta S\over\delta\psi^\dagger} = 
-i\bar\sigma^\mu\d_\mu\psi + m\psid \;,
\label{eqmpsi}
\end{equation}
Restoring the spinor indices, this reads
\begin{equation}
0=-i\bar\sigma^{\mu\adot c}\d_\mu\psi_c + m\psi^{\dagger\adot} \;.
\label{eqmpsi2}
\end{equation}
Taking the hermitian conjugate (or, equivalently, computing
$-\delta S/\delta\psi$), we get
\begin{eqnarray}
0 &=& +i(\bar\sigma^{\mu a\ccdot})^*\,\d_\mu\psid_\ccdot + m\psi^a 
\nonumber \\
\noalign{\smallskip}
  &=& +i\bar\sigma^{\mu \ccdot a}\d_\mu\psid_\ccdot + m\psi^a 
\nonumber \\
\noalign{\smallskip}
  &=& -i\sigma^\mu_{a \ccdot}\,\d_\mu\psi^{\dagger\ccdot} + m\psi_a  \;.
\label{eqmpsi3}
\end{eqnarray}
In the second line, we used the hermiticity of the matrices 
$\bar\sigma^\mu=(I,-\vec\sigma)$. 
In the third, we lowered the undotted index, and switched
${}^\ccdot{\,}_\ccdot$ to ${}_\ccdot{\,}^\ccdot$, 
which gives an extra minus sign.

\Eqs{eqmpsi3} and (\ref{eqmpsi2}) can be combined to read 
\begin{equation}
\pmatrix{m\delta_a{}^c & -i\sigma^\mu_{a \ccdot}\,\d_\mu \cr
\noalign{\medskip}
-i\bar\sigma^{\mu\adot c}\,\d_\mu & m\delta^\adot{}_\ccdot \cr}
\pmatrix{ \psi_c \cr 
\noalign{\medskip}
\psi^{\dagger\ccdot} \cr} = 0\;.
\label{dirac1}
\end{equation}
We can write this more compactly by introducing the $4\times4$ 
{\it gamma matrices\/}
\begin{equation}
\gamma^\mu \equiv
\pmatrix{0 & \sigma^\mu_{a \ccdot} \cr
\noalign{\medskip}
\bar\sigma^{\mu\adot c} & 0 \cr} .
\label{gamma}
\end{equation}
Using the sigma-matrix relations,
\begin{eqnarray}
(\sigma^\mu\bar\sigma^\nu + 
\sigma^\nu\bar\sigma^\mu)_a{}^c
&=& -2g^{\mu\nu}\delta_a{}^c  \;,
\nonumber \\
\noalign{\smallskip}
(\bar\sigma^\mu\sigma^\nu + 
\bar\sigma^\nu\sigma^\mu)^\adot{}_\ccdot 
&=& -2g^{\mu\nu}\delta^\adot{}_\ccdot \;,
\label{anticommsigma}
\end{eqnarray}
which are most easily derived from the numerical formulae
$\sigma^\mu_{a\adot}=(I,\vec\sigma)$ and 
$\bar\sigma^{\mu\adot a}=(I,-\vec\sigma)$, we see that the
gamma matrices obey
\begin{equation}
\{\gamma^\mu,\gamma^\nu\}=-2g^{\mu\nu}\;,
\label{anticommgamma}
\end{equation}
where $\{A,B\}\equiv AB+BA$ denotes the anticommutator,
and there is an understood $4\times4$ identity matrix on the 
right-hand side.
We also introduce a four-component {\it Majorana field}
\begin{equation}
\Psi \equiv
\pmatrix{ \psi_c \cr 
\noalign{\medskip}
\psi^{\dagger\ccdot} \cr} \;.
\label{maj}
\end{equation}
Then \eq{dirac1} becomes
\begin{equation}
(-i\gamma^\mu\d_\mu + m)\Psi = 0\;.
\label{dirac2}
\end{equation}
This is the {\it Dirac equation}.  We first encountered it in 
section 1, where the gamma matrices were given different names
($\beta=\gamma^0$ and $\alpha^k = \gamma^0\gamma^k$).
Also, in section 1 we were trying (and failing) to interpret
$\Psi$ as a wave function, rather than as a quantum field.

Now consider a theory of two left-handed spinor fields
with an SO(2) symmetry, 
\begin{equation}
\L = i\psi_i^\dagger\bar\sigma^\mu\d_\mu\psi_i - \half m\psi_i\psi_i
     -\half m\psid_i\psid_i \;,
\label{ellpsi2}
\end{equation} 
where the spinor indices are suppressed and $i=1,2$ 
is implicitly summed.   As in the analogous case of two
scalar fields discussed in sections 22 and 23, this lagrangian
is invariant under the SO(2) transformation
\begin{equation}
\pmatrix{ 
\psi_1 \cr 
\noalign{\medskip}
\psi_2 \cr}
\to
\pmatrix{ \phantom{-}\cos\alpha & \sin\alpha \cr
\noalign{\medskip}
           {-}\sin\alpha & \cos\alpha \cr}
\pmatrix{ 
\psi_1 \cr 
\noalign{\medskip}
\psi_2 \cr}.
\label{psi1psi2}
\end{equation}
We can write the lagrangian so that the SO(2) symmetry
appears as a U(1) symmetry instead; let
\begin{eqnarray}
\chi &=& {\ts{1\over\sqrt2}}(\psi_1 + i\psi_2) \;,
\label{chipsi3} \\
\noalign{\smallskip}
\xi &=& {\ts{1\over\sqrt2}}(\psi_1 - i\psi_2) \;.
\label{chixidef}
\end{eqnarray}
In terms of these fields, we have
\begin{equation}
\L = i\chi^\dagger\bar\sigma^\mu\d_\mu\chi 
+i\xi^\dagger\bar\sigma^\mu\d_\mu\xi
- m\chi\xi - m\xi^\dagger\chi^\dagger \;.
\label{ellchixi}
\end{equation} 
\Eq{ellchixi} is invariant under the U(1) version of \eq{psi1psi2},
\begin{eqnarray}
\chi &\to& e^{-i\alpha}\chi \;,
\nonumber \\
\noalign{\smallskip}
\xi &\to& e^{+i\alpha}\xi \;,
\nonumber \\
\noalign{\smallskip}
\chi^\dagger &\to& e^{+i\alpha}\chi^\dagger \;,
\nonumber \\
\noalign{\smallskip}
\xi^\dagger &\to& e^{-i\alpha}\xi^\dagger \;.
\label{chixiu1}
\end{eqnarray}

Next, let us derive the equations of motion that we get from
\eq{ellchixi}, following the same procedure that ultimately led to
\eq{dirac1}.  The result is
\begin{equation}
\pmatrix{m\delta_a{}^c & -i\sigma^\mu_{a \ccdot}\,\d_\mu \cr
\noalign{\medskip}
-i\bar\sigma^{\mu\adot c}\,\d_\mu & m\delta^\adot{}_\ccdot \cr}
\pmatrix{ \chi_c \cr 
\noalign{\medskip}
\xi^{\dagger\ccdot} \cr} = 0\;.
\label{dirac3}
\end{equation}
We can now define a four-component {\it Dirac field}
\begin{equation}
\Psi \equiv
\pmatrix{ \chi_c \cr 
\noalign{\medskip}
\xi^{\dagger\ccdot} \cr} \;,
\label{dir}
\end{equation}
which obeys the {\it Dirac equation}, \eq{dirac2}.
(We have annoyingly used the same symbol $\Psi$
to denote both a Majorana field and a Dirac field; these are
different objects, and so we must always announce
which is meant when we write $\Psi$.) 

We can also write the lagrangian, \eq{ellchixi}, in terms of
the Dirac field $\Psi$, \eq{dir}.  First we take the hermitian
conjugate of $\Psi$ to get
\begin{equation}
\Psi^\dagger =
(\chi^\dagger_\adot\,,\; \xi^a) \;.
\label{dirdag}
\end{equation}
Introduce the matrix
\begin{equation}
\beta \equiv
\pmatrix{ 0 & \delta^\adot{}_\ccdot \cr
\noalign{\medskip}
\delta_a{}^c & 0 \cr} \;.
\label{beta}
\end{equation}
Numerically, $\beta=\gamma^0$.  However, the spinor index
structure of $\beta$ and $\gamma^0$ is different, and so we will
distinguish them.  Given $\beta$, we define
\begin{equation}
\Psibar \equiv \Psi^\dagger\beta =
(\xi^a,\, \chi^\dagger_\adot) \;.
\label{dirbar}
\end{equation}
Then we have
\begin{equation}
\Psibar\Psi = \xi^a\chi_a + \chi^\dagger_\adot \xi^{\dagger\adot}\;.
\label{psibarpsi}
\end{equation}
Also,
\begin{equation}
\Psibar\gamma^\mu\d_\mu\Psi 
= \xi^a\sigma^\mu_{a\ccdot}\,\d_\mu\xi^{\dagger\ccdot} 
  + \chi^\dagger_\adot\,\bar\sigma^{\mu\adot c}\,\d_\mu\chi_c \;.
\label{psibargpsi}
\end{equation}
Using $A\d B=-(\d A)B+\d(AB)$, 
the first term on the right-hand side of \eq{psibargpsi} can be rewritten as
\begin{equation}
\xi^a\sigma^\mu_{a\ccdot}\,\d_\mu\xi^{\dagger\ccdot} 
= -(\d_\mu\xi^a)\sigma^\mu_{a\ccdot}\,\xi^{\dagger\ccdot} 
+ \d_\mu(\xi^a\sigma^\mu_{a\ccdot}\,\xi^{\dagger\ccdot}) \;.
\label{psibargpsi2}
\end{equation}
The first term on the right-hand side of \eq{psibargpsi2} can be rewritten as
\begin{equation}
-(\d_\mu\xi^a)\sigma^\mu_{a\ccdot}\,\xi^{\dagger\ccdot} 
= +\xi^{\dagger\ccdot}\sigma^\mu_{a\ccdot}\,\d_\mu\xi^a
= +\xi^\dagger_\ccdot\,\bar\sigma^{\mu\ccdot a}\d_\mu\xi_a \;.
\label{psibargpsi3}
\end{equation}
Here we used anticommutation of the fields to get the first equality,
and switched ${}^\ccdot{\,}_\ccdot$ to ${}_\ccdot{\,}^\ccdot$ 
and ${}_a{\,}^a$ to ${}^a{\,}_a$ (thus generating two minus signs)
to get the second.
Combining eqs.$\,$(\ref{psibargpsi}--\ref{psibargpsi3}), we get
\begin{equation}
\Psibar\gamma^\mu\d_\mu\Psi 
= \chi^\dagger\bar\sigma^\mu\d_\mu\chi 
+ \xi^\dagger\bar\sigma^\mu\d_\mu\xi 
+ \d_\mu(\xi\sigma^\mu\xi^\dagger) \;.
\label{psibargpsi4}
\end{equation}
Therefore, up to an irrelevant total divergence, we have
\begin{equation}
\L = i\Psibar\gamma^\mu\d_\mu\Psi - m\Psibar\Psi \;.
\label{elldir}
\end{equation}
This form of the lagrangian is invariant under the 
U(1) transformation
\begin{eqnarray}
\Psi &\to& e^{-i\alpha}\,\Psi \;,
\nonumber \\
\noalign{\smallskip}
\Psibar &\to& e^{+i\alpha}\,\Psibar \;,
\label{diru1}
\end{eqnarray}
which, given \eq{dir}, is the same as \eq{chixiu1}. 
The Noether current associated with this symmetry is
\begin{equation}
j^\mu = \Psibar\gamma^\mu\Psi
= \chi^\dagger\bar\sigma^\mu\chi 
- \xi^\dagger\bar\sigma^\mu\xi \;.
\label{noetherdirac}
\end{equation}
In quantum electrodynamics, the electromagnetic current
is $e\Psibar\gamma^\mu\Psi$, where $e$ is the charge
of the electron.

As in the case of a complex scalar field with a U(1) symmetry,
there is an additional discrete symmetry, called
{\it charge conjugation}, that enlarges SO(2) to O(2).
Charge conjugation simply exchanges $\chi$ and $\xi$.
We can define a unitary {\it charge conjugation operator\/} 
$C$ that implements this,
\begin{eqnarray}
C^{-1}\chi_a(x)C &=& \xi_a(x) \;,
\nonumber \\
\noalign{\smallskip}
C^{-1}\xi_a(x)C &=& \chi_a(x) \;,
\label{cchic}
\end{eqnarray}
where, for the sake of precision, we have restored the spinor index and 
spacetime argument.  We then have $C^{-1}\L(x)C=\L(x)$.

To express \eq{cchic} in terms of the Dirac field, \eq{dir},
we first introduce the {\it charge conjugation matrix\/}
\begin{equation}
\C\equiv \pmatrix{\e_{ac} & 0 \cr 
            \noalign{\medskip}
                   0 & \e^{\dot a\dot c} \cr }.
\label{cee0}
\end{equation}
Next we notice that, if we take the transpose of $\Psibar$,
\eq{dirbar}, we get
\begin{equation}
\Psibart =
\pmatrix{\xi^a \cr \noalign{\smallskip} \chi^\dagger_\adot\cr} \;.
\label{dirbart}
\end{equation}
Then, if we multiply by $\C$, we get a field that we will call
$\Psi^{\rm\sss C}$, the {\it charge conjugate\/} of $\Psi$,
\begin{equation}
\Psi^{\rm\sss C} \equiv \C\Psibart =
\pmatrix{\xi_a \cr \noalign{\smallskip} \chi^{\dagger\adot}\cr} \;.
\label{cdirbart}
\end{equation}
We see that $\Psi^{\rm\sss C}$
is the same as the original field $\Psi$, \eq{dir},
except that the roles of $\chi$ and $\xi$ have been switched.
 
The charge conjugation matrix has a number of useful properties.
As a numerical matrix, it obeys
\begin{equation}
\C^{\rm\sss T}=\C^\dagger=\C^{-1}=-\C \;,
\label{cees}
\end{equation}
and we can also write it as
\begin{equation}
\C = \pmatrix{-\e^{ac} & 0 \cr 
\noalign{\medskip}
              0 & -\e_{\dot a\dot c} \cr }.
\label{cee2}
\end{equation}
A result that we will need later is
\begin{eqnarray}
\C^{-1}\gamma^\mu\C &=&
\pmatrix{\e^{ab} & 0 \cr 
\noalign{\medskip}
         0 & \e_{\dot a\dot b} \cr }
\pmatrix{0 & \sigma^\mu_{b \dot c} \cr
\noalign{\medskip}
\bar\sigma^{\mu\bdot c} & 0 \cr} 
\pmatrix{\e_{ce} & 0 \cr 
\noalign{\medskip}
         0 & \e^{\dot c\dot e} \cr }
\nonumber \\
\noalign{\medskip}
&=&
\pmatrix{0 & \e^{ab} \sigma^\mu_{b\ccdot} \e^{\ccdot\edot} \cr
\noalign{\medskip}
\e_{\adot\bdot} \bar\sigma^{\mu\bdot c} \e_{ce} & 0 \cr} 
\nonumber \\
\noalign{\medskip}
&=&
\pmatrix{0 & -\bar\sigma^{\mu a \dot e} \cr
\noalign{\medskip}
-\sigma^\mu_{\adot e} & 0 \cr} .
\label{cgcg}
\end{eqnarray}
The minus signs in the last line come from raising or lowering
an index by contracting with the first (rather than the second)
index of an $\e$ symbol.  Comparing with 
\begin{equation}
\gamma^\mu =
\pmatrix{0 & \sigma^\mu_{e \adot} \cr
\noalign{\medskip}
\bar\sigma^{\mu\edot a} & 0 \cr} ,
\label{gamma2}
\end{equation}
we see that
\begin{equation}
\C^{-1}\gamma^\mu\C = -(\g^\mu)^{\rm\sss T}\;.
\label{cgcg2}
\end{equation}

Now let us return to the Majorana field, \eq{maj}.
It is obvious that a Majorana field is its own charge conjugate,
that is, $\Psi^{\rm\sss C}=\Psi$.  This condition is analogous
to the condition $\ph^\dagger=\ph$ that is satisfied by a real
scalar field.  A Dirac field, with its U(1) symmetry, is analogous
to a complex scalar field, while a Majorana field, which has no
U(1) symmetry, is analogous to a real scalar field.

We can write the lagrangian for a single left-handed spinor field,
\eq{ellpsi}, in terms of a Majorana field, \eq{maj}, by retracing
eqs.$\,$(\ref{dirdag}--\ref{elldir}) with $\chi\to\psi$ and $\xi\to\psi$.
The result is
\begin{equation}
\L ={\ts{i\over2}}\Psibar\gamma^\mu\d_\mu\Psi - \half m\Psibar\Psi \;.
\label{ellmaj}
\end{equation}
However, this expression is not yet useful for deriving the equations
of motion, because it does not yet incorporate 
the Majorana condition $\Psi^{\rm\sss C}=\Psi$.  To remedy this,
we use \eq{cees} to write the Majorana condition $\Psi=\C\Psibart$ 
as $\Psibar = \Psit\C$.
Then we can replace $\Psibar$ in \eq{ellmaj} by 
$\Psit\C$ to get
\begin{equation}
\L = {\ts{i\over2}}\Psit\C\gamma^\mu\d_\mu\Psi 
- \half m\Psit\C\Psi \;.
\label{ellmaj2}
\end{equation}
The equation of motion that follows from this lagrangian is
once again the Dirac equation.

We can also recover the Weyl components of a Dirac or Majorana field
by means of a suitable projection matrix.  Define
\begin{equation}
\gamma_5 \equiv \pmatrix{-\delta_a{}^c & 0 \cr
\noalign{\medskip}
                         0 & +\delta^\adot{}_\ccdot \cr},
\label{gamma5}
\end{equation}
where the subscript 5 is simply part of the traditional name of this matrix,
rather than the value of some index.
Then we can define left and right projection matrices
\begin{eqnarray}
P_{\rm\sss L} &\equiv& \half(1-\g_5) 
= \pmatrix{\delta_a{}^c & 0 \cr
\noalign{\medskip}
           0 & 0 \cr},
\nonumber \\
\noalign{\smallskip}
P_{\rm\sss R} &\equiv& \half(1+\g_5)
= \pmatrix{0 & 0 \cr
\noalign{\medskip}
           0 & \delta^\adot{}_\ccdot \cr}.
\label{plpr}
\end{eqnarray}
Thus we have, for a Dirac field,
\begin{eqnarray}
P_{\rm\sss L}\Psi
&=& \pmatrix{\chi_c \cr
\noalign{\medskip}
             0  \cr},
\nonumber \\
\noalign{\medskip}
P_{\rm\sss R}\Psi
&=& \pmatrix{0 \cr
\noalign{\smallskip}
             \xi^{\dagger\ccdot}  \cr}.
\label{plprPsi}
\end{eqnarray}
The matrix $\g_5$ can also be expressed as
\begin{eqnarray}
\gamma_5 &=& i\g^0\g^1\g^2\g^3
\nonumber \\
&=& -{\ts{i\over24}}\e_{\mu\nu\rho\sigma}\g^\mu\g^\nu\g^\rho\g^\sigma \;,
\label{g500}
\end{eqnarray}
where $\e_{0123}=-1$.  

Finally, let us consider the behavior of a Dirac or Majorana field
under a Lorentz transformation.  Recall that left- and right-handed
spinor fields transform according to
\begin{eqnarray}
U(\Lam)^{-1}\psi_a(x)U(\Lam) &=& L(\Lam)_a{}^c\,\psi_c(\Lam^{-1}x) \;,
\label{uleftu} \\
\noalign{\medskip}
U(\Lam)^{-1}\psi^\dagger_\adot(x)U(\Lam) 
&=& R(\Lam)_\adot{}^\ccdot\,\psi^\dagger_\ccdot(\Lam^{-1}x) \;,
\label{urightu}
\end{eqnarray}
where, for an infinitesimal transformation 
$\Lam^\mu{}_\nu=\delta^\mu{}_\nu + \delta\w^\mu{}_\nu$,
\begin{eqnarray}
L(1{+}\delta\w)_a{}^c &=& \delta_a{}^c 
+ {\ts{i\over2}}\delta\w_{\mu\nu}(S^{\mu\nu}_{\rm\sss L})_a{}^c \;,
\label{dwleft} \\
\noalign{\medskip}
R(1{+}\delta\w)_\adot{}^\ccdot &=& 
\delta_\adot{}^\ccdot  
+ {\ts{i\over2}}\delta\w_{\mu\nu}
(S^{\mu\nu}_{\rm\sss R})_\adot{}^\ccdot \;,
\label{dwright}
\end{eqnarray}
and where
\begin{eqnarray}
(S_{\rm\sss L}^{\mu\nu})_a{}^c &=&
+{\ts{i\over 4}}(\sigma^\mu\bar\sigma^\nu
                     -\sigma^\nu\bar\sigma^\mu)_a{}^c \;,
\label{sleft} \\
\noalign{\medskip}
(S_{\rm\sss R}^{\mu\nu})^\adot{}_\ccdot &=&
-{\ts{i\over 4}}(\bar\sigma^\mu\sigma^\nu
                     -\bar\sigma^\nu\sigma^\mu)^\adot{}_\ccdot\;.
\label{sright}
\end{eqnarray}
From these formulae, and the definition of $\gamma^\mu$, \eq{gamma},
we can see that
\begin{equation}
{\ts{i\over 4}}[\gamma^\mu,\gamma^\nu] = 
\pmatrix{+(S_{\rm\sss L}^{\mu\nu})_a{}^c & 0 \cr
\noalign{\medskip}
0 & -(S_{\rm\sss R}^{\mu\nu})^\adot{}_\ccdot \cr} 
\equiv S^{\mu\nu} \;.
\label{dispin}
\end{equation}
Then, for either a Dirac or Majorana field $\Psi$, we can write
\begin{equation}
U(\Lambda)^{-1}\Psi(x)U(\Lambda) =  D(\Lambda)\Psi(\Lambda^{-1}x)\;,
\label{Psilam0}
\end{equation}
where, for an infinitesimal transformation, 
the $4\times4$ matrix $D(\Lam)$ is 
\begin{equation}
D(1{+}\delta\omega)=1 + {\ts{i\over2}}\delta\omega_{\mu\nu} S^{\mu\nu}\;,
\label{dlam0}
\end{equation}
with $S^{\mu\nu}$ given by \eq{dispin}.
The minus sign in front of $S^{\mu\nu}_{\rm\sss R}$ in \eq{dispin}
is compensated by the switch from a 
${}^\ccdot{\,}_\ccdot$ contraction in \eq{dwright} to a 
${}_\ccdot{\,}^\ccdot$ contraction in \eq{Psilam0}.

\vskip0.5in

\begin{center}
Problems
\end{center}

\vskip0.25in

36.1) 
Using the results of problem 2.8, show that, for a rotation 
by an angle $\theta$ about the $z$ axis, we have
\begin{equation}
D(\Lam) = \exp(-i\theta S^{12}) \;,
\label{lamrot35}
\end{equation}
and that, for a boost by rapidity $\eta$ in the $z$ direction, we have
\begin{equation}
D(\Lam) = \exp(+i\eta S^{30}) \;.
\label{lamboost35}
\end{equation}

36.2) 
Show that $\overline{D(\Lam)}\gamma^\mu D(\Lam)=\Lam^\mu{}_\nu\gamma^\nu$.

\vfill\eject

\noindent Quantum Field Theory  \hfill   Mark Srednicki

\vskip0.5in

\begin{center}
\large{37: Canonical Quantization of Spinor Fields I}
\end{center}
\begin{center}
Prerequisite: 36
\end{center}

\vskip0.5in

Consider a left-handed Weyl field $\psi$ with lagrangian
\begin{equation}
\L = i\psid\bar\sigma^\mu\d_\mu\psi 
     - \half m(\psi\psi+\psid\psid) \;.
\label{ellpsi35}
\end{equation}
The canonically conjugate momentum to the field $\psi_a(x)$ is then
\begin{eqnarray}
\pi^a(x) &\equiv& {\d\L\over\d(\d_0\psi_a(x))}
\nonumber \\
\noalign{\medskip}
  &=& i\psid_\adot(x)\bar\sigma^{0\adot a} \;.
\label{piconj}
\end{eqnarray}
[Here we have glossed over a subtlety about differentiating with
respect to an anticommuting object; we will take up this topic
in section 44, and here simply assume that \eq{piconj} is correct.]
The hamiltonian is
\begin{eqnarray}
\H &=& \pi^a\d_0\psi_a - \L
\nonumber \\
\noalign{\medskip}
   &=& i\psid_\adot\bar\sigma^{0\adot a}\psi_a - \L
\nonumber \\
\noalign{\medskip}
   &=& -i\psid\bar\sigma^i\d_i\psi +
       \half m(\psi\psi+\psid\psid) \;.
\label{psih}
\end{eqnarray}
The appropriate canonical {\it anticommutation\/} relations are
\begin{eqnarray}
\{\psi_a(\x,t),\psi_c(\y,t)\} &=& 0 \;, 
\nonumber \\
\noalign{\medskip}
\{\psi_a(\x,t),\pi^c(\y,t)\} &=& i\delta_a{}^c\,\delta^3(\x-\y)\;.
\label{psicomm}
\end{eqnarray}
Substituting in \eq{piconj} for $\pi^c$, we get
\begin{equation}
\{\psi_a(\x,t),\psid_\ccdot(\y,t)\}\bar\sigma^{0\ccdot c} 
= \delta_a{}^c\,\delta^3(\x-\y)\;.
\label{psicomm2}
\end{equation}
Then, using $\bar\sigma^0=\sigma^0=I$, we have
\begin{equation}
\{\psi_a(\x,t),\psid_\ccdot(\y,t)\}
= \sigma^0_{a\ccdot}\,\delta^3(\x-\y)\;,
\label{psicomm3}
\end{equation}
or, equivalently,
\begin{equation}
\{\psi^a(\x,t),\psi^{\dagger\ccdot}(\y,t)\}
= \bar\sigma^{0\ccdot a}\,\delta^3(\x-\y)\;.
\label{psicomm4}
\end{equation}

We can also translate this into four-component notation for either
a Dirac or a Majorana field.  A Dirac field is defined in terms of
two left-handed Weyl fields $\chi$ and $\xi$ via
\begin{equation}
\Psi \equiv
\pmatrix{ \chi_c \cr
\noalign{\medskip}
\xi^{\dagger\ccdot} \cr} \;.
\label{dir2}
\end{equation}
We also define 
\begin{equation}
\Psibar \equiv \Psi^\dagger\beta =
(\xi^a,\, \chi^\dagger_\adot) \;,
\label{dirbar2}
\end{equation}
where
\begin{equation}
\beta \equiv
\pmatrix{ 0 & \delta^\adot{}_\ccdot \cr
\noalign{\medskip}
\delta_a{}^c & 0 \cr} \;.
\label{beta3}
\end{equation}
The lagrangian is
\begin{eqnarray}
\L &=& i\chi^\dagger\bar\sigma^\mu\d_\mu\chi
      +i\xi^\dagger\bar\sigma^\mu\d_\mu\xi
     - m(\chi\xi + \xi^\dagger\chi^\dagger)
\nonumber \\
\noalign{\medskip}
&=& i\Psibar\gamma^\mu\d_\mu\Psi-m\Psibar\Psi\;.
\label{ellPsi}
\end{eqnarray}
The fields $\chi$ and $\xi$ each obey the canonical anticommutation
relations of \eq{psicomm}.  This translates into
\begin{eqnarray}
\{\Psi_\alpha(\x,t),\Psi_\beta(\y,t)\} &=& 0\;,
\label{Psicomm1} \\
\noalign{\medskip}
\{\Psi_\alpha(\x,t),\Psibar_\beta(\y,t)\}
&=& (\gamma^0)_{\alpha\beta}\,\delta^3(\x-\y)\;,
\label{Psicomm2}
\end{eqnarray}
where $\alpha$ and $\beta$ are four-component spinor indices, and
\begin{equation}
\gamma^\mu \equiv
\pmatrix{0 & \sigma^\mu_{a \ccdot} \cr
\noalign{\medskip}
\bar\sigma^{\mu\adot c} & 0 \cr} \;.
\label{gamma0}
\end{equation}
\Eqs{Psicomm1} and (\ref{Psicomm2})
can also be derived directly from the four-component form
of the lagrangian, \eq{ellPsi}, by noting that 
the canonically conjugate momentum to the field $\Psi$ is 
$\d\L/\d(\d_0\Psi) = i\Psibar\gamma^0$, and that $(\g^0)^2=1$.

A Majorana field is defined in terms of a single
left-handed Weyl field $\psi$ via
\begin{equation}
\Psi \equiv
\pmatrix{ \psi_c \cr
\noalign{\medskip}
\psi^{\dagger\ccdot} \cr} \;.
\label{maj2}
\end{equation}
We also define 
\begin{equation}
\Psibar \equiv \Psi^\dagger\beta =
(\psi^a,\, \psi^\dagger_\adot) \;.
\label{majbar2}
\end{equation}
A Majorana field obeys the Majorana condition
\begin{equation}
\Psibar=\Psit\C\;, 
\label{Majcond}
\end{equation}
where
\begin{equation}
\C\equiv \pmatrix{-\e^{ac} & 0 \cr
            \noalign{\medskip}
                   0 & -\e_{\adot\ccdot} \cr }
\label{cee1}
\end{equation}
is the charge conjugation matrix.  The lagrangian is 
\begin{eqnarray}
\L &=& i\psid\bar\sigma^\mu\d_\mu\psi 
     - \half m(\psi\psi+\psid\psid) 
\nonumber \\
\noalign{\medskip}
&=& {\ts{i\over2}}\Psibar\gamma^\mu\d_\mu\Psi - \half m\Psibar\Psi 
\nonumber \\
\noalign{\medskip}
&=& {\ts{i\over2}}\Psit\C\gamma^\mu\d_\mu\Psi
- \half m\Psit\C\Psi \;.
\label{ellMaj}
\end{eqnarray}
The field $\psi$ obeys the canonical anticommutation
relations of \eq{psicomm}.  This translates into
\begin{eqnarray}
\{\Psi_\alpha(\x,t),\Psi_\beta(\y,t)\}
&=& (\C\gamma^0)_{\alpha\beta}\,\delta^3(\x-\y)\;,
\label{Majcomm1} \\
\noalign{\medskip}
\{\Psi_\alpha(\x,t),\Psibar_\beta(\y,t)\}
&=& (\gamma^0)_{\alpha\beta}\,\delta^3(\x-\y)\;,
\label{Majcomm2}
\end{eqnarray}
where $\alpha$ and $\beta$ are four-component spinor indices.
To derive \eqs{Majcomm1} and (\ref{Majcomm2})
directly from the four-component form of the lagrangian, \eq{ellMaj},
requires new formalism for the quantization of {\it constrained systems}.
This is because the canonically conjugate momentum to the field 
$\Psi$ is $\d\L/\d(\d_0\Psi) = {i\over2}\Psit\C\gamma^0$, 
and this is linearly related to $\Psi$ itself; this relation
constitutes a constraint that must be solved before imposition
of the anticommutation relations.  In this case, solving the constraint
simply returns us to the Weyl formalism with which we began.

The equation of motion that follows from either \eq{ellPsi}
or \eq{ellMaj} is the Dirac equation,
\begin{equation}
(-i\dsl + m)\Psi=0 \;.
\label{deq}
\end{equation}
Here we have introduced the {\it Feynman slash\/}:
given any four-vector $a^\mu$, we define
\begin{equation}
\asl \equiv a_\mu\gamma^\mu \;.
\label{sl}
\end{equation}

To solve the Dirac equation, we first note that if we act on it
with $i\dsl+m$, we get
\begin{eqnarray}
0 &=& (i\dsl+m)(-i\dsl + m)\Psi
\nonumber \\
  &=& (\dsl\dsl + m^2)\Psi
\nonumber \\
  &=& (-\d^2 + m^2)\Psi \;.
\label{dkg}
\end{eqnarray}
Here we have used
\begin{eqnarray}
\asl\asl &=& a_\mu a_\nu\gamma^\mu\gamma^\nu 
\nonumber \\
&=& a_\mu a_\nu\Bigl(\half\{\gamma^\mu,\gamma^\nu\} + 
                     \half [\gamma^\mu,\gamma^\nu]\Bigr)
\nonumber \\
&=& a_\mu a_\nu\Bigl(-g^{\mu\nu} + 
                     \half [\gamma^\mu,\gamma^\nu]\Bigr)
\nonumber \\
&=& -a_\mu a_\nu g^{\mu\nu} + 0 
\nonumber \\
&=& -a^2 \;.
\label{aslasl}
\end{eqnarray}
From \eq{dkg}, we see that $\Psi$ obeys the Klein-Gordon equation.
Therefore, the Dirac equation has plane-wave solutions.
Let us consider a specific solution of the form
\begin{equation}
\Psi(x) = u(\p)e^{ipx} + v(\p)e^{-ipx} \;.
\label{puv}
\end{equation}
where $p^0=\w\equiv(\p^2+m^2)^{1/2}$, and 
$u(\p)$ and $v(\p)$ are four-component constant spinors.
Plugging \eq{puv} into the \eq{deq}, we get
\begin{equation}
(\psl + m)u(\p)e^{ipx} + (-\psl + m)v(\p)e^{-ipx} = 0\;.
\label{deqpuv}
\end{equation}
Thus we require
\begin{eqnarray}
(\psl+m)u(\p) &=& 0 \;, 
\nonumber \\
(-\psl+m)v(\p) &=& 0 \;.
\label{equv0}
\end{eqnarray}
Each of these equations has two linearly independent solutions
that we will call $u_\pm(\p)$ and $v_\pm(\p)$;
their detailed properties will be worked out
in the next section.
The general solution of the Dirac equation can then be written as
\begin{equation}
\Psi(x) = \sum_{s=\pm}\int\dpt\,\left[b_s(\p)  u_s(\p)e^{ipx}
                              +d_s^\dagger(\p)v_s(\p)e^{-ipx} \right]\;,
\label{Psi}
\end{equation}
where the integration measure is
\begin{equation}
\dpt \equiv {\dtp\over(2\pi)^3 2\w}\;.
\label{dp}
\end{equation}

\vfill\eject

\noindent Quantum Field Theory  \hfill   Mark Srednicki

\vskip0.5in

\begin{center}
\large{38: Spinor Technology}
\end{center}
\begin{center}
Prerequisite: 37
\end{center}

\vskip0.5in

The four-component spinors $u_s(\p)$ and $v_s(\p)$ obey the equations
\begin{eqnarray}
(\psl+m)u_s(\p) &=& 0 \;, 
\nonumber \\
(-\psl+m)v_s(\p) &=& 0 \;.
\label{equv}
\end{eqnarray}
Each of these equations has two solutions,
which we label via $s=+$ and $s=-$.
For $m\ne0$, we can go to the rest frame, $\p={\bf 0}$.  We will
then distinguish the two solutions by the eigenvalue of the spin matrix
\begin{equation}
S_z={\ts{i\over 4}}[\g^1,\g^2]={\ts{i\over 2}}\g^1\g^2
=\pmatrix{ \half\sigma_3 & 0 \cr 
\noalign{\medskip}
          0 & \half\sigma_3 \cr}. 
\label{sz}
\end{equation}
Specifically, we will require
\begin{eqnarray}
S_z u_\pm({\bf 0}) &=& \pm\half u_\pm({\bf 0})\;,
\nonumber \\
S_z v_\pm({\bf 0}) &=& \mp\half v_\pm({\bf 0})\;.
\label{sz2}
\end{eqnarray}
The reason for the opposite sign for the $v$ spinor is that
this choice results in
\begin{eqnarray}
[J_z, b^\dagger_\pm({\bf 0})] &=& \pm\half b^\dagger_\pm({\bf 0})\;,
\nonumber \\
{[}J_z, d^\dagger_\pm({\bf 0}){]} &=& \pm\half d^\dagger_\pm({\bf 0}) \;,
\label{jbd}
\end{eqnarray}
so that $b^\dagger_+({\bf 0})$ and $d^\dagger_+({\bf 0})$ each creates
a particle with spin up along the $z$ axis.

For $\p={\bf 0}$, we have $\psl = -m\g^0$, where
\begin{equation}
\g^0=\pmatrix{ 0 & I \cr 
\noalign{\medskip} 
               I & 0 \cr }.
\label{g00}
\end{equation}
\Eqs{equv} and (\ref{sz2}) are then easy to solve.
Choosing (for later convenience) a specific normalization
and phase for each of $u_\pm({\bf 0})$ and $v_\pm({\bf 0})$, we get
\begin{eqnarray}
u_+({\bf 0}) = \sqrt{m}\pmatrix{1 \cr 0 \cr 1 \cr 0},
&\qquad&
u_-({\bf 0}) = \sqrt{m}\pmatrix{0 \cr 1 \cr 0 \cr 1},
\nonumber \\
\noalign{\bigskip}
v_+({\bf 0}) = \sqrt{m}\pmatrix{0 \cr 1 \cr 0 \cr -1},
&\qquad&
v_-({\bf 0}) = \sqrt{m}\pmatrix{-1 \cr 0 \cr 1 \cr 0}.
\label{uv0}
\end{eqnarray}
For later use we also compute the barred spinors
\begin{eqnarray}
\ubar_s(\p) &\equiv& u_s^\dagger(\p)\beta \;,
\nonumber \\
\vbar_s(\p) &\equiv& v_s^\dagger(\p)\beta \;,
\label{uvbar00}
\end{eqnarray}
where
\begin{equation}
\beta=\g^0=\pmatrix{ 0 & I \cr 
\noalign{\medskip} 
                     I & 0 \cr }
\label{beta00}
\end{equation}
satisfies 
\begin{equation}
\beta^{\rm\sss T}=\beta^\dagger=\beta^{-1}=\beta \;.
\label{bi}
\end{equation}
We get
\begin{eqnarray}
\ubar_+({\bf 0}) &=& \sqrt{m}\;(1,\,0,\,1,\,0)\;,
\nonumber \\
\ubar_-({\bf 0}) &=& \sqrt{m}\;(0,\,1,\,0,\,1)\;,
\nonumber \\
\vbar_+({\bf 0}) &=& \sqrt{m}\;(0,\,-1,\,0,\,1)\;,
\nonumber \\
\vbar_-({\bf 0}) &=& \sqrt{m}\;(1,\,0,\,-1,\,0)\;.
\label{uvbar0}
\end{eqnarray}
We can now find the spinors corresponding to an arbitrary three-momentum
$\p$ by applying to $u_s({\bf 0})$ and $v_s({\bf 0})$ 
the matrix $D(\Lam)$ that corresponds to an appropriate boost. 
This is given by
\begin{equation}
D(\Lam) = \exp(i\eta\,\hat\p\cd {\bf K})\;,
\label{dlamboost}
\end{equation}
where $\hat\p$ is a unit vector in the $\p$ direction,
$K^j={i\over4}[\g^j,\g^0]={i\over2}\g^j\g^0$ is the boost matrix,
and $\eta\equiv\sinh^{-1}(|\p|/m)$ is the {\it rapidity\/} (see problem 2.8).
Thus we have
\begin{eqnarray}
u_s(\p) &=& \exp(i\eta\,\hat\p\cd {\bf K})u_s({\bf 0})\;,
\nonumber \\
v_s(\p) &=& \exp(i\eta\,\hat\p\cd {\bf K})v_s({\bf 0})\;.
\label{uvp}
\end{eqnarray}
We also have
\begin{eqnarray}
\ubar_s(\p) &=& 
\ubar_s({\bf 0})\exp(-i\eta\,\hat\p\cd {\bf K}) \;,
\nonumber \\
\vbar_s(\p) &=& 
\vbar_s({\bf 0})\exp(-i\eta\,\hat\p\cd {\bf K}) \;.
\label{uvbarp}
\end{eqnarray}
This follows from $\overline{K^j}=K^j$,
where for any general combination of gamma matrices,
\begin{equation}
\overline{A}\equiv \beta A^\dagger\beta \;.
\label{abar}
\end{equation}
In particular, it turns out that
\begin{eqnarray}
\overline{\hbox{$\g^\mu$}} &=& \g^\mu \;,
\nonumber \\
\overline{\hbox{$S^{\mu\nu}$}} &=& S^{\mu\nu} \;,
\nonumber \\
\overline{\hbox{$i\g_5$}} &=& i\g_5 \;,
\nonumber \\
\overline{\hbox{$\g^\mu\g_5$}} &=& \g^\mu\g_5 \;,
\nonumber \\
\overline{\hbox{$i\g_5 S^{\mu\nu}$}} &=& i\g_5 S^{\mu\nu} \;.
\label{gbar}
\end{eqnarray}
The barred spinors satisfy the equations
\begin{eqnarray}
\ubar_s(\p)(\psl+m) &=& 0 \;, 
\nonumber \\
\vbar_s(\p)(-\psl+m) &=& 0 \;.
\label{equvbar}
\end{eqnarray}

It is not very hard to work out
$u_s(\p)$ and $v_s(\p)$ from \eq{uvp}, but it is even easier
to use various tricks that will sidestep any need for the explicit formulae.
Consider, for example, 
$\ubar_{s'}(\p)u_s(\p)$; from \eqs{uvp} and (\ref{uvbarp}), we see that
$\ubar_{s'}(\p)u_s(\p) = \ubar_{s'}({\bf 0}) u_s({\bf 0})$,
and this is easy to compute from \eqs{uv0} and (\ref{uvbar0}).
We find
\begin{eqnarray}
\ubar_{s'}(\p)u_s(\p) &=& +2m\,\delta_{s's} \;,
\nonumber \\
\vbar_{s'}(\p)v_s(\p) &=& -2m\,\delta_{s's} \;,
\nonumber \\
\ubar_{s'}(\p)v_s(\p) &=& 0 \;,
\nonumber \\
\vbar_{s'}(\p)u_s(\p) &=& 0\;.
\label{uvnorm}
\end{eqnarray}
Also useful are the {\it Gordon identities},
\begin{eqnarray}
2m\,\ubar_{s'}(\p')\g^\mu u_s(\p)
&=& \ubar_{s'}(\p')
  \Bigl[(p'+p)^\mu - 2i S^{\mu\nu}(p'-p)_\nu \Bigr] u_s(\p) \;,
\nonumber \\
-2m\,\vbar_{s'}(\p')\g^\mu v_s(\p)
&=& \vbar_{s'}(\p')
  \Bigl[  (p'+p)^\mu - 2i S^{\mu\nu}(p'-p)_\nu \Bigr] v_s(\p) \;. \qquad
\label{gordon}
\end{eqnarray}
To derive them, start with
\begin{eqnarray}
\g^\mu\psl &=&
\half\{\g^\mu,\psl\} + \half [\g^\mu,\psl]
\nonumber \\
&=& -p^\mu-2iS^{\mu\nu}p_\nu \;.
\label{gmid}
\end{eqnarray}
Similarly,
\begin{eqnarray}
\psl{\,}'\g^\mu &=&
\half\{\g^\mu,\psl{\,}'\} - \half [\g^\mu,\psl{\,}']
\nonumber \\
&=& -p^{\prime\mu}+2iS^{\mu\nu}p'_\nu \;.
\label{gmid2}
\end{eqnarray}
Add \eqs{gmid} and (\ref{gmid2}), 
sandwich them between $\ubarp$ and $u$ spinors
(or $\vbarp$ and $v$ spinors), 
and use \eqs{equv} and (\ref{equvbar}). 
An important special case is $p'=p$; then, using \eq{uvnorm}, we find
\begin{eqnarray}
\ubar_{s'}(\p)\g^\mu u_s(\p) &=& 2p^\mu\delta_{s's} \;,
\nonumber \\
\vbar_{s'}(\p)\g^\mu v_s(\p) &=& 2p^\mu\delta_{s's} \;.
\label{gordonp}
\end{eqnarray}
With a little more effort, we can also show
\begin{eqnarray}
\ubar_{s'}(\p)\g^0 v_s(-\p) &=& 0 \;,
\nonumber \\
\vbar_{s'}(\p)\g^0 u_s(-\p) &=& 0 \;.
\label{gordonpuv}
\end{eqnarray}
We will need \eqs{gordonp} and (\ref{gordonpuv}) in the next section.

Consider now the spin sums $\sum_{s=\pm}u_s(\p)\ubar_s(\p)$ and
$\sum_{s=\pm}v_s(\p)\vbar_s(\p)$, each of which is a $4\times4$ matrix.  
The sum over eigenstates of 
$S_z$ should remove any memory of the spin-quantization axis, and
so the result should be expressible in terms of the four-vector $p^\mu$ and 
various gamma matrices, with all vector indices contracted.  
In the rest frame, $\psl=-m\g^0$, and
it is easy to check that
$\sum_{s=\pm}u_s({\bf 0})\ubar_s({\bf 0})=m\g^0+m$ and
$\sum_{s=\pm}v_s({\bf 0})\vbar_s({\bf 0})=m\g^0-m$.  
We therefore conclude that
\begin{eqnarray}
\sum_{s=\pm}u_s(\p)\ubar_s(\p) &=& -\psl+m \;,
\nonumber \\
\sum_{s=\pm}v_s(\p)\vbar_s(\p) &=& -\psl-m \;.
\label{ssum}
\end{eqnarray}
We will make extensive use of \eq{ssum} when we calculate scattering 
cross sections for spin-one-half particles.

From \eq{ssum}, we can get $u_+(\p)\ubar_+(\p)$, etc, by applying appropriate
spin projection matrices.  In the rest frame, we have
\begin{eqnarray}
\half(1+2sS_z)u_{s'}({\bf 0}) &=& \delta_{ss'}\,u_{s'}({\bf 0}) \;,
\nonumber \\
\half(1-2sS_z)v_{s'}({\bf 0}) &=& \delta_{ss'}\,v_{s'}({\bf 0}) \;. 
\label{sproj0}
\end{eqnarray}
In order to boost these projection matrices to a more general frame, 
we first recall that
\begin{equation}
\g_5\equiv i\g^0\g^1\g^2\g^3 = \pmatrix{ -I & 0 \cr \noalign{\medskip}
                                          0 & I \cr}.
\label{g5again}
\end{equation}
This allows us to write $S_z={i\over2}\g^1\g^2$
as $S_z=-\half\g_5\g^3\g^0$.   
In the rest frame, we can write $\g^0$ as $-\psl/m$,
and $\g^3$ as $\zsl$, where $z^\mu=(0,\hat{\bf z})$; 
thus we have 
\begin{equation}
S_z={\ts{1\over 2m}}\g_5\zsl\psl \;.
\label{szagain}
\end{equation}
Now we can boost $S_z$ to any other frame simply by replacing
$\zsl$ and $\psl$ with their values in that frame.  
(Note that, in any frame, $z^\mu$ satisfies $z^2=1$ and $z\cd p=0$.)
Boosting \eq{sproj0} then yields 
\begin{eqnarray}
\half(1-s\g_5\zsl)u_{s'}(\p) &=& \delta_{ss'}\,u_{s'}(\p) \;,
\nonumber \\
\half(1-s\g_5\zsl)v_{s'}(\p) &=& \delta_{ss'}\,v_{s'}(\p) \;, 
\label{sprojp}
\end{eqnarray}
where we have used \eq{equv} to eliminate $\psl$.  
Combining \eqs{ssum} and (\ref{sprojp}) we get
\begin{eqnarray}
u_s(\p)\ubar_s(\p) &=& \half(1-s\g_5\zsl)(-\psl+m) \;,
\nonumber \\
v_s(\p)\vbar_s(\p) &=& \half(1-s\g_5\zsl)(-\psl-m) \;.
\label{uvproj}
\end{eqnarray}

It is interesting to consider the extreme relativistic limit
of this formula.  Let us take the three-momentum to be in the $z$ direction,
so that it is parallel to the spin-quantization axis.
The component of the spin in the direction of the three-momentum is called
the {\it helicity}.  A fermion with helicity $+1/2$ is said to be
{\it right-handed}, and a fermion with helicity $-1/2$ is said to be
{\it left-handed}.  For rapidity $\eta$, we have 
\begin{eqnarray}
{\ts{1\over m}}p^\mu &=& (\cosh\eta,0,0,\sinh\eta) \;,
\nonumber \\
z^\mu &=& (\sinh\eta,0,0,\cosh\eta) \;.
\label{pz}
\end{eqnarray}
The first equation is simply the definition of $\eta$,
and the second follows from $z^2=1$ and $p\cd z=0$ (along with
the knowledge that a boost of a four-vector in the $z$ direction
does not change its $x$ and $y$ components).
In the limit of large $\eta$, we see that 
\begin{equation}
z^\mu = {\ts{1\over m}}p^\mu + O(e^{-\eta})\;.
\label{z=p}
\end{equation}
Hence, in \eq{uvproj}, we can replace $\zsl$ with $\psl/m$, and then use 
$(\psl/m)(-\psl\pm m)=\mp(-\psl\pm m)$, which holds for $p^2=-m^2$.
For consistency, we should then also drop the $m$ relative to $\psl$,
since it is down by a factor of $O(e^{-\eta})$.  We get
\begin{eqnarray}
u_s(\p)\ubar_s(\p) &\to& \half(1+s\g_5)(-\psl) \;,
\nonumber \\
v_s(\p)\vbar_s(\p) &\to& \half(1-s\g_5)(-\psl)\;.
\label{uvproj2}
\end{eqnarray}
The spinor corresponding to a right-handed fermion (helicity $+1/2$)
is $u_+(\p)$ for a $b$-type  particle and $v_-(\p)$ for a $d$-type particle.  
According to \eq{uvproj2}, either of these is projected by 
$\half(1+\g_5)=\hbox{diag}(0,0,1,1)$ onto the lower two components only.  
In terms of the Dirac field $\Psi(x)$, this is the part that corresponds to
the right-handed Weyl field.  Similarly, left-handed fermions are 
projected (in the extreme relativistic limit) onto the upper two spinor
components only, corresponding to the left-handed Weyl field.  

The case of a massless particle follows from the extreme relativistic
limit of a massive particle.  In particular, \eqs{equv}, (\ref{equvbar}), 
(\ref{uvnorm}), (\ref{gordonp}), (\ref{gordonpuv}),
and (\ref{ssum}) are all valid with $m=0$,
and \eq{uvproj2} becomes exact.

Finally, for our discussion of parity, time reversal, 
and charge conjugation in section 40,
we will need a number of relationships among the $u$ and $v$ spinors.
First, note that $\beta u_s({\bf 0})= +u_s({\bf 0})$ and
$\beta v_s({\bf 0})= -v_s({\bf 0})$.  Also,
$\beta K^j=-K^j\beta$.  We then have
\begin{eqnarray}
u_s(-\p) &=& +\beta u_s(\p) \;,
\nonumber \\
v_s(-\p) &=& -\beta v_s(\p) \;.
\label{uvP}
\end{eqnarray}
Next, we need the charge conjugation matrix
\begin{equation}
\C=           \pmatrix{ 0 & -1 &  0 &  0 \cr
                       +1 &  0 &  0 &  0 \cr
                        0 &  0 &  0 & +1 \cr
                        0 &  0 & -1 &  0 \cr},
\label{c}
\end{equation}
which obeys
\begin{equation}
\C^{\rm\sss T}=\C^\dagger=\C^{-1}=-\C \;,
\label{ci}
\end{equation}
\begin{equation}
\beta\C = - \C\beta \;,
\label{cbeta}
\end{equation}
\begin{equation}
\C^{-1}\g^\mu \C=-(\g^\mu)^{\rm\sss T} \;.
\label{cgc2}
\end{equation}
Using \eqs{uv0}, (\ref{uvbar0}), and (\ref{c}), we get
$\C\ubar_s({\bf 0})^{\rm\sss T} = v_s({\bf 0})$ and
$\C\vbar_s({\bf 0})^{\rm\sss T} = u_s({\bf 0})$.
Also, \eq{cgc2} implies $\C^{-1}K^j \C = -(K^j)^{\rm\sss T}$.  
From this we can conclude that
\begin{eqnarray}
\C\ubar_s(\p)^{\rm\sss T} &=& v_s(\p) \;,
\nonumber \\
\C\vbar_s(\p)^{\rm\sss T} &=& u_s(\p) \;.
\label{uvC}
\end{eqnarray}
Taking the complex conjugate of \eq{uvC}, and
using $\ubar^{{\rm\sss T}*}=\ubar^\dagger=\beta u$, we get
\begin{eqnarray}
u^*_s(\p) &=& \C\beta v_s(\p) \;,
\nonumber \\
v^*_s(\p) &=& \C\beta u_s(\p) \;.
\label{uv*}
\end{eqnarray}
Next, note that 
$\g_5 u_s({\bf 0})= +s\,v_{-s}({\bf 0})$ and
$\g_5 v_s({\bf 0})= -s\,u_{-s}({\bf 0})$, and that
$\g_5 K^j = K^j \g_5$.  Therefore
\begin{eqnarray}
\g_5 u_s(\p) &=& +s\,v_{-s}(\p) \;,
\nonumber \\
\g_5 v_s(\p) &=& -s\,u_{-s}(\p) \;.
\label{uv5}
\end{eqnarray}
Combining \eqs{uvP}, (\ref{uv*}), and (\ref{uv5}) results in
\begin{eqnarray}
u^*_{-s}(-\p) &=& -s\,\C\g_5 u_s(\p) \;,
\nonumber \\
v^*_{-s}(-\p) &=& -s\,\C\g_5 v_s(\p) \;.
\label{uvT}
\end{eqnarray}
We will need \eq{uvP} in our discussion of parity,
\eq{uvC} in our discussion of charge conjugation, and
\eq{uvT} in our discussion of time reversal.

\vskip0.5in

\begin{center}
Problems
\end{center}

\vskip0.25in

38.1) 
Use \eq{uvp} to compute $u_s(\p)$ and $v_s(\p)$ explicity.
Hint: show that the matrix $2i{\bf\hat p\cd K}$ has eigenvalues $\pm1$,
and that, for any matrix $A$ with eigenvalues $\pm1$,
$e^{c A} = (\cosh c) + (\sinh c)A$,
where $c$ is an arbitrary complex number.

38.2) Verify \eq{gbar}.

38.3) Verify \eq{gordonpuv}.
  
\vfill\eject

\noindent Quantum Field Theory  \hfill   Mark Srednicki

\vskip0.5in

\begin{center}
\large{39: Canonical Quantization of Spinor Fields II}
\end{center}
\begin{center}
Prerequisite: 38
\end{center}

\vskip0.5in

A Dirac field $\Psi$ with lagrangian
\begin{equation}
\L = i\Psibar\dsl\Psi-m\Psibar\Psi
\label{ellPsi2}
\end{equation}
obeys the canonical anticommutation relations
\begin{eqnarray}
\{\Psi_\alpha(\x,t),\Psi_\beta(\y,t)\} &=& 0\;,
\label{PsiPsi} \\
\noalign{\smallskip}
\{\Psi_\alpha(\x,t),\Psibar_\beta(\y,t)\}
&=& (\gamma^0)_{\alpha\beta}\,\delta^3(\x-\y)\;,
\label{PsiPsibar}
\end{eqnarray}
and has the Dirac equation
\begin{equation}
(-i\dsl + m)\Psi=0
\label{deq2}
\end{equation}
as its equation of motion.  The general solution is
\begin{equation}
\Psi(x) = \sum_{s=\pm}\int\dpt\,\left[b_s(\p)  u_s(\p)e^{ipx}
                              +d_s^\dagger(\p)v_s(\p)e^{-ipx} \right]\;,
\label{Psi2}
\end{equation}
where $b_s(\p)$ and $d_s^\dagger(\p)$ are operators; the properties
of the four-component spinors $u_s(\p)$ and $v_s(\p)$ were belabored
in the previous section.

Let us express $b_s(\p)$ and $d_s^\dagger(\p)$ in terms of $\Psi(x)$
and $\Psibar(x)$.  We begin with
\begin{equation}
\int \dtx\;e^{-ipx}\Psi(x) = 
\sum_{s'=\pm}\left[
{\ts{1\over2\w}}b_{s'}(\p)u_{s'}(\p)
 + {\ts{1\over2\w}}e^{2i\w t}d^\dagger_{s'}(-\p)v_{s'}(-\p)\right].
\label{0bd}
\end{equation}
Next, multiply on the left by $\ubar_s(\p)\g^0$, and use 
$\ubar_{s}(\p)\g^0 u_{s'}(\p) = 2\w\delta_{ss'}$ and
$\ubar_{s}(\p)\g^0 v_{s'}(-\p) = 0$ from section 38.  The result is
\begin{equation}
b_s(\p) = \int \dtx\;e^{-ipx}\,\ubar_s(\p)\g^0\Psi(x) \;.
\label{b}
\end{equation}
Note that $b_s(\p)$ is time independent.

To get $b^\dagger_s(\p)$, take the hermitian conjugate of \eq{b}, using
\begin{eqnarray}
\Bigl[\ubar_s(\p)\g^0\Psi(x)\Bigr]^\dagger 
&=& \overline{\hbox{$\ubar_s(\p)\g^0\Psi(x)$}}
\nonumber \\
&=& \Psibar(x)\overline{\hbox{$\g^0$}}u_s(\p)
\nonumber \\
\noalign{\smallskip}
&=& \Psibar(x)\g^0 u_s(\p) \;,
\label{barring}
\end{eqnarray}
where, for any general combination of gamma matrices $A$,
\begin{equation}
\overline{A}\equiv \beta A^\dagger\beta \;.
\label{abar2}
\end{equation}
Thus we find
\begin{equation}
b^\dagger_s(\p) = \int \dtx\;e^{ipx}\,\Psibar(x)\g^0 u_s(\p) \;.
\label{bd}
\end{equation}

To extract $d^\dagger_s(\p)$ from $\Psi(x)$, we start with
\begin{equation}
\int \dtx\;e^{ipx}\Psi(x) = 
\sum_{s'=\pm}\left[
{\ts{1\over2\w}}e^{-2i\w t}b_{s'}(-\p)u_{s'}(-\p)
 + {\ts{1\over2\w}}d^\dagger_{s'}(\p)v_{s'}(\p)\right].
\label{0bdd}
\end{equation}
Next, multiply on the left by $\vbar_s(\p)\g^0$, and use 
$\vbar_{s}(\p)\g^0 v_{s'}(\p) = 2\w\delta_{ss'}$ and
$\vbar_{s}(\p)\g^0 u_{s'}(-\p) = 0$ from section 38.  The result is
\begin{equation}
\dd_s(\p) = \int \dtx\;e^{ipx}\,\vbar_s(\p)\g^0\Psi(x) \;.
\label{dd}
\end{equation}
To get $d_s(\p)$, take the hermitian conjugate of \eq{dd}, which yields
\begin{equation}
d_s(\p) = \int \dtx\;e^{-ipx}\,\Psibar(x)\g^0 v_s(\p) \;.
\label{d}
\end{equation}

Next, let us work out the anticommutation relations of the $b$
and $d$ operators (and their hermitian conjugates).  
From \eq{PsiPsi}, it is immediately clear that
\begin{eqnarray}
\{b_s(\p),b_{s'}(\p')\} &=& 0 \;,
\nonumber \\
\{d_s(\p),d_{s'}(\p')\} &=& 0 \;, 
\nonumber \\
\{b_s(\p),d^\dagger_{s'}(\p')\} &=& 0 \;,
\label{bd0}
\end{eqnarray}
because these involve only the anticommutator of $\Psi$ with itself,
and this vanishes.  Of course, hermitian conjugation also yields
\begin{eqnarray}
\{\bd_s(\p),\bd_{s'}(\p')\} &=& 0 \;,
\nonumber \\
\{\dd_s(\p),\dd_{s'}(\p')\} &=& 0 \;, 
\nonumber \\
\{\bd_s(\p),d_{s'}(\p')\} &=& 0 \;.
\label{bdd0}
\end{eqnarray}

Now consider
\begin{eqnarray}
\{b_s(\p),b^\dagger_{s'}(\p')\} 
&=& \int\dtx\,\dty\,e^{-ipx+ip'y}\,\ubar_s(\p)\g^0
\{\Psi(x),\Psibar(y)\}\g^0 u_{s'}(\p')
\nonumber \\
&=& \int\dtx\,e^{-i(p-p')x}\,\ubar_s(\p)\g^0\g^0\g^0 u_{s'}(\p')
\nonumber \\
\noalign{\smallskip}
&=& (2\pi)^3 \delta^3(\p-\p')\,\ubar_s(\p)\g^0 u_{s'}(\p)
\nonumber \\
\noalign{\medskip}
&=& (2\pi)^3 \delta^3(\p-\p')\,2\w\delta_{ss'}\;.
\label{bbdcomm}
\end{eqnarray}
In the first line, we are free to set $x^0=y^0$ because
$b_s(\p)$ and $\bd_{s'}(\p')$ are actually time independent.
In the third, we used $(\g^0)^2=1$, and in the fourth,
$\ubar_s(\p)\g^0 u_{s'}(\p)=2\w\delta_{ss'}$.

Similarly,
\begin{eqnarray}
\{\dd_s(\p),d_{s'}(\p')\} 
&=& \int\dtx\,\dty\,e^{ipx-ip'y}\,\vbar_s(\p)\g^0
\{\Psi(x),\Psibar(y)\}\g^0 v_{s'}(\p')
\nonumber \\
&=& \int\dtx\,e^{i(p-p')x}\,\vbar_s(\p)\g^0\g^0\g^0 v_{s'}(\p')
\nonumber \\
\noalign{\smallskip}
&=& (2\pi)^3 \delta^3(\p-\p')\,\vbar_s(\p)\g^0 v_{s'}(\p)
\nonumber \\
\noalign{\medskip}
&=& (2\pi)^3 \delta^3(\p-\p')\,2\w\delta_{ss'}\;.
\label{dddcomm}
\end{eqnarray}
And finally,
\begin{eqnarray}
\{b_s(\p),d_{s'}(\p')\} 
&=& \int\dtx\,\dty\,e^{-ipx-ip'y}\,\ubar_s(\p)\g^0
\{\Psi(x),\Psibar(y)\}\g^0 v_{s'}(\p')
\nonumber \\
&=& \int\dtx\,e^{-i(p+p')x}\,\ubar_s(\p)\g^0\g^0\g^0 v_{s'}(\p')
\nonumber \\
\noalign{\medskip}
&=& (2\pi)^3 \delta^3(\p+\p')\,\ubar_s(\p)\g^0 v_{s'}(-\p)
\nonumber \\
\noalign{\medskip}
&=& 0 \;.
\label{bdcomm2}
\end{eqnarray}
According to the discussion in section 3,
eqs.$\,$(\ref{bd0}--\ref{bdcomm2}) are exactly what we need
to describe the creation and annihilation of fermions.
In this case, we have two different kinds:
$b$-type and $d$-type, each with two possible spin states,
$s=+$ and $s=-$.

Next, let us evaluate the hamiltonian
\begin{equation}
H=\int\dtx\;\Psibar(-i\g^i\d_i + m)\Psi
\label{Psiham}
\end{equation}
in terms of the $b$ and $d$ operators.  We have
\begin{eqnarray}
(-i\g^i\d_i + m)\Psi 
&=& \smash{\sum_{s=\pm}}
\int\dpt\;\Bigl(-i\g^i\d_i + m\Bigr)\!
\Bigl(\,b_s(\p)u_s(\p)e^{ipx}
\nonumber \\
&& \qquad \qquad \qquad \qquad \quad 
     {}+\dd_s(\p)v_s(\p)e^{-ipx} \Bigr)
\nonumber \\
\noalign{\medskip}
&=& \smash{\sum_{s=\pm}}
\int\dpt\;\Bigl[\,
     b_s(\p)(+\g^i p_i+m)u_s(\p)e^{ipx}
\nonumber \\
&& \qquad\quad\;
  {}+\dd_s(\p)(-\g^i p_i+m)v_s(\p)e^{-ipx} \,\Bigr]
\nonumber \\
\noalign{\medskip}
&=& \smash{\sum_{s=\pm}}
\int\dpt\;\Bigl[\,
     b_s(\p)(\g^0\w)u_s(\p)e^{ipx}
\nonumber \\
&& \qquad\quad\;
  {}+\dd_s(\p)(-\g^0\w)v_s(\p)e^{-ipx} \,\Bigr].
\label{Psiham2}
\end{eqnarray}
Therefore
\begin{eqnarray}
H &=& \smash{\sum_{s,s'}}
\int\dpt\;\dpt{\,}'\;\dtx
\left( \bd_{s'}(\p')\ubar_{s'}(\p')e^{-ip'x}
       +d_{s'}(\p')\vbar_{s'}(\p')e^{ip'x} \right)
\nonumber \\
      && \qquad\qquad\quad {}\times
\w\left( b_s(\p)\g^0 u_s(\p)e^{ipx}
      -\dd_s(\p)\g^0 v_s(\p)e^{-ipx} \right) 
\nonumber \\
\noalign{\medskip}
&=& \smash{\sum_{s,s'}}
\int\dpt\;\dpt{\,}'\;\dtx\;\w\,\Bigl[\;
\bd_{s'}(\p')b_s(\p)\;\ubar_{s'}(\p')\g^0 u_s(\p)\;e^{-i(p'-p)x}
\nonumber \\
&& \qquad\qquad\qquad\quad\;\,
{} - \bd_{s'}(\p')\dd_s(\p)\;\ubar_{s'}(\p')\g^0 v_s(\p)\;e^{-i(p'+p)x}
\nonumber \\
\noalign{\medskip}
&& \qquad\qquad\qquad\quad\;\,
{} + d_{s'}(\p')b_s(\p)\;\vbar_{s'}(\p')\g^0 u_s(\p)\;e^{+i(p'+p)x}
\nonumber \\
\noalign{\smallskip}
&& \qquad\qquad\qquad\quad\;\,
{} - d_{s'}(\p')\dd_s(\p)\;\vbar_{s'}(\p')\g^0 v_s(\p)\;e^{+i(p'-p)x}\, \Bigr]
\nonumber \\
\noalign{\medskip}
&=& \smash{\sum_{s,s'}}
\int\dpt\;\;\half\Bigl[\;
\bd_{s'}(\p)b_s(\p)\;\ubar_{s'}(\p)\g^0 u_s(\p)
\nonumber \\
&& \qquad\qquad\,
{} - \bd_{s'}(-\p)\dd_s(\p)\;\ubar_{s'}(-\p)\g^0 v_s(\p)\;e^{+2i\w t}
\nonumber \\
\noalign{\medskip}
&& \qquad\qquad\,
{} + d_{s'}(-\p)b_s(\p)\;\vbar_{s'}(-\p)\g^0 u_s(\p)\;e^{-2i\w t}
\nonumber \\
\noalign{\smallskip}
&& \qquad\qquad\,
{} - d_{s'}(\p)\dd_s(\p)\;\vbar_{s'}(\p)\g^0 v_s(\p)\,\Bigr]
\nonumber \\
\noalign{\smallskip}
&=& \smash{\sum_s}
\int\dpt\;\,\w\,\Bigl[\,
\bd_s(\p)b_s(\p) - d_s(\p)\dd_s(\p)\,\Bigr].
\label{Psiham3}
\end{eqnarray}
Using \eq{dddcomm}, we can rewrite this as
\begin{equation}
H = \sum_{s=\pm}\int\dpt\;\,\w\,\Bigl[\,
\bd_s(\p)b_s(\p) + \dd_s(\p)d_s(\p)\,\Bigr] - 4{\cal E}_0 V\;,
\label{Psiham4}
\end{equation}
where ${\cal E}_0 = \half\int \dtk\;\omega$ is the 
zero-point energy per unit volume that we found for a real scalar
field in section 3,
and $V=(2\pi)^3\delta^3({\bf 0})=\int\dtx$ is the volume of space.
That the zero-point energy is negative rather than positive is
characteristic of fermions; that it is larger in magnitude by a factor
of four is due to the four types of particles that are associated with
a Dirac field.
We can cancel off this constant energy by
including a constant term $-4{\cal E}_0$
in the original lagrangian density;
from here on, we will assume that this has been done.

The ground state of the hamiltonian (\ref{Psiham4}) is the {\it vacuum state\/}
$|0\ra$ that is annihilated by every $b_s(\p)$ and $d_s(\p)$,
\begin{equation}
b_s(\p)|0\ra = d_s(\p)|0\ra = 0\;.
\label{Psivac}
\end{equation}
Then, we can interpret the $\bd_s(\p)$ operator
as creating a $b$-type particle with momentum $\p$, 
energy $\w=(\p^2+m^2)^{1/2}$, and spin $S_z=\half s$, and the
$\dd_s(\p)$ operator as creating a $d$-type particle with the same properties.
The $b$-type and $d$-type particles are distinguished by the value of the
charge $Q=\int\dtx\,j^0$, where $j^\mu=\Psibar\g^\mu\Psi$ is the
Noether current associated with the invariance of $\L$ under the U(1)
transformation $\Psi\to e^{-i\alpha}\Psi$, $\Psibar\to e^{+i\alpha}\Psibar$.
Following the same procedure that we used for the hamiltonian, we can show that 
\begin{eqnarray}
Q &=& \int\dtx\;\Psibar\g^0\Psi
\nonumber \\
&=& \sum_{s=\pm}\int\dpt\;\Bigl[\,
\bd_s(\p)b_s(\p) + d_s(\p)\dd_s(\p)\,\Bigr]
\nonumber \\
&=& \sum_{s=\pm}\int\dpt\;\Bigl[\,
\bd_s(\p)b_s(\p) - \dd_s(\p)d_s(\p)\,\Bigr] + \hbox{constant}\;;
\label{PsiQ}
\end{eqnarray}
Thus the conserved charge $Q$ counts the total number of $b$-type particles
minus the total number of $d$-type particles.
(We are free to shift the overall value of $Q$ to remove the constant term,
and so we shall.)
In quantum electrodynamics, we will identify the $b$-type particles as
electrons and the $d$-type particles as positrons.

Now consider a Majorana field $\Psi$ with lagrangian
\begin{equation}
\L = {\ts{i\over2}}\Psit\C\dsl\Psi-\half m\Psit\C\Psi \;.
\label{ellmaj3}
\end{equation}
The equation of motion for $\Psi$
is once again the Dirac equation, and so the general solution is 
once again given by \eq{Psi2}.  
However, $\Psi$ must also obey the Majorana condition 
$\Psi=\C\Psibart$.  Starting from the barred form of
\eq{Psi2},
\begin{equation}
\Psibar(x) = \sum_{s=\pm}\int\dpt\,\left[\bd_s(\p)\ubar_s(\p)e^{-ipx}
                              +d_s(\p)\vbar_s(\p)e^{ipx} \right],
\label{Psibar2}
\end{equation}
we have
\begin{equation}
\C\Psibart(x)
= \sum_{s=\pm}\int\dpt\,\left[\bd_s(\p)\,\C\ubar^{\rm\sss T}_s(\p)e^{-ipx}
                      + d_s(\p)\,\C\vbar^{\rm\sss T}_s(\p)e^{ipx} \right].
\label{PsiC}
\end{equation}
From section 38, we have
\begin{eqnarray}
\C\ubar_s(\p)^{\rm\sss T} &=& v_s(\p) \;,
\nonumber \\
\C\vbar_s(\p)^{\rm\sss T} &=& u_s(\p) \;,
\label{uvC2}
\end{eqnarray}
and so
\begin{equation}
\C\Psibart(x)
= \sum_{s=\pm}\int\dpt\,\left[\bd_s(\p)v_s(\p)e^{-ipx}
                      + d_s(\p)u_s(\p)e^{ipx} \right]\;.
\label{PsiC2}
\end{equation}
Comparing \eqs{Psi2} and (\ref{PsiC2}), we see that we will have
$\Psi=\C\Psibart$ if
\begin{equation}
d_s(\p)=b_s(\p) \;.
\label{d=b}
\end{equation}
Thus a free Majorana field can be written as
\begin{equation}
\Psi(x) = \sum_{s=\pm}\int\dpt\,\left[b_s(\p)  u_s(\p)e^{ipx}
                              +b_s^\dagger(\p)v_s(\p)e^{-ipx} \right]\;.
\label{Psimaj}
\end{equation}
The anticommutation relations for a Majorana field,
\begin{eqnarray}
\{\Psi_\alpha(\x,t),\Psi_\beta(\y,t)\}
&=& (\C\gamma^0)_{\alpha\beta}\,\delta^3(\x-\y)\;,
\label{Majcomm1a} \\
\noalign{\medskip}
\{\Psi_\alpha(\x,t),\Psibar_\beta(\y,t)\}
&=& (\gamma^0)_{\alpha\beta}\,\delta^3(\x-\y)\;,
\label{Majcomm2a}
\end{eqnarray}
can be used to show that
\begin{eqnarray}
\{b_s(\p),b_{s'}(\p')\} &=& 0 \;,
\nonumber \\
\noalign{\medskip}
\{b_s(\p),\bd_{s'}(\p')\} 
&=& (2\pi)^3 \delta^3(\p-\p')\,2\w\delta_{ss'}\;,
\label{bdMaj}
\end{eqnarray}
as we would expect.

The hamiltonian for the Majorana field $\Psi$ is 
\begin{eqnarray}
H&=&\half\int\dtx\;\Psit\C(-i\g^i\d_i + m)\Psi 
\nonumber \\
&=&\half\int\dtx\;\Psibar(-i\g^i\d_i + m)\Psi \;,
\label{Majham}
\end{eqnarray}
and we can work through
the same manipulations that led to \eq{Psiham3}; the only differences are 
an extra overall factor of one-half, and $d_s(\p)=b_s(\p)$.  Thus we get
\begin{equation}
H = \half\sum_{s=\pm}\int\dpt\;\,\w\,\Bigl[\,
\bd_s(\p)b_s(\p) - b_s(\p)\bd_s(\p)\,\Bigr].
\label{Majham2}
\end{equation}
Note that this would reduce to a constant 
if we tried to use commutators rather than
anticommutators in \eq{bdMaj}, a reflection of the
spin-statistics theorem.  Using \eq{bdMaj} as it is, we find
\begin{equation}
H = \sum_{s=\pm}\int\dpt\;\w\,\bd_s(\p)b_s(\p) - 2{\cal E}_0 V.
\label{Majham3}
\end{equation}
Again, we can (and will) cancel off the zero-point energy by including
a term $-2{\cal E}_0$ in the original lagrangian density.

The Majorana lagrangian has no U(1) symmetry.  Thus there
is no associated charge, 
and only one kind of particle (with two possible spin states).

\vskip0.5in

\begin{center}
Problems
\end{center}

\vskip0.25in

39.1) Verify \eq{PsiQ}.

39.2) Show that
\begin{eqnarray}
U(\Lam)^{-1}\bd_s(\p)U(\Lam) &=& \bd_s(\Lam^{-1}\p) \;,
\nonumber \\
\noalign{\medskip}
U(\Lam)^{-1}\dd_s(\p)U(\Lam) &=& \dd_s(\Lam^{-1}\p) \;,
\label{ulambdulam}
\end{eqnarray}
and hence that
\begin{equation}
U(\Lam)|p,s,q\ra = |\Lam p,s,q\ra\;,
\label{ulampsq}
\end{equation}
where 
\begin{eqnarray}
|p,s,{+}\ra &\equiv& \bd_s(\p)|0\ra\;,
\nonumber \\
\noalign{\smallskip}
|p,s,{-}\ra &\equiv& \dd_s(\p)|0\ra
\label{pqpm}
\end{eqnarray}
are single-particle states.

\vfill\eject

\noindent Quantum Field Theory  \hfill   Mark Srednicki

\vskip0.5in

\begin{center}
\large{40: Parity, Time Reversal, and Charge Conjugation}
\end{center}
\begin{center}
Prerequisite: 39
\end{center}

\vskip0.5in

Recall that, under a Lorentz transformation $\Lambda$ implemented 
by the unitary operator $U(\Lambda)$, a Dirac (or Majorana)
field transforms as
\begin{equation}
U(\Lambda)^{-1}\Psi(x)U(\Lambda) =  D(\Lambda)\Psi(\Lambda^{-1}x)\;.
\label{Psilam}
\end{equation}
For an infinitesimal transformation 
$\Lambda^\mu{}_\nu=\delta^\mu{}_\nu+\delta\omega^\mu{}_\nu$, the matrix
$D(\Lambda)$ is given by
\begin{equation}
D(1{+}\delta\omega)=I + {\ts{i\over2}}\delta\omega_{\mu\nu} S^{\mu\nu}\;,
\label{dlam}
\end{equation}
where the Lorentz generator matrices are
\begin{equation}
S^{\mu\nu}={\ts{i\over4}}[\g^\mu,\g^\nu]\;.
\label{smunu}
\end{equation}
In this section, we will consider the two Lorentz transformations that
cannot be reached via a sequence of infinitesimal transformations
away from the identity: parity and time reversal.  We begin with parity.

Define the parity transformation
\begin{equation}
\P^\mu{}_\nu = (\P^{-1}){}^\mu{}_\nu 
             =\pmatrix{ +1 & & & \cr
                        & -1 & & \cr
                        & & -1 & \cr
                        & & & -1 \cr} 
\label{p4}
\end{equation}
and the corresponding unitary operator
\begin{equation}
P \equiv U(\P)\;.
\label{pop}
\end{equation}
Now we have
\begin{equation}
P^{-1}\Psi(x)P =  D(\P)\Psi(\P x)\;.
\label{ppsip}
\end{equation}
The question we wish to answer is, what is the matrix $D(\P)$?

First of all, if we make a second parity transformation, we get
\begin{equation}
P^{-2}\Psi(x)P^2 =  D(\P)^2\Psi(x)\;,
\label{ppsip2}
\end{equation}
and it is tempting to conclude that we should have
$D(\P)^2=1$, so that we return to the original field.
This is correct for scalar fields, since they are themselves
observable.  With fermions, however, it takes an even number of
fields to construct an observable.
Therefore we need only require the weaker condition $D(\P)^2=\pm 1$.

We will also require the particle creation and annihilation operators
to transform in a simple way.  Because
\begin{eqnarray}
P^{-1}{\bf P} P &=& -{\bf P}\;,
\label{ppip} \\
P^{-1}{\bf J} P &=& +{\bf J}\;,
\label{pjip}
\end{eqnarray}
where $\bf P$ is the total three-momentum operator and $\bf J$ 
is the total angular momentum operator, a parity transformation should
reverse the three-momentum while leaving the spin direction unchanged.
We therefore require
\begin{eqnarray}
P^{-1}b^\dagger_s(\p)P &=& \eta\,b^\dagger_s(-\p) \;,
\nonumber \\
\noalign{\medskip}
P^{-1}d^\dagger_s(\p)P &=& \eta\,d^\dagger_s(-\p) \;,
\label{pbdp}
\end{eqnarray}
where $\eta$ is a possible phase factor that (by the previous
argument about observables) should satisfy
$\eta^2=\pm 1$.  We could in principle assign different phase
factors to the $b$ and $d$ operators, but we choose them to be
the same so that the parity transformation is compatible with
the Majorana condition $d_s(\p)=b_s(\p)$.
Writing the mode expansion of the free field
\begin{equation}
\Psi(x)=
\sum_{s=\pm}\int\dpt\left[b_s(\p)u_s(\p)e^{ipx}
                        +d^\dagger_s(\p)v_s(\p)e^{-ipx}\right],
\label{psi}
\end{equation}
the parity transformation reads
\begin{eqnarray}
P^{-1}\Psi(x)P &=& \sum_{s=\pm}\int\dpt\left[
                   \Bigl(P^{-1}b_s(\p)P\Bigr)u_s(\p)e^{ipx}
                 + \Bigl(P^{-1}d^\dagger_s(\p)P\Bigr)v_s(\p)e^{-ipx}\right]
\nonumber \\
&=& \sum_{s=\pm}\int\dpt\left[
                   \eta^* b_s(-\p)u_s(\p)e^{ipx}
                 + \eta d^\dagger_s(-\p)v_s(\p)e^{-ipx}\right]
\nonumber \\
&=& \sum_{s=\pm}\int\dpt\left[
                   \eta^* b_s(\p)u_s(-\p)e^{ip\P x}
                 + \eta d^\dagger_s(\p)v_s(-\p)e^{-ip\P x}\right].
\label{ppsip3}
\end{eqnarray}
In the last line, we have changed the integration variable from
$\p$ to $-\p$.  We now use a result from section 38, namely that
\begin{eqnarray}
u_s(-\p) &=& +\beta u_s(\p) \;,
\nonumber \\
v_s(-\p) &=& -\beta v_s(\p) \;,
\label{uvP2}
\end{eqnarray}
where
\begin{equation}
\beta=\pmatrix{ 0 & I \cr  
\noalign{\medskip}
                I & 0 \cr}.
\label{beta2}
\end{equation}
Then, if we choose $\eta=-i$, \eq{ppsip3} becomes
\begin{eqnarray}
P^{-1}\Psi(x)P &=& \sum_{s=\pm}\int\dpt\left[
                   i b_s(\p)\beta u_s(\p)e^{ip\P x}
                 + i d^\dagger_s(\p)\beta v_s(\p)e^{-ip\P x}\right]
\nonumber \\
\noalign{\medskip}
&=& i\beta\,\Psi(\P x)\;.
\label{ppsip4}
\end{eqnarray}
Thus we see that $D(\P)=i\beta$.  (We could also have chosen $\eta=i$,
resulting in $D(\P)=-i\beta$; either choice is acceptable.)

The factor of $i$ has an interesting physical consequence.  Consider
a state of an electron and positron with zero center-of-mass momentum,
\begin{equation}
|\phi\ra = \int \dpt\;\phi(\p)b^\dagger_s(\p)d^\dagger_{s'}(-\p)|0\ra\;;
\label{state}
\end{equation}
here $\phi(\p)$ is the momentum-space wave function.
Let us assume that the vacuum is  parity invariant:
$P|0\ra = P^{-1}|0\ra = |0\ra$.  Let us also assume that
the wave function has definite parity: $\phi(-\p)=(-)^\ell\phi(\p)$.
Then, applying the inverse parity operator on $|\phi\ra$, we get
\begin{eqnarray}
P^{-1}|\phi\ra 
&=& \int\dpt\;\phi(\p)\Bigl(P^{-1}b^\dagger_s(\p)P\Bigr)
\bigl(P^{-1}d^\dagger_{s'}(-\p)P\Bigr)P^{-1}|0\ra\;.
\nonumber \\
&=& (-i)^2\int\dpt\;\phi(\p)b^\dagger_s(-\p)d^\dagger_{s'}(\p)|0\ra
\nonumber \\
&=& (-i)^2\int\dpt\;\phi(-\p)b^\dagger_s(\p)d^\dagger_{s'}(-\p)|0\ra
\nonumber \\
\noalign{\smallskip}
&=& -(-)^\ell |\phi\ra \;.
\label{pstate}
\end{eqnarray}
Thus, the parity of this state is opposite to that of its wave function;
an electron-positron pair has an {\it intrinsic parity\/} of $-1$.
This also applies to a pair of Majorana fermions.
This influences the selection rules for fermion pair annihilation
in theories which conserve parity.
(A pair of electrons also has negative intrinsic parity, 
but this is not interesting because 
the electrons are prevented from annihilating by charge conservation.)

It is interesting to see what \eq{ppsip4} implies for the two Weyl fields
that comprise the Dirac field.  Recalling that
\begin{equation}
\Psi=\pmatrix{ \chi_a \cr 
\noalign{\medskip}
               \xi^{\dagger\dot a} \cr },
\label{psi2}
\end{equation}
we see from \eqs{beta2} and (\ref{ppsip4}) that
\begin{eqnarray}
P^{-1}\chi_a(x)P &=& i\xi^\dagger{}^{\dot a}(\P x)\;,
\nonumber \\
\noalign{\medskip}
P^{-1}\xi^\dagger{}^{\dot a}(x)P &=& i\chi_a(\P x)\;.
\label{pchip}
\end{eqnarray}
Thus a parity transformation exchanges a left-handed field 
for a right-handed one.  

If we take the hermitian conjugate of \eq{pchip}, then raise the index
on one side while lowering it on the other (and remember that this
introduces a relative minus sign!), we get
\begin{eqnarray}
P^{-1}\chi^\dagger{}^{\dot a}(x)P &=& i\xi_a(\P x)\;,
\nonumber \\
\noalign{\medskip}
P^{-1}\xi_a(x)P &=& i\chi^\dagger{}^{\dot a}(\P x)\;.
\label{pchip2}
\end{eqnarray}
Comparing \eqs{pchip} and (\ref{pchip2}), we see that they are compatible
with the Majorana condition $\chi_a(x)=\xi_a(x)$.

Next we take up time reversal.  Define the time-reversal transformation
\begin{equation}
\T^\mu{}_\nu = (\T^{-1}){}^\mu{}_\nu 
             =\pmatrix{ -1 & & & \cr
                        & +1 & & \cr
                        & & +1 & \cr
                        & & & +1 \cr} 
\label{tmu3}
\end{equation}
and the corresponding operator
\begin{equation}
T \equiv U(\T)\;.
\label{top}
\end{equation}
Now we have
\begin{equation}
T^{-1}\Psi(x)T =  D(\T)\Psi(\T x)\;.
\label{tpsit}
\end{equation}
The question we wish to answer is, what is the matrix $D(\T)$?

As with parity, we can conclude that $D(\T)^2=\pm 1$,
and we will require the particle creation and annihilation operators
to transform in a simple way.  Because
\begin{eqnarray}
T^{-1}{\bf P} T &=& -{\bf P}\;,
\label{tpit} \\
T^{-1}{\bf J} T &=& -{\bf J}\;,
\label{tjit}
\end{eqnarray}
where $\bf P$ is the total three-momentum operator and $\bf J$ 
is the total angular momentum operator, a time-reversal transformation should
reverse the direction of both the three-momentum and the spin.
We therefore require
\begin{eqnarray}
T^{-1}b^\dagger_s(\p)T &=& \zeta_s\,b^\dagger_{-s}(-\p) \;,
\nonumber \\
\noalign{\medskip}
T^{-1}d^\dagger_s(\p)T &=& \zeta_s\,d^\dagger_{-s}(-\p) \;.
\label{tbdt}
\end{eqnarray}
This time we allow for possible $s$-dependence of the phase factor,
which should satisfy $\zeta^*_s\zeta_{-s}=\pm 1$.
Also, we recall from section 23 that $T$ must be an antiunitary operator,
so that $T^{-1}iT=-i$.  Then we have
\begin{eqnarray}
T^{-1}\Psi(x)T &=& \sum_{s=\pm}\int\dpt\left[
                   \Bigl(T^{-1}b_s(\p)T\Bigr)u^*_s(\p)e^{-ipx}
                 + \Bigl(T^{-1}d^\dagger_s(\p)T\Bigr)v^*_s(\p)e^{ipx}\right]
\nonumber \\
&=& \sum_{s=\pm}\int\dpt\left[
                   \zeta_s^* b_{-s}(-\p)u^*_s(\p)e^{-ipx}
                 + \zeta_s d^\dagger_{-s}(-\p)v^*_s(\p)e^{ipx}\right]
\label{tpsit2} \\
&=& \sum_{s=\pm}\int\dpt\left[
                   \zeta_{-s}^* b_s(\p)u^*_{-s}(-\p)e^{ip\T x}
                 + \zeta_{-s} d^\dagger_s(\p)v^*_{-s}(-\p)e^{-ip\T x}\right].
\nonumber
\end{eqnarray}
In the last line, we have changed the integration variable from
$\p$ to $-\p$, and the summation variable from $s$ to $-s$.  
We now use a result from section 38, namely that
\begin{eqnarray}
u^*_{-s}(-\p) &=& -s\,\C\g_5 u_s(\p) \;,
\nonumber \\
v^*_{-s}(-\p) &=& -s\,\C\g_5 v_s(\p) \;.
\label{uvT2}
\end{eqnarray}
Then, if we choose $\zeta_s=s$, \eq{tpsit2} becomes
\begin{equation}
T^{-1}\Psi(x)T = \C\g_5\Psi(\T x)\;.
\label{tpsit3}
\end{equation}
Thus we see that $D(\T)=\C\g_5$.  (We could also have chosen $\zeta_s=-s$,
resulting in $D(\T)=-\C\g_5$; either choice is acceptable.)

As with parity, we can consider the effect of time reversal on the
Weyl fields.  Using \eqs{psi2}, (\ref{tpsit3}), 
\begin{equation}
\C=\pmatrix{-\e^{ab} & 0 \cr 
\noalign{\medskip}
            0 & -\e_{\dot a\dot b} \cr },
\label{cee}
\end{equation}
and 
\begin{equation}
\g_5=\pmatrix{-\delta_a{}^c &  0 \cr 
\noalign{\medskip}
               0 & +\delta^\adot{}_\ccdot \cr},
\label{g5000}
\end{equation}
we see that
\begin{eqnarray}
T^{-1}\chi_a(x)T &=& +\chi^a(\T x)\;,
\nonumber \\
\noalign{\medskip}
T^{-1}\xi^\dagger{}^{\dot a}(x)T &=& -\xi^\dagger_{\dot a}(\T x)\;.
\label{tchit}
\end{eqnarray}
Thus left-handed Weyl fields transform into left-handed Weyl fields
(and right-handed into right-handed) under time reversal.

If we take the hermitian conjugate of \eq{tchit}, then raise the index
on one side while lowering it on the other (and remember that this
introduces a relative minus sign!), we get
\begin{eqnarray}
T^{-1}\chi^\dagger{}^{\dot a}(x)T &=& -\chi^\dagger_{\dot a}(\T x)\;,
\nonumber \\
\noalign{\medskip}
T^{-1}\xi_a(x)T &=& +\xi^a(\T x)\;.
\label{tchit2}
\end{eqnarray}
Comparing \eqs{tchit} and (\ref{tchit2}), we see that they are compatible
with the Majorana condition $\chi_a(x)=\xi_a(x)$.

It is interesting and important to evaluate the transformation
properties of fermion bilinears of the form $\Psibar A\Psi$, 
where $A$ is some combination of gamma matrices.  We will consider $A$'s 
that satisfy $\overline A=A$, where $\overline A \equiv \beta A^\dagger\beta$;
in this case, $\Psibar A\Psi$ is hermitian.

Let us begin with parity transformations.  From 
$\Psibar = \Psi^\dagger\beta$ and \eq{ppsip4} we get
\begin{equation}
P^{-1}\Psibar(x)P = -i\Psibar(\P x)\beta \;,
\label{ppsibarp}
\end{equation}
Combining \eqs{ppsip4} and (\ref{ppsibarp}) we find
\begin{equation}
P^{-1}\Bigl(\Psibar A\Psi\Bigr)P = \Psibar\Bigl(\beta A\beta\Bigr)\Psi \;,
\label{bilinp}
\end{equation}
where we have suppressed the spacetime arguments (which transform
in the obvious way).
For various particular choices of $A$ we have
\begin{eqnarray}
\beta 1\beta &=& +1 \;,
\nonumber \\
\beta i\g_5\beta &=& -i\g_5 \;,
\nonumber \\
\beta \g^0\beta &=& +\g^0 \;,
\nonumber \\
\beta \g^i\beta &=& -\g^i \;,
\nonumber \\
\beta \g^0\g_5\beta &=& -\g^0\g_5 \;,
\nonumber \\
\beta \g^i\g_5\beta &=& +\g^i\g_5 \;.
\label{pgp}
\end{eqnarray}
Therefore, the corresponding hermitian bilinears transform as 
\begin{eqnarray}
P^{-1}\Bigl(\Psibar\Psi\Bigr)P &=& +\,\Psibar\Psi \;,
\nonumber \\
P^{-1}\Bigl(\Psibar i\g_5\Psi\Bigr)P &=& -\,\Psibar i\g_5\Psi \;,
\nonumber \\
P^{-1}\Bigl(\Psibar\g^\mu\Psi\Bigr)P &=& +\,\P^\mu{}_\nu\Psibar\g^\nu\Psi \;,
\nonumber \\
P^{-1}\Bigl(\Psibar\g^\mu\g_5\Psi\Bigr)P &=& 
                                      -\,\P^\mu{}_\nu\Psibar\g^\nu\g_5\Psi\;,
\label{bilinp2}
\end{eqnarray}
Thus we see that $\Psibar\Psi$ and $\Psibar\g^\mu\Psi$ 
are even under a parity transformation, while 
$\Psibar i\g_5\Psi$ and $\Psibar\g^\mu\g_5\Psi$ are odd.  We say that
$\Psibar\Psi$ is a scalar, $\Psibar\g^\mu\Psi$ is a vector or {\it
polar vector}, 
$\Psibar i\g_5\Psi$ is a pseudoscalar,
and $\Psibar\g^\mu\g_5\Psi$ is a {\it pseudovector} or {\it axial vector}.

Turning to time reversal, from \eq{tpsit3} we get
\begin{equation}
T^{-1}\Psibar(x)T = \Psibar(\T x)\g_5\C^{-1} \;.
\label{tpsibart}
\end{equation}
Combining \eqs{tpsit3} and (\ref{tpsibart}), along with
$T^{-1}AT=A^*$, we find
\begin{equation}
T^{-1}\Bigl(\Psibar A\Psi\Bigr)T 
= \Psibar\Bigl(\g_5\C^{-1}\! A^* \C\g_5\Bigr)\Psi\;,
\label{bilint}
\end{equation}
where we have suppressed the spacetime arguments
(which transform in the obvious way).
Recall that $\C^{-1}\g^\mu\C=-(\g^\mu)^{\rm\sss T}$ and that
$\C^{-1}\g_5\C=\g_5$.  Also, 
$\g^0$ and $\g_5$ are real, hermitian, and square to one,
while $\g^i$ is antihermitian.  
Finally, $\g_5$ anticommutes with $\g^\mu$.
Using all of this info, we find
\begin{eqnarray}
\g_5\C^{-1} 1^* \C\g_5 &=& +1 \;,
\nonumber \\
\g_5\C^{-1} (i\g_5)^* \C\g_5 &=& -i\g_5 \;,
\nonumber \\
\g_5\C^{-1} (\g^0)^* \C\g_5 &=& +\g^0 \;,
\nonumber \\
\g_5\C^{-1} (\g^i)^* \C\g_5 &=& -\g^i \;,
\nonumber \\
\g_5\C^{-1} (\g^0\g_5)^* \C\g_5 &=& +\g^0\g_5 \;,
\nonumber \\
\g_5\C^{-1} (\g^i\g_5)^* \C\g_5 &=& -\g^i\g_5 \;.
\label{tgt}
\end{eqnarray}
Therefore,
\begin{eqnarray}
T^{-1}\Bigl(\Psibar\Psi\Bigr)T &=& +\,\Psibar\Psi \;,
\nonumber \\
T^{-1}\Bigl(\Psibar i\g_5\Psi\Bigr)T &=& -\,\Psibar i\g_5\Psi \;,
\nonumber \\
T^{-1}\Bigl(\Psibar\g^\mu\Psi\Bigr)T &=& -\,\T^\mu{}_\nu\Psibar\g^\nu\Psi \;,
\nonumber \\
T^{-1}\Bigl(\Psibar\g^\mu\g_5\Psi\Bigr)T &=& 
                           -\,\T^\mu{}_\nu\Psibar\g^\nu\g_5\Psi\;.
\label{bilint2}
\end{eqnarray}
Thus we see that $\Psibar\Psi$ is even under
time reversal, while $\Psibar i\g_5\Psi$,
$\Psibar\g^\mu\Psi$, and $\Psibar\g^\mu\g_5\Psi$ are odd.

For completeness we will also consider the transformation properties
of bilinears under charge conjugation.  Recall that
\begin{eqnarray}
C^{-1}\Psi(x)C &=& \C\Psibart\!(x) \;,
\nonumber \\
\noalign{\medskip}
C^{-1}\Psibar(x)C &=& \Psit\!(x)\C \;.
\label{cpsic2}
\end{eqnarray}
The bilinear $\Psibar A\Psi$ therefore transforms as
\begin{equation}
C^{-1}\Bigl(\Psibar A\Psi\Bigr)C = \Psit\C A\C\Psibart\;.
\label{bilinc}
\end{equation}
Since all indices are contracted, we can rewrite the right-hand side as
its transpose, with an extra minus sign for exchanging the order of the
two fermion fields.  We get
\begin{equation}
C^{-1}\Bigl(\Psibar A\Psi\Bigr)C 
= -\Psibar\C^{\rm\sss T}\!A^{\rm\sss T}\C^{\rm\sss T}\Psi \;.
\label{bilinc2}
\end{equation}
Recalling that $\C^{\rm\sss T}=\C^{-1}=-\C$, we have
\begin{equation}
C^{-1}\Bigl(\Psibar A\Psi\Bigr)C 
= \Psibar\Bigl(\C^{-1}\! A^{\rm\sss T}\C\Bigr)\Psi \;.
\label{bilinc3}
\end{equation}
Once again we can go through the list:
\begin{eqnarray}
\C^{-1}1^{\rm\sss T}\C &=& +1 \;,
\nonumber \\
\C^{-1}(i\g_5)^{\rm\sss T}\C &=& +i\g_5 \;,
\nonumber \\
\C^{-1}(\g^\mu)^{\rm\sss T}\C &=& -\g^\mu \;,
\nonumber \\
\C^{-1}(\g^\mu\g_5)^{\rm\sss T}\C &=& +\g^\mu\g_5 \;.
\label{cgc}
\end{eqnarray}
Therefore,
\begin{eqnarray}
C^{-1}\Bigl(\Psibar\Psi\Bigr)C &=& +\,\Psibar\Psi \;,
\nonumber \\
C^{-1}\Bigl(\Psibar i\g_5\Psi\Bigr)C &=& +\,\Psibar i\g_5\Psi \;,
\nonumber \\
C^{-1}\Bigl(\Psibar\g^\mu\Psi\Bigr)C &=& -\,\Psibar\g^\mu\Psi \;,
\nonumber \\
C^{-1}\Bigl(\Psibar\g^\mu\g_5\Psi\Bigr)C &=& +\,\Psibar\g^\mu\g_5\Psi\;.
\label{bilinc4}
\end{eqnarray}
Thus we see that $\Psibar\Psi$, $\Psibar i\g_5\Psi$, 
and $\Psibar\g^\mu\g_5\Psi$ are even under charge conjugation, 
while $\Psibar\g^\mu\Psi$ is odd.
For a Majorana field, this implies $\Psibar\g^\mu\Psi=0$.

Let us consider the combined effects of the three transformations
($C$, $P$, and $T$) on the bilinears.
From \eqs{bilinp2}, (\ref{bilint2}), and (\ref{bilinc4}),
we have
\begin{eqnarray}
(CPT)^{-1}\Bigl(\Psibar\Psi\Bigr)CPT &=& +\,\Psibar\Psi \;,
\nonumber \\
(CPT)^{-1}\Bigl(\Psibar i\g_5\Psi\Bigr)CPT &=& +\,\Psibar i\g_5\Psi \;,
\nonumber \\
(CPT)^{-1}\Bigl(\Psibar\g^\mu\Psi\Bigr)CPT &=& -\,\Psibar\g^\mu\Psi \;,
\nonumber \\
(CPT)^{-1}\Bigl(\Psibar\g^\mu\g_5\Psi\Bigr)CPT &=& -\,\Psibar\g^\mu\g_5\Psi\;,
\label{bilincpt}
\end{eqnarray}
where we have used $\P^\mu{}_\nu\T^\nu{}_\rho=-\delta^\mu{}_\rho$.
We see that $\Psibar\Psi$ and $\Psibar i\g_5\Psi$
are both even under $CPT$, while
$\Psibar\g^\mu\Psi$ and $\Psibar\g^\mu\g_5\Psi$ are both odd.
These are (it turns out) examples of a more general rule:
a fermion bilinear with $n$ vector
indices (and no uncontracted spinor indices) is even (odd)
under $CPT$ if $n$ is even (odd).
This also applies if we allow
derivatives acting on the fields, since
each factor of $\d_\mu$ is odd under the combination $PT$ 
and even under $C$.  

For scalar and vector fields, it is 
always possible to choose the phase factors in the $C$, $P$,
and $T$ transformations so that, overall, they obey the same rule:
a hermitian combination of fields and derivatives is even or odd depending on
the total number of uncontracted vector indices.
Putting this together with our result for fermion bilinears,
we see that any hermitian combination
of any set of fields (scalar, vector, Dirac, Majorana)
and their derivatives that is a Lorentz scalar (and so carries no indices)
is even under $CPT$.  Since the lagrangian must be formed out
of such combinations, we have $\L(x)\to\L(-x)$ under $CPT$,
and so the action $S=\int\dfx\,\L$ is invariant.  This is the $CPT$ theorem.

\vskip0.5in

\begin{center}
Problems
\end{center}

\vskip0.25in

40.1) Find the transformation properties of 
$\Psibar S^{\mu\nu} \Psi$ and 
$\Psibar iS^{\mu\nu}\g_5 \Psi$ under $P$, $T$, and $C$.
Verify that they are both even under $CPT$, as claimed.

\vfill\eject

\noindent Quantum Field Theory  \hfill   Mark Srednicki

\vskip0.5in

\begin{center}
\large{41: LSZ Reduction for Spin-One-Half Particles}
\end{center}
\begin{center}
Prerequisite: 39
\end{center}

\vskip0.5in

Let us now consider how to construct appropriate initial and final states for
scattering experiments.  
We will first consider the case of a Dirac field $\Psi$,
and assume that its interactions respect the U(1) symmetry that gives rise
to the conserved current $j^\mu=\Psibar\g^\mu\Psi$ and 
its associated charge $Q$.

In the free theory, we can create a state of one particle
by acting on the vacuum state with a creation operator:
\begin{eqnarray}
|p,s,{+}\ra &=& \bd_s(\p)|0\ra\;,
\label{1b} \\
\noalign{\medskip}
|p,s,{-}\ra &=& \dd_s(\p)|0\ra\;,
\label{1d}
\end{eqnarray}
where the label $\pm$ on the ket indicates the value of the U(1) charge $Q$, and
\begin{eqnarray}
b^\dagger_s(\p) &=& \int \dtx\;e^{ipx}\,\Psibar(x)\g^0 u_s(\p) \;,
\label{bd2} \\
\dd_s(\p) &=& \int \dtx\;e^{ipx}\,\vbar_s(\p)\g^0\Psi(x) \;.
\label{dd2}
\end{eqnarray}
Recall that $\bd_s(\p)$ and $\dd_s(\p)$ are time independent in the free theory.
The states $|p,s,{\pm}\ra$ have the Lorentz-invariant normalization
\begin{equation}
\la p,s,q|p',s',q'\ra 
= (2\pi)^3\,2\w\,\delta^3(\p-\p')\,\delta_{ss'}\,\delta_{qq'}\;,
\label{norm40}
\end{equation}
where $\w=(\p^2+m^2)^{1/2}$.

Let us consider an operator that (in the free theory) creates a particle 
with definite spin and charge, localized in
momentum space near $\p_1$, and localized in position space near the origin:
\begin{equation}
\bd_1 \equiv \int \dtp \; f_1(\p) \bd_{s_1}\!(\p)\;,
\label{bdag1}
\end{equation}
where
\begin{equation}
f_1(\p) \propto \exp[-(\p-\p_1)^2/4\sigma^2]
\label{f12}
\end{equation}
is an appropriate wave packet, and $\sigma$ is its width in momentum space.
If we time evolve (in the Schr\"odinger picture)
the state created by this time-independent operator,
then the wave packet will propagate (and spread out).  
The particle will thus be localized
far from the origin as $t\to\pm\infty$.  
If we consider instead an initial state of the form
$|i\ra = \bd_1\bd_2|0\ra$, where $\p_1\ne\p_2$, 
then we have two particles that are widely separated in the far past.  

Let us guess that this still works in the interacting theory.
One complication is that $\bd_s(\p)$ will no longer be time independent,
and so $\bd_1$, \eq{bdag1}, becomes time dependent as well.
Our guess for a suitable initial state for a scattering experiment is then
\begin{equation}
|i\ra = \lim_{t\to-\infty}\bd_1(t)\bd_2(t)|0\ra \;.
\label{init2}
\end{equation}
By appropriately normalizing the wave packets, we can make $\la i|i\ra=1$,
and we will assume that this is the case.
Similarly, we can consider a final state
\begin{equation}
|f\ra = \lim_{t\to+\infty}\bd_{1'}(t)\bd_{2'}(t)|0\ra \;,
\label{final2}
\end{equation}
where $\p'_1\ne\p'_2$, and $\la f|f\ra=1$.
This describes two widely separated particles in the far future.
(We could also consider acting with more creation operators, if 
we are interested in the production of some
extra particles in the collision of two, or using $\dd$ operators
instead of $\bd$ operators for some or all of the initial and final particles.)
Now the scattering amplitude is simply given by $\la f|i\ra$.

We need to find a more useful expression for $\la f|i\ra$.  
To this end, let us note that
\begin{eqnarray}
\bd_1(-\infty) &-& \bd_1(+\infty)
\nonumber \\
&=& - \int_{-\infty}^{+\infty} dt\;\d_0\bd_1(t)
\nonumber \\
&=& - \int \dtp\;f_1(\p) \int \dfx\;
\d_0\Bigl(e^{ipx}\,\Psibar(x)\g^0 u_{s_1}(\p)\Bigr)\;.\qquad
\nonumber \\
&=& -\int \dtp\;f_1(\p) \int \dfx\; 
\Psibar(x)\Bigl(\g^0{\buildrel\leftarrow\over\d_0}-i\g^0 p^0\Bigr)
        u_{s_1}(\p)e^{ipx}
\nonumber \\
&=& -\int \dtp\;f_1(\p) \int \dfx\; 
\Psibar(x)\Bigl(\g^0{\buildrel\leftarrow\over\d_0}-i\g^i p_i -im\Bigr)
        u_{s_1}(\p)e^{ipx}
\nonumber \\
&=& -\int \dtp\;f_1(\p) \int \dfx\; 
\Psibar(x)\Bigl(\g^0{\buildrel\leftarrow\over\d_0}
-\g^i{\buildrel\rightarrow\over\d_i}  -im\Bigr)
        u_{s_1}(\p)e^{ipx}
\nonumber \\
&=& -\int \dtp\;f_1(\p) \int \dfx\; 
\Psibar(x)\Bigl(\g^0{\buildrel\leftarrow\over\d_0}
+\g^i{\buildrel\leftarrow\over\d_i}  -im\Bigr)
        u_{s_1}(\p)e^{ipx}
\nonumber \\
&=& i\int \dtp\;f_1(\p) \int \dfx\; 
 \Psibar(x)(+i{\buildrel\leftarrow\over\dsl} + m)u_s(\p)e^{ipx}
        u_{s_1}(\p)e^{ipx} \;.
\label{bddiff}
\end{eqnarray}
The first equality is just the fundamental theorem of calculus.
To get the second, we substituted the definition of $\bd_1(t)$,
and combined the $\dtx$ from this definition with the $dt$ to get $\dfx$.
The third comes from straightforward evaluation 
of the time derivatives.  The fourth uses $(\psl+m)u_s(\p)=0$.
The fifth writes $i p_i$ as $\d_i$ acting on $e^{ipx}$.
The sixth uses integration by parts to move the $\d_i$ onto
the field $\Psibar(x)$; here the wave packet is needed to avoid
a surface term.  The seventh simply identifies $\g^0\d_0+\g^i\d_i$ as $\dsl$.  

In free-field theory, the right-hand side of \eq{bddiff} is zero,
since $\Psi(x)$ obeys the Dirac equation, which, after barring it, reads 
\begin{equation}
\Psibar(x)(+i{\buildrel\leftarrow\over\dsl} + m)=0\;.
\label{diracbar}
\end{equation}
In an interacting theory, however,
the right-hand side of \eq{bddiff} will not be zero.

We will also need the hermitian conjugate of \eq{bddiff}, which (after
some slight rearranging) reads
\begin{eqnarray}
b_1(+\infty) &-& b_1(-\infty) 
\nonumber \\
&=& i\int \dtp\;f_1(\p) \int \dfx\; 
e^{-ipx}\;\ubar_{s_1}(\p)(-i\dsl+m)\Psi(x)\;,
\label{bdiff}
\end{eqnarray}
and the analogous formulae for the $d$ operators,
\begin{eqnarray}
\dd_1(-\infty) &-& \dd_1(+\infty)
\nonumber \\
&=& -i\int \dtp\;f_1(\p) \int \dfx\; 
e^{ipx}\;\vbar_{s_1}(\p)(-i\dsl+m)\Psi(x)\;,
\label{dddiff} \\
\noalign{\medskip}
d_1(+\infty) &-& d_1(-\infty)
\nonumber \\
&=&- i\int \dtp\;f_1(\p) \int \dfx\; 
\Psibar(x)(+i{\buildrel\leftarrow\over\dsl} + m)v_{s_1}(\p)e^{-ipx} \;.
\label{ddiff}
\end{eqnarray}

Let us now return to the scattering amplitude we were considering,
\begin{equation}
\la f|i\ra = \la0|b_{2'}(+\infty)b_{1'}(+\infty)
                  \bd_1(-\infty)\bd_2(-\infty)|0\ra \;.
\label{fi2}
\end{equation}
Note that the operators are in time order.  Thus, if we feel like it, we can 
put in a {\it time-ordering symbol\/} without changing anything:
\begin{equation}
\la f|i\ra = \la0|\,{\rm T}\,b_{2'}(+\infty)b_{1'}(+\infty)
                  \bd_1(-\infty)\bd_2(-\infty)|0\ra \;.
\label{fit2}
\end{equation}
The symbol T means the product of operators
to its right is to be ordered, not as written, but with operators at
later times to the left of those at earlier times.  
However, {\it there is an extra minus sign if this rearrangement involves
an odd number of exchanges of these anticommuting operators}.

Now let us use \eqs{bddiff} and (\ref{bdiff}) in \eq{fit2}.  The time-ordering
symbol automatically moves all $b_{i'}(-\infty)$'s to the right, where they
annihilate $|0\ra$.  Similarly, all $\bd_i(+\infty)$'s move to the left,
where they annihilate $\la 0|$.  

The wave packets no longer play a key role, and we can take the $\sigma\to0$
limit in \eq{f12}, so that $f_1(\p)=\delta^3(\p-\p_1)$.  
The initial and final states now have a delta-function normalization,
the multiparticle generalization of \eq{norm40}.  We are left with the
{\it Lehmann-Symanzik-Zimmerman reduction formula\/} for 
spin-one-half particles,
\begin{eqnarray}
\la f|i\ra &=& 
i^{4}\int \dfx_1\,\dfx_2\,\dfx_{1'}\,\dfx_{2'}
\nonumber \\
&&\qquad {} \times
e^{-ip'_1 x'_1}\,
[\ubar_{s_{1'}}(\p_{1'})(-i\dsl_{1'}+m)]_{\alpha_{1'}}
\nonumber \\
\noalign{\medskip}
&&\qquad {} \times
e^{-ip'_2 x'_2}\,
[\ubar_{s_{2'}}(\p_{2'})(-i\dsl_{2'}+m)]_{\alpha_{2'}}
\nonumber \\
\noalign{\medskip}
&&\qquad {} \times
\la 0 |\,{\rm T}\,
\Psi_{\alpha_{2'}}(x_{2'})
\Psi_{\alpha_{1'}}(x_{1'})
\Psibar_{\alpha_1}(x_1)
\Psibar_{\alpha_2}(x_2)
|0\ra
\nonumber \\
\noalign{\medskip}
&&\qquad {} \times
[(+i\smash{\buildrel\leftarrow\over\dsl}_1+m)u_{s_1}(\p_1)]_{\alpha_1}
\,e^{ip_1 x_1}
\nonumber \\
\noalign{\medskip}
&&\qquad {} \times
[(+i\smash{\buildrel\leftarrow\over\dsl}_2+m)u_{s_2}(\p_2)]_{\alpha_2}
\,e^{ip_2 x_2}\;.
\label{lsz0b}
\end{eqnarray}
The generalization of the LSZ formula to other processes should be clear;
insert a time-ordering symbol, and make the following replacements:
\begin{eqnarray}
\bd_s(\p)_{\rm in} &\to& + i\int\dfx\;
\Psibar(x)(+i{\buildrel\leftarrow\over\dsl} + m)u_s(\p)\,e^{+ipx}\;,
\label{bdsub} \\
\noalign{\medskip}
b_s(\p)_{\rm out} &\to& + i\int\dfx\;
e^{-ipx}\;\ubar_s(\p)(-i\dsl+m)\Psi(x)\;,
\label{bsub} \\
\noalign{\medskip}
\dd_s(\p)_{\rm in} &\to& - i\int\dfx\;
e^{+ipx}\;\vbar_s(\p)(-i\dsl+m)\Psi(x)\;,
\label{ddsub} \\
\noalign{\medskip}
d_s(\p)_{\rm out} &\to& - i\int\dfx\;
\Psibar(x)(+i{\buildrel\leftarrow\over\dsl} + m)v_s(\p)\,e^{-ipx} \;,
\label{dsub}
\end{eqnarray}
where we have used the subscripts ``in'' and ``out'' to denote
$t\to-\infty$ and $t\to+\infty$, respectively.

All of this holds for a Majorana field as well.  
In that case, $d_s(\p)=b_s(\p)$, 
and we can use {\it either\/} \eq{bdsub} {\it or\/} \eq{ddsub} 
for the incoming particles,
and {\it either\/} \eq{bsub} {\it or\/} \eq{dsub} 
for the outgoing particles,
whichever is more convenient.
The Majorana condition $\Psibar=\Psi^{\rm\sss T}\C$ guarantees 
that the results will be equivalent.

As in the case of a scalar field, 
we cheated a little in our derivation of the LSZ formula, because
we assumed that the creation operators of {\it free\/} field theory would
work comparably in the {\it interacting\/} theory.  
After performing an analysis that is entirely analogous to what we
did for the scalar in section 5, we come to the same conclusion:
the LSZ formula holds provided the field is properly
normalized.  For a Dirac field, we must require
\begin{eqnarray}
\la 0|\Psi(x)|0\ra &=& 0 \;,
\label{0psi0} \\
\noalign{\smallskip}
\la p,s,{+}|\Psi(x)|0\ra &=& 0 \;,
\label{+psi0} \\
\noalign{\smallskip}
\la p,s,{-}|\Psi(x)|0\ra &=& v_s(p)e^{-ipx} \;,
\label{-psi0} \\
\noalign{\smallskip}
\la p,s,{+}|\Psibar(x)|0\ra &=& \ubar_s(p)e^{-ipx} \;,
\label{+psibar0} \\
\noalign{\smallskip}
\la p,s,{-}|\Psibar(x)|0\ra &=&  0 \;,
\label{-psibar0}
\end{eqnarray}
where $\la 0|0\ra=1$, and 
the one-particle states are normalized according to \eq{norm40}.  

The zeros on the right-hand sides of \eqs{+psi0} and (\ref{-psibar0})
are required by charge conservation.  To see this, start with
$[Q,\Psi(x)]=-\Psi(x)$, take the matrix elements indicated,
and use $Q|0\ra=0$ and $Q|p,s,{\pm}\ra=\pm|p,s,{\pm}\ra$.

The zero on the right-hand side of \eq{0psi0} is required by Lorentz
invariance.  To see this, start with
$[M^{\mu\nu},\Psi(0)]=S^{\mu\nu}\Psi(0)$,
and take the expectation value in the vacuum state $|0\ra$. 
If $|0\ra$ is Lorentz invariant (as we will assume), then
it is annihilated by the Lorentz generators $M^{\mu\nu}$, which means
that we must have $S^{\mu\nu}\la0|\Psi(0)|0\ra=0$; 
this is possible for all $\mu$ and $\nu$
only if $\la0|\Psi(0)|0\ra=0$, which (by translation invariance) is possible
only if $\la0|\Psi(x)|0\ra=0$.

The right-hand sides of \eqs{-psi0} and (\ref{+psibar0}) are similarly
fixed by Lorentz invariance: only the overall scale might be different
in an interacting theory.  However, 
the LSZ formula is correctly normalized if and only if 
\eqs{-psi0} and (\ref{+psibar0}) hold as written.
We will enforce this by rescaling 
(or, one might say, {\it renormalizing\/}) $\Psi(x)$ by an overall constant.
This is just a change of the name of the operator of interest,
and does not affect the physics.  However, the rescaled
$\Psi(x)$ will obey \eqs{-psi0} and (\ref{+psibar0}). 
(These two equations are related by charge conjugation,
and so actually constitute only one condition on $\Psi$.)

For a Majorana field, there is no conserved charge, and we have
\begin{eqnarray}
\la 0|\Psi(x)|0\ra &=& 0 \;,
\label{0maj0} \\
\noalign{\smallskip}
\la p,s|\Psi(x)|0\ra &=& v_s(p)e^{-ipx} \;,
\label{maj0} \\
\noalign{\smallskip}
\la p,s|\Psibar(x)|0\ra &=& \ubar_s(p)e^{-ipx} \;,
\label{0maj}
\end{eqnarray}
instead of eqs.$\,$(\ref{0psi0}--\ref{-psibar0}).
  
The renormalization of $\Psi$ necessitates including appropriate $Z$
factors in the lagrangian.  Consider, for example, 
\begin{equation}
\L = 
i Z \Psibar\dsl\Psi - Z_m m\Psibar\Psi -{\ts{1\over4}}Z_g g(\Psibar\Psi)^2\;,
\label{ellg}
\end{equation}
where $\Psi$ is a Dirac field, and $g$ is a coupling constant.  We must
choose the three constants $Z$, $Z_m$, and $Z_g$ so that the following three
conditions are satisfied: $m$ is the mass of a 
single particle; $g$ is fixed by some appropriate scattering cross section;
and \eq{-psi0} and is obeyed. 
[\Eq{+psibar0} then follows by charge conjugation.]

Next, we must develop the tools needed to compute the correlation functions
$\la 0 |{\rm T}\Psi_{\alpha_{1'}}(x_{1'})\ldots
\Psibar_{\alpha_1}(x_1)\ldots |0\ra$ in an interacting quantum field theory.

\vfill\eject

\noindent Quantum Field Theory  \hfill   Mark Srednicki

\vskip0.5in

\begin{center}
\large{42: The Free Fermion Propagator}
\end{center}
\begin{center}
Prerequisite: 39
\end{center}

\vskip0.5in

Consider a free Dirac field
\begin{eqnarray}
\Psi(x) &=& \sum_{s=\pm}\int\dpt\,\left[b_s(\p)  u_s(\p)e^{ipx}
                              +d_s^\dagger(\p)v_s(\p)e^{-ipx} \right]\;,
\label{Psi3} \\
\Psibar(y) &=& \sum_{s'=\pm}\int\dpt{\,}'
             \,\left[\bd_{s'}(\p')\ubar_{s'}(\p')e^{-ip'y}
                              +d_{s'}(\p')\vbar_{s'}(\p')e^{ip'y} \right]\;,
\label{Psibar3}
\end{eqnarray}
where
\begin{equation}
b_s(\p)|0\ra = d_s(\p)|0\ra = 0\;,
\label{bd0ra}
\end{equation}
and
\begin{eqnarray}
\{b_s(\p),b^\dagger_{s'}(\p')\} &=& (2\pi)^3 \delta^3(\p-\p')\,2\w\delta_{ss'}\;,
\label{bbdcomm2} \\
\noalign{\smallskip}
\{d_s(\p),\dd_{s'}(\p')\} &=& (2\pi)^3 \delta^3(\p-\p')\,2\w\delta_{ss'}\;,
\label{dddcomm2}
\end{eqnarray}
and all the other possible anticommutators between $b$ and $d$ operators
(and their hermitian conjugates) vanish.

We wish to compute the Feynman propagator
\begin{equation}
S(x-y)_{\alpha\beta}\equiv i\la 0 |{\rm T}\Psi_\alpha(x)\Psibar_\beta(y)|0\ra\;,
\label{dirprop}
\end{equation}
where T denotes the time-ordered product,
\begin{equation}
{\rm T}\Psi_\alpha(x)\Psibar_\beta(y) \equiv
\theta(x^0-y^0) \Psi_\alpha(x)\Psibar_\beta(y) -
\theta(y^0-x^0) \Psibar_\beta(y) \Psi_\alpha(x) \;,
\label{tprod}
\end{equation}
and $\theta(t)$ is the unit step function.
Note the minus sign in the second term; this is needed because
$\Psi_\alpha(x)\Psibar_\beta(y) =-\Psibar_\beta(y)\Psi_\alpha(x)$
when $x^0\ne y^0$.

We can now compute 
$\la 0 |\Psi_\alpha(x)\Psibar_\beta(y)|0\ra$
and
$\la 0 |\Psibar_\beta(y) \Psi_\alpha(x)|0\ra$
by inserting \eqs{Psi3} and (\ref{Psibar3}), and then using
eqs.$\,$(\ref{bd0ra}--\ref{dddcomm2}).  We get
\vfill\eject
\begin{eqnarray}
&& \la 0 |\Psi_\alpha(x)\Psibar_\beta(y)|0\ra
\nonumber \\
\noalign{\medskip}
&& \qquad {} =
\sum_{s,s'}\int\dpt\,\dpt{\,}'\, e^{ipx}\, e^{-ip'y}\,
u_s(\p)_\alpha \ubar_{s'}(\p')_\beta \,
\la0|b_s(\p)\bd_{s'}(\p')|0\ra
\nonumber \\
&& \qquad {} =
\sum_{s,s'}\int\dpt\,\dpt{\,}'\, e^{ipx}\, e^{-ip'y}\,
u_s(\p)_\alpha \ubar_{s'}(\p')_\beta \,
(2\pi)^3 \delta^3(\p-\p')\,2\w\delta_{ss'}
\nonumber \\
&& \qquad {} =
\sum_s\int\dpt\;e^{ip(x-y)}\,
u_s(\p)_\alpha \ubar_s(\p)_\beta 
\nonumber \\
&& \qquad {} =
\int\dpt\;e^{ip(x-y)}\,(-\psl+m)_{\alpha\beta} \;. 
\label{dirprop2}
\end{eqnarray}
To get the last line, we used a result from section 38.
Similarly,
\begin{eqnarray}
&& \la 0 |\Psibar_\beta(y) \Psi_\alpha(x) |0\ra
\nonumber \\
\noalign{\medskip}
&& \qquad {} =
\sum_{s,s'}\int\dpt\,\dpt{\,}'\, e^{-ipx}\, e^{ip'y}\,
v_s(\p)_\alpha \vbar_{s'}(\p')_\beta \,
\la0|d_{s'}(\p')\dd_s(\p)|0\ra
\nonumber \\
&& \qquad {} =
\sum_{s,s'}\int\dpt\,\dpt{}'\, e^{-ipx}\, e^{ip'y}\,
v_s(\p)_\alpha \vbar_{s'}(\p')_\beta \,
(2\pi)^3 \delta^3(\p-\p')\,2\w\delta_{ss'}
\nonumber \\
&& \qquad {} =
\sum_s\int\dpt\;e^{-ip(x-y)}\,
v_s(\p)_\alpha \vbar_s(\p)_\beta \,
\nonumber \\
&& \qquad {} =
\int\dpt\;e^{-ip(x-y)}\,(-\psl-m)_{\alpha\beta} \;. 
\label{dirprop3}
\end{eqnarray}
We can combine \eqs{dirprop2} and (\ref{dirprop3}) 
into a compact formula for the time-ordered product by means of the identity
\begin{eqnarray}
\int {\dfp\over(2\pi)^4}\,{e^{ip(x-y)}f(p)\over p^2+m^2-i\eps}
&=& i\theta(x^0{-}y^0)\int\dpt\,e^{ip(x-y)}\,f(p)
\nonumber \\
&& {}
+i\theta(y^0{-}x^0)\int\dpt\,e^{-ip(x-y)}\,f(-p)\;,\qquad
\label{cont9}
\end{eqnarray}
where $f(p)$ is a polynomial in $p$;
the derivation of \eq{cont9} was sketched in section 8.
We get
\begin{equation}
\la 0 |{\rm T}\Psi_\alpha(x)\Psibar_\beta(y)|0\ra
= {1\over i} \int {\dfp\over(2\pi)^4}\,e^{ip(x-y)}\,{(-\psl+m)_{\alpha\beta}
                                        \over p^2+m^2-i\eps}\;,
\label{dirprop4}
\end{equation}
and so
\begin{equation}
S(x-y)_{\alpha\beta}
= \int {\dfp\over(2\pi)^4}\,e^{ip(x-y)}\,{(-\psl+m)_{\alpha\beta}
                                        \over p^2+m^2-i\eps}\;.
\label{dirprop5}
\end{equation}
Note that $S(x-y)$ is a Green's function for the Dirac wave operator:
\begin{eqnarray}
(-i\dslx + m)_{\alpha\beta}S(x-y)_{\beta\gamma}
&=& \int {\dfp\over(2\pi)^4}\, e^{ip(x-y)}\,
{(\psl + m)_{\alpha\beta}(-\psl+m)_{\beta\gamma} \over p^2+m^2-i\eps}
\nonumber \\
\noalign{\smallskip}
&=& \int {\dfp\over(2\pi)^4}\,
e^{ip(x-y)}\,{(p^2 + m^2)\delta_{\alpha\gamma} \over p^2+m^2-i\eps}
\nonumber \\
\noalign{\medskip}
&=& \delta^4(x-y)\delta_{\alpha\gamma} \;.
\label{greens}
\end{eqnarray}
Similarly,
\begin{eqnarray}
S(x-y)_{\alpha\beta} (+i {\buildrel\leftarrow\over\dsly} + m)_{\beta\gamma}
&=& \int {\dfp\over(2\pi)^4}\, e^{ip(x-y)}\,
{(-\psl + m)_{\alpha\beta}(\psl+m)_{\beta\gamma} \over p^2+m^2-i\eps}
\nonumber \\
\noalign{\smallskip}
&=& \int {\dfp\over(2\pi)^4}\,
e^{ip(x-y)}\,{(p^2 + m^2)\delta_{\alpha\gamma} \over p^2+m^2-i\eps}
\nonumber \\
\noalign{\medskip}
&=& \delta^4(x-y)\delta_{\alpha\gamma} \;.
\label{greens2}
\end{eqnarray}
We can also consider
$\la 0 |{\rm T}\Psi_\alpha(x)\Psi_\beta(y)|0\ra$ 
and $\la 0 |{\rm T}\Psibar_\alpha(x)\Psibar_\beta(y)|0\ra$,
but it is easy to see that now there is no way to pair up a
$b$ with a $\bd$ or a $d$ with a $\dd$, and so
\begin{eqnarray}
\la 0 |{\rm T}\Psi_\alpha(x)\Psi_\beta(y)|0\ra &=& 0\;,
\label{pp0} \\
\noalign{\medskip}
\la 0 |{\rm T}\Psibar_\alpha(x)\Psibar_\beta(y)|0\ra &=& 0 \;.
\label{pppp0}
\end{eqnarray}

Next, consider a Majorana field
\begin{eqnarray}
\Psi(x) &=& \sum_{s=\pm}\int\dpt\,\left[b_s(\p)  u_s(\p)e^{ipx}
                              +b_s^\dagger(\p)v_s(\p)e^{-ipx} \right]\;,
\label{Maj3} \\
\Psibar(y) &=& \sum_{s'=\pm}\int\dpt{\,}'
             \,\left[\bd_{s'}(\p')\ubar_{s'}(\p')e^{-ip'y}
                              +b_{s'}(\p')\vbar_{s'}(\p')e^{ip'y} \right]\;.
\label{Majbar3}
\end{eqnarray}
It is easy to see that 
$\la 0 |{\rm T}\Psi_\alpha(x)\Psibar_\beta(y)|0\ra$ 
is the same as it is in the Dirac case;
the only difference in the calculation is that we would have 
$b$ and $\bd$ in place of
$d$ and $\dd$ in the second line of \eq{dirprop3}, and this does not
change the final result.  Thus,
\begin{equation}
i\la 0 |{\rm T}\Psi_\alpha(x)\Psibar_\beta(y)|0\ra
= S(x-y)_{\alpha\beta}\;,
\label{majprop}
\end{equation}
where $S(x-y)$ is given by \eq{dirprop5}.

However, \eqs{pp0}
and (\ref{pppp0}) no longer hold for a Majorana field.  
Instead, the Majorana condition
$\Psibar=\Psit\C$, which can be rewritten as 
$\Psit=\Psibar\C^{-1}$, implies
\begin{eqnarray}
i\la 0 |{\rm T}\Psi_\alpha(x)\Psi_\beta(y)|0\ra 
&=& i\la 0 |{\rm T}\Psi_\alpha(x)\Psibar_\gamma(y)|0\ra(\C^{-1})_{\gamma\beta} 
\nonumber \\
\noalign{\medskip}
&=& [S(x-y)\C^{-1}]_{\alpha\beta} \;.
\label{pp0m}
\end{eqnarray}
Similarly, using $\C^{\rm\sss T}=\C^{-1}$, we can write the Majorana condition as
$\Psibart=\C^{-1}\Psi$, and so 
\begin{eqnarray}
i\la 0 |{\rm T}\Psibar_\alpha(x)\Psibar_\beta(y)|0\ra 
&=& i(\C^{-1})_{\alpha\gamma}\la 0 |{\rm T}\Psi_\gamma(x)\Psibar_\beta(y)|0\ra 
\nonumber \\
\noalign{\medskip}
&=& [\C^{-1}S(x-y)]_{\alpha\beta} \;.
\label{pppp0m}
\end{eqnarray}
Of course, $\C^{-1}=-\C$, but it will prove more convenient to leave
\eqs{pp0m} and (\ref{pppp0m}) as they are.

We can also consider the vacuum expectation value of a time-ordered product
of more than two fields.  In the Dirac case, we must have an equal number of
$\Psi$'s and $\Psibar$'s to get a nonzero result; and then, the 
$\Psi$'s and $\Psibar$'s must pair up to form propagators.  There
is an extra minus sign if the ordering of the fields in their pairs
is an odd permutation of the original ordering.  For example,
\begin{eqnarray}
i^2\la 0 |{\rm T}\Psi_\alpha(x)\Psibar_\beta(y)
\Psi_\gamma(z)\Psibar_\delta(w)|0\ra
&=& {} + S(x-y)_{\alpha\beta}\;S(z-w)_{\gamma\delta}
\nonumber\\
\noalign{\smallskip}
&&  {} - S(x-w)_{\alpha\delta}\;S(z-y)_{\gamma\beta} \;.\qquad
\label{pppbpb}
\end{eqnarray}
In the Majorana case, we may as well let all the fields be $\Psi$'s
(since we can always replace a $\Psibar$ with $\Psit\C$).  Then we
must pair them up in all possible ways.  There
is an extra minus sign if the ordering of the fields in their pairs
is an odd permutation of the original ordering.  For example,
\begin{eqnarray}
i^2\la 0 |{\rm T}\Psi_\alpha(x)\Psi_\beta(y)
\Psi_\gamma(z)\Psi_\delta(w)|0\ra
&=& {} + [S(x-y)\C^{-1}]_{\alpha\beta}\;
              [S(z-w)\C^{-1}]_{\gamma\delta}
\nonumber\\
\noalign{\smallskip}
&& {} - [S(x-z)\C^{-1}]_{\alpha\gamma}\;
              [S(y-w)\C^{-1}]_{\beta\delta}
\nonumber\\
\noalign{\smallskip}
&& {} + [S(x-w)\C^{-1}]_{\alpha\delta}\;
              [S(y-z)\C^{-1}]_{\beta\gamma} \;. 
\nonumber\\
\label{pppp}
\end{eqnarray}
Note that the ordering within a pair does not matter, since
\begin{equation}
[S(x-y)\C^{-1}]_{\alpha\beta} = -[S(y-x)\C^{-1}]_{\beta\alpha} \;.
\label{majprop5}
\end{equation}
This follows from anticommutation of the fields and \eq{pp0m};
it can also be proven directly using $\C\g^\mu\C^{-1}=-(\g^\mu)^{\rm\sss T}$
and $\C^{-1}=\C^{\rm\sss T}=-\C$.

\vfill\eject

\noindent Quantum Field Theory  \hfill   Mark Srednicki

\vskip0.5in

\begin{center}
\large{43: The Path Integral for Fermion Fields}
\end{center}
\begin{center}
Prerequisite: 9, 42
\end{center}

\vskip0.5in

We would like to write down a path integral formula for the
vacuum-expectation value of a time-ordered product of free Dirac
or Majorana fields.
Recall that for a real scalar field with
\begin{eqnarray}
\L_0 &=& -\half\d^\mu\ph\d_\mu\ph - \half m^2\ph^2 
\nonumber \\
\noalign{\medskip}
&=& -\half\ph(-\d^2 + m^2)\ph - \half\d_\mu(\ph\d^\mu\ph)\;,
\label{ellph}
\end{eqnarray}
we have
\begin{equation}
\la 0|{\rm T}\ph(x_1)\ldots|0\ra =
{1\over i}\,{\delta\over\delta J(x_1)}\ldots 
Z_0(J)\Bigr|_{J=0}\;,
\label{phph42}
\end{equation}
where
\begin{equation}
Z_0(J) = \int\D\ph\;\exp\!\left[i\int\dfx\,(\L_0+J\ph)\right].
\label{z0420}
\end{equation}
In this formula, we use the epsilon trick (see section 6) of replacing
$m^2$ with $m^2-i\epsilon$ to construct the vacuum as the
initial and final state.  Then we get
\begin{equation}
Z_0(J) = \exp\!\left[{i\over2}\int \dfx\,\dfy\,J(x)\Delta(x-y)J(y)\right],
\label{z042}
\end{equation}
where the Feynman propagator 
\begin{equation}
\Delta(x-y)=\int{\dfk\over(2\pi)^4}\;{e^{ik(x-y)}\over k^2+m^2-i\epsilon}
\label{feyndelt42}
\end{equation}
is the inverse of the Klein-Gordon wave operator:
\begin{equation}
(-\d_x^2 + m^2)\Delta(x-y) = \delta^4(x-y)\;.
\label{feyn42}
\end{equation}

For a complex scalar field with
\begin{eqnarray}
\L_0 &=& -\d^\mu\ph^\dagger\d_\mu\ph - m^2\ph^\dagger\ph
\nonumber \\
\noalign{\medskip}
&=& -\ph^\dagger(-\d^2 + m^2)\ph - \d_\mu(\ph^\dagger\d^\mu\ph)\;,
\label{ellphd}
\end{eqnarray}
we have instead
\begin{equation}
\la 0|{\rm T}\ph(x_1)\ldots\ph^\dagger(y_1)\ldots|0\ra =
{1\over i}\,{\delta\over\delta J^\dagger(x_1)}\ldots 
{1\over i}\,{\delta\over\delta J(y_1)}\ldots 
Z_0(J^\dagger,J)\Bigr|_{J=J^\dagger=0}\;,
\label{phphd42}
\end{equation}
where
\begin{eqnarray}
Z_0(J^\dagger,J) &=& \int\D\ph^\dagger\D\ph
\;\exp\!\left[i\int\dfx\,(\L_0+J^\dagger\ph+\ph^\dagger J)\right]
\nonumber \\
\noalign{\medskip}
&=& \exp\!\left[i\int \dfx\,\dfy\,J^\dagger(x)\Delta(x-y)J(y)\right].
\label{zd042}
\end{eqnarray}
We treat $J$ and $J^\dagger$ as independent variables when evaluating \eq{phphd42}.

In the case of a fermion field, we should have something similar, except
that we need to account for the extra minus signs from anticommutation.
For this to work out, a functional derivative with respect to an anticommuting
variable must itself be treated as anticommuting.  Thus if we define an
anticommuting source $\eta(x)$ for a Dirac field, we can write
\begin{eqnarray}
{\delta\over\delta\eta(x)}
\int\dfy\,\Bigl[\etabar(y)\Psi(y)+\Psibar(y)\eta(y)\Bigr]
&=& -\Psibar(x) \;,
\label{ddeta} \\
{\delta\over\delta\etabar(x)}
\int\dfy\,\Bigl[\etabar(y)\Psi(y)+\Psibar(y)\eta(y)\Bigr]
&=& +\Psi(x) \;.
\label{ddetabar}
\end{eqnarray}
The minus sign in \eq{ddeta} arises because the $\delta/\delta\eta$ must pass
through $\Psibar$ before reaching $\eta$. 

Thus, consider a free Dirac field with
\begin{eqnarray}
\L_0 &=& i\Psibar\dsl\Psi - m\Psibar\Psi
\nonumber \\
\noalign{\medskip}
&=& -\Psibar(-i\dsl+m)\Psi \;.
\label{ellPsi0}
\end{eqnarray}
A natural guess for the appropriate path-integral formula, 
based on analogy with \eq{zd042}, is 
\begin{eqnarray}
&& \la 0 |{\rm T}\Psi_{\alpha_1}(x_1)\ldots\Psibar_{\beta_1}(y_1)\ldots |0\ra 
\nonumber \\
\noalign{\smallskip}
&& \qquad {} = 
{1\over i}\,{\delta\over\delta\etabar_{\alpha_1}(x_1)}\; \ldots \;
i\,{\delta\over\delta \eta_{\beta_1}(y_1)}\; \ldots \;
Z_0(\etabar,\eta)\Bigr|_{\eta=\etabar=0}\;,
\label{psipsibar}
\end{eqnarray}
where
\begin{eqnarray}
Z_0(\etabar,\eta)
&=& \int\D\Psi\,\D\Psibar\;\exp\!\left[
i\int\dfx\,(\L_0+\etabar\Psi+\Psibar\eta)\right]
\nonumber \\
\noalign{\smallskip}
&=& \exp\!\left[i\int \dfx\,\dfy\;\etabar(x)S(x-y)\eta(y)\right],
\label{zpsid042}
\end{eqnarray}
and the Feynman propagator 
\begin{equation}
S(x-y)=\int{\dfp\over(2\pi)^4}\;{(-\psl+m)e^{ip(x-y)}
                 \over p^2+m^2-i\epsilon} 
\label{feyns}
\end{equation}
is the inverse of the Dirac wave operator:
\begin{equation}
(-i\dsl_x + m)S(x-y) = \delta^4(x-y)\;.
\label{feyndir}
\end{equation}
Note that each $\delta/\delta\eta$ in \eq{psipsibar} comes with a factor
of $i$ rather than the usual $1/i$; this reflects the extra minus sign of \eq{ddeta}.
We treat $\eta$ and $\etabar$ as independent variables when evaluating \eq{psipsibar}.
It is straightforward to check (by working out a few examples)
that eqs.$\,$(\ref{psipsibar}--\ref{feyndir}) do indeed reproduce the
result of section 42 for the vacuum expectation value of a time-ordered product
of Dirac fields.

This is really all we need to know.  Recall that, for a complex scalar
field with interactions specified by $\L_1(\ph^\dagger,\ph)$, we have
\begin{equation}
Z(J^\dagger,J) \;\propto\;
\exp\!\left[\,i\int\dfx\;\L_1\!\!\left({1\over i}{\delta\over\delta J(x)},
                 {1\over i}{\delta\over\delta J^\dagger(x)} \right)\right]
Z_0(J^\dagger,J)\;,
\label{zd1}
\end{equation}
where the overall normalization is fixed by $Z(0,0)=1$.
Thus, for a Dirac field 
with interactions specified by $\L_1(\Psibar,\Psi)$, we have
\begin{equation}
Z(\etabar,\eta) \;\propto\;
\exp\!\left[\,i\int\dfx\;\L_1\!\!\left(i{\delta\over\delta\eta(x)},
                        {1\over i}{\delta\over\delta\etabar(x)}\right)\right]
Z_0(\etabar,\eta)\;,
\label{zppb1}
\end{equation}
where again the overall normalization is fixed by $Z(0,0)=1$.
Vacuum expectation values of time-ordered products of Dirac fields in an
interacting theory will now be given by \eq{psipsibar}, 
but with $Z_0(\etabar,\eta)$ 
replaced by $Z(\etabar,\eta)$.  Then, just as for a scalar field, this will
lead to a Feynman-diagram expansion for $Z(\etabar,\eta)$.  There are two extra
complications: we must keep track of the spinor indices, and we must
keep track of the extra minus signs from anticommutation.  Both tasks are
straightforward; we will take them up in section 45.  

Next, let us consider a Majorana field with
\begin{eqnarray}
\L_0 &=& {\ts{i\over2}}\Psit\C\dsl\Psi- \half m\Psit\C\Psi
\nonumber \\
\noalign{\medskip}
&=& -\half\Psit\C(-i\dsl + m)\Psi \;.
\label{ellMaj0}
\end{eqnarray}
A natural guess for the appropriate path-integral formula, 
based on analogy with \eq{phph42}, is
\begin{equation}
\la 0 |{\rm T}\Psi_{\alpha_1}(x_1)\ldots|0\ra =
{1\over i}\, {\delta\over\delta \eta_{\alpha_1}(x_1)}\; \ldots \;
Z_0(\eta)\Bigr|_{\eta=0}\;,
\label{psipsi}
\end{equation}
where
\begin{eqnarray}
Z_0(\eta)
&=& \int\D\Psi\;\exp\!\left[i\int\dfx\,(\L_0+\eta^{\rm\sss T}\Psi)\right]
\nonumber \\
\noalign{\medskip}
&=& \exp\!\left[-{i\over2}\int \dfx\,\dfy\;
    \eta^{\rm\sss T}(x)S(x-y)\C^{-1}\eta(y)\right].
\label{zmajd042}
\end{eqnarray}
The Feynman propagator $S(x-y)\C^{-1}$ 
is the inverse of the Majorana wave operator $\C(-i\dsl+m)$:
\begin{equation}
\C(-i\dsl_x+m)S(x-y)\C^{-1} = \delta^4(x-y)\;.
\label{feynmaj}
\end{equation}
It is straightforward to check (by working out a few examples)
that eqs.$\,$(\ref{psipsi}--\ref{feynmaj}) do indeed reproduce the
result of section 42 for the vacuum expectation value of a time-ordered product
of Majorana fields.
The extra minus sign in \eq{zmajd042}, as compared with \eq{zpsid042},
arises because all functional derivative in \eq{psipsi} are accompanied
by $1/i$, rather than half by $1/i$ and half by $i$, as in \eq{psipsibar}.

\vfill\eject

\noindent Quantum Field Theory  \hfill   Mark Srednicki

\vskip0.5in

\begin{center}
\large{44: Formal Development of Fermionic Path Integrals}
\end{center}
\begin{center}
Prerequisite: 43
\end{center}

\vskip0.5in

In section 43, we formally defined the fermionic path integral
for a free Dirac field $\Psi$ via
\begin{eqnarray}
Z_0(\etabar,\eta)
&=& \int\D\Psi\,\D\Psibar\;\exp\!\left[
\,i\int\dfx\,\Psibar(i\dsl-m)\Psi
+\etabar\Psi+\Psibar\eta\,\right]
\nonumber \\
\noalign{\smallskip}
&=& \exp\!\left[\,i\int \dfx\,\dfy\;\etabar(x)S(x-y)\eta(y)\,\right],
\label{zpsid043}
\end{eqnarray}
where the Feynman propagator $S(x-y)$ 
is the inverse of the Dirac wave operator:
\begin{equation}
(-i\dsl_x + m)S(x-y) = \delta^4(x-y)\;.
\label{feyndir43}
\end{equation}
We would like to find a mathematical framework that allows us to
derive this formula, rather than postulating it by analogy.

Consider a set of {\it anticommuting numbers\/} or
{\it Grassmann variables\/} $\psi_i$ that obey
\begin{equation}
\{\psi_i,\psi_j\}=0\;,
\label{pp043}
\end{equation}
where $i=1,\ldots,n$.  Let us begin with the very simplest
case of $n=1$, and thus a single anticommuting number $\psi$
that obeys $\psi^2=0$.  We can define a function $f(\psi)$
of such an object via a Taylor expansion; because $\psi^2=0$,
this expansion ends with the second term:
\begin{equation}
f(\psi) = a + \psi b\;.
\label{fpsi}
\end{equation}
The reason for writing the coefficient $b$ to the right of the
variable $\psi$ will become clear in a moment. 

Next we would like to define the derivative of $f(\psi)$
with respect to $\psi$.  Before we can do so, we must
decide if $f(\psi)$ itself is to be commuting or anticommuting;
generally we will be interested in functions that are themselves
commuting.  In this case, $a$ in \eq{fpsi} should be treated
as an ordinary commuting number, but $b$ should be treated
as an anticommuting number: $\{b,b\}=\{b,\psi\}=0$.
In this case,
$f(\psi)=a+\psi b=a-b\psi$.

Now we can define two kinds of derivatives.  The
{\it left derivative} of $f(\psi)$ with respect to $\psi$
is given by the coefficient of $\psi$ when $f(\psi)$ is
written with the $\psi$ always on the far left: 
\begin{equation}
\d_\psi f(\psi) = +b\;.
\label{dfpsil}
\end{equation}
Similarly, the 
{\it right derivative} of $f(\psi)$ with respect to $\psi$
is given by the coefficient of $\psi$ when $f(\psi)$ is
written with the $\psi$ always on the far right:
\begin{equation}
f(\psi){\buildrel\leftarrow\over\d}_\psi = -b \;.
\label{dfpsir}
\end{equation}
Generally, when we write a derivative with respect
to a Grassmann variable, we mean the
left derivative.  However, in section 37, when we
wrote the canonical momentum for a fermionic field
$\psi$ as $\pi=\d\L/\d(\d_0\psi)$, we actually meant the
right derivative.  (This is a standard, though
rarely stated, convention.)  Correspondingly, we
wrote the hamiltonian density as 
$\H=\pi\d_0\psi-\L$, with the $\d_0\psi$ to the
right of $\pi$.  

Finally, we would like to define a definite integral, analogous
to integrating a real variable $x$ from minus to plus infinity.
The key features of such an integral over $x$
(when it converges) are linearity,
\begin{equation}
\int_{-\infty}^{+\infty}dx\,cf(x)=
c\int_{-\infty}^{+\infty}dx\,f(x)\;,
\label{ffa}
\end{equation}
and invariance under shifts of the dependent variable
$x$ by a constant:
\begin{equation}
\int_{-\infty}^{+\infty}dx\,f(x+a)=
\int_{-\infty}^{+\infty}dx\,f(x)\;.
\label{ffb}
\end{equation}
Up to an overall numerical factor that is the same for every $f(\psi)$,
the only possible nontrivial definition of
$\int d\psi\,f(\psi)$ that is both linear and shift invariant is
\begin{equation}
\int d\psi\,f(\psi) = b \;.
\label{intf}
\end{equation}

Now let us generalize this to $n>1$.  We have
\begin{equation}
f(\psi) = a + \psi_i b_i+\half\psi_{i_1}\!\psi_{i_2}c_{i_1 i_2}+ \ldots 
+{\ts{1\over n!}}\psi_{i_1}\ldots\psi_{i_n}d_{{i_1}\ldots{i_n}} \;,
\label{fpsin}
\end{equation}
where the indices are implicitly summed.
Here we have written the coefficients to the right of the variables
to facilitate left-differentiation.  
These coefficients are completely antisymmetric on
exchange of any two indices.  
The left derivative of $f(\psi)$ with respect to $\psi_j$ is
\begin{equation}
{\ts{\d\over\d\psi_j}}f(\psi) = b_j + \psi_i c_{ji} + \ldots
+{\ts{1\over(n-1)!}}\psi_{i_2}\ldots\psi_{i_n}d_{j{i_2}\ldots{i_n}} \;.
\label{dfdpsi}
\end{equation}

Next we would like to find a linear, shift-invariant definition of the
integral of $f(\psi)$.  
Note that the antisymmetry of the coefficients implies that
\begin{equation}
d_{{i_1}\ldots{i_n}} = d\,\e_{{i_1}\ldots{i_n}} \;.
\label{de}
\end{equation}
where $d$ is a just a number (ordinary if $f$ is commuting and 
$n$ is even, Grassmann if $f$ is commuting and $n$ is odd, etc.), 
and $\e_{{i_1}\ldots{i_n}}$ is the completely antisymmetric
Levi-Civita symbol with $\e_{1\ldots n}=+1$.
This number $d$ is a candidate (in fact, 
up to an overall numerical factor, the only candidate!) 
for the integral of $f(\psi)$:
\begin{equation}
\int d^n\!\psi\,f(\psi) = d \;.
\label{intfn}
\end{equation}
Although \eq{intfn} really tells us everything we need to know
about $\int d^n\!\psi$, we can, if we like, write
$d^n\!\psi=d\psi_n\ldots d\psi_1$ (note the backwards ordering),
and treat the individual differentials as anticommuting: 
$\{d\psi_i,d\psi_j\}=0$, $\{d\psi_i,\psi_j\}=0$.  Then we take
$\int d\psi_i=0$ and
$\int d\psi_i\,\psi_j=\delta_{ij}$ as our basic formulae,
and use them to derive \eq{intfn}.

Let us work out some consequences of \eq{intfn}.  Consider what happens
if we make a linear change of variable,
\begin{equation}
\psi_i = J_{ij} \psi'_j \;,
\label{pjp43}
\end{equation}
where $J_{ji}$ is a matrix of commuting numbers (and therefore can
be written on either the left or right of $\psi'_j$).  We now have
\begin{equation}
f(\psi) = a + \ldots +  
{\ts{1\over n!}}(J_{i_1 j_1}\psi'_{j_1})\ldots(J_{i_n j_n}\psi'_{j_n})
\e_{{i_1}\ldots{i_n}} d \;.
\label{fpsipn}
\end{equation}
Next we use
\begin{equation}
\e_{{i_1}\ldots{i_n}} J_{i_1 j_1}\ldots J_{i_n j_n}
= (\det J)\e_{{j_1}\ldots{j_n}} \;,
\label{detj}
\end{equation}
which holds for any $n\times n$ matrix $J$, to get
\begin{equation}
f(\psi) = a + \ldots +  
{\ts{1\over n!}}\psi'_{i_1}\ldots\psi'_{i_n}
\e_{{i_1}\ldots{i_n}}(\det J)d \;.
\label{fpsipn2}
\end{equation}
If we now integrate $f(\psi)$ over $d^n\!\psi'$, \eq{intfn} tells us
that the result is $(\det J)d$.  Thus, 
\begin{equation}
\int d^n\!\psi\,f(\psi) = (\det J)^{-1} \int d^n\!\psi'\,f(\psi)\;.
\label{intfnj}
\end{equation}
Recall that, for integrals over commuting real numbers $x_i$ with 
$x_i=J_{ij}x'_j$, we have instead
\begin{equation}
\int d^n\!x\,f(x) = (\det J)^{+1} \int d^n\!x'\,f(x)\;.
\label{intfnj2}
\end{equation}
Note the opposite sign on the power of the determinant.

Now consider a quadratic
form $\psit\!M\psi = \psi_i M_{ij}\psi_j$, 
where $M$ is an antisymmetric matrix of commuting numbers
(possibly complex).
Let's evaluate the gaussian integral 
$\int d^n\!\psi\,\exp(\half\psit\!M\psi)$.
For example, for $n=2$, we have
\begin{equation}
M = \pmatrix{  0 & +m \cr
\noalign{\medskip}
             - m &  0 \cr},
\label{m2}
\end{equation}
and $\psit\!M\psi = 2m\psi_1\psi_2$.  Thus
$\exp(\half\psit\!M\psi)=1+m\psi_1\psi_2$, and so
\begin{equation}
\int d^n\!\psi\,\exp(\half\psit\!M\psi)=m \;.  
\label{intg2}
\end{equation}
For larger $n$, we use the fact that a complex antisymmetric matrix can
be brought to a block-diagonal form via
\begin{equation}
U^{\rm\sss T}MU =
\pmatrix{  0 & +m_1 & \cr
\noalign{\medskip}
       - m_1 &    0 & \cr
\noalign{\medskip}
             &         & \ddots \cr},
\label{rtmr}
\end{equation}
where $U$ is a unitary matrix, and each $m_I$ is real and positive.
(If $n$ is odd there is a final row and column of all zeroes;
from here on, we assume $n$ is even.)
We can now let $\psi_i = U_{ij}\psi'_j$; then, we have
\begin{equation}
\int d^n\!\psi\,\exp(\half\psit\!M\psi)
= (\det U)^{-1}\prod_{I=1}^{n/2}
    \int d^2\!\psi_I\,\exp(\half\psit\!M_I\psi) \;,
\label{intgn}
\end{equation}
where $M_I$ represents one of the $2\times2$ blocks in \eq{rtmr}.
Each of these two-dimensional integrals can be evaluated using
\eq{intg2}, and so
\begin{equation}
\int d^n\!\psi\,\exp(\half\psit\!M\psi)
= (\det U)^{-1}\prod_{I=1}^{n/2}m_I \;.
\label{intgnl}
\end{equation}
Taking the determinant of \eq{rtmr}, we get
\begin{equation}
(\det U)^2(\det M) = \prod_{I=1}^{n/2}m^2_I \;.
\label{detu2m}
\end{equation}
We can therefore rewrite the right-hand side of \eq{intgnl} as
\begin{equation}
\int d^n\!\psi\,\exp(\half\psit\!M\psi)
= (\det M)^{1/2} \;.
\label{intgnm}
\end{equation}
In this form, there is a sign ambiguity associated with the
square root; it is resolved by \eq{intgnl}.
However, the overall sign (more generally, any overall
numerical factor) will never be of concern to us, so
we can use \eq{intgnm} without worrying about the 
correct branch of the square root.

It is instructive to compare \eq{intgnm} with the corresponding
gaussian integral for commuting real numbers,
\begin{equation}
\int d^n\!x\,\exp(-\half x^{\rm\sss T}\!Mx)
= (2\pi)^{n/2}(\det M)^{-1/2} \;.
\label{intgnmcomm}
\end{equation}
Here $M$ is a complex symmetric matrix.
Again, note the opposite sign on the power of the determinant.

Now let us introduce the notion of 
{\it complex\/} Grassmann variables via
\begin{eqnarray}
\chi &\equiv& {\ts{1\over\sqrt2}}(\psi_1+i\psi_2) \;,
\nonumber \\
\noalign{\medskip}
\chibar &\equiv& {\ts{1\over\sqrt2}}(\psi_1-i\psi_2) \;.
\label{chichibar}
\end{eqnarray}
We can invert this to get
\begin{equation}
\pmatrix{ \psi_1 \cr
\noalign{\medskip}
          \psi_2 \cr } =
{\ts{1\over\sqrt2}}
\pmatrix{ 1 & 1 \cr
\noalign{\medskip}
          i & -i \cr }
\pmatrix{ \chibar \cr
\noalign{\medskip}
          \chi \cr }.
\label{psi1psi243}
\end{equation}
The determinant of this transformation matrix is $-i$, and so 
\begin{equation}
d^2\!\psi = d\psi_2 d\psi_1 = (-i)^{-1}d\chi\,d\chibar\;.
\label{dpsi1psi2}
\end{equation}
Also, $\psi_1\psi_2 = -i\chibar\chi$.  Thus we have
\begin{equation}
\int d\chi\,d\chibar\,\chibar\chi 
= (-i)(-i)^{-1}\int d\psi_2d\psi_1\,\psi_1\psi_2 = 1\;.
\label{dppchch}
\end{equation}
Thus, if we have a function
\begin{equation}
f(\chi,\chibar) = a + \chi b + \chibar c + \chibar\chi d \;,
\label{fchi}
\end{equation}
its integral is
\begin{equation}
\int d\chi\,d\chibar\,f(\chi,\chibar) = d \;.
\label{intfchi}
\end{equation}
In particular,
\begin{equation}
\int d\chi\,d\chibar\,\exp(m\chibar\chi) = m \;.
\label{intgchi}
\end{equation}

Let us now consider $n$ complex Grassmann variables $\chi_i$
and their complex conjugates, $\chibar_i$.  We define
\begin{equation}
d^n\!\chi\,d^n\!\chibar \equiv
d\chi_n d\chibar_n \ldots d\chi_1 d\chibar_1 \;.
\label{dchi}
\end{equation}
Then under a change of variable, 
$\chi_i = J_{ij}\chi'_j$ and
$\chibar_i = K_{ij}\chibar'_j$, we have
\begin{equation}
d^n\!\chi\,d^n\!\chibar = 
(\det J)^{-1}(\det K)^{-1}\,d^n\!\chi'\,d^n\!\chibar'\;.
\label{dchip}
\end{equation}
Note that we need not require $K_{ij}=J_{ij}^*$, 
because, as far as the integral is concerned, it is does
not matter whether or not $\chibar_i$ is the complex conjugate
of $\chi_i$.

We now have enough information to evaluate
$\int d^n\!\chi\,d^n\!\chibar\,\exp(\chi^\dagger M\chi)$,
where $M$ is a general complex matrix.  
We make the change of variable $\chi=U\chi'$
and $\chi^\dagger = \chi^{\prime\dagger}V$, where $U$ and $V$
are unitary matrices with the property that
$VMU$ is diagonal with positive real entries $m_i$.
Then we get
\begin{eqnarray}
\int d^n\!\chi\,d^n\!\chibar\,\exp(\chi^\dagger M\chi)
&=& (\det U)^{-1}(\det V)^{-1}
\prod_{i=1}^n \int d\chi_i d\chibar_i\,\exp(m_i\chibar_i\chi_i)
\nonumber \\
&=& (\det U)^{-1}(\det V)^{-1}\prod_{i=1}^n m_i
\nonumber \\
\noalign{\medskip} 
&=& \det M \;.
\label{intgchibar}
\end{eqnarray}
This can be compared to the analogous integral for commuting
complex variables $z_i=(x_i+iy_i)/\sqrt2$ and 
$\bar z=(x_i-iy_i)/\sqrt2$,
with $d^n\!z\,d^n\!\bar z=d^n\!x\,d^n\!y$, namely
\begin{equation}
\int d^n\!z\,d^n\!\bar z\,\exp(-z^\dagger Mz)
=(2\pi)^n(\det M)^{-1}\;.
\label{intgzbar}
\end{equation}

We can now generalize \eqs{intgnm} and (\ref{intgchibar}) by
shifting the integration variables, and using shift invariance
of the integrals.  Thus, 
by making the replacement $\psi\to\psi-M^{-1}\eta$ 
in \eq{intgnm}, we get
\begin{equation}
\int d^n\!\psi\,\exp(\half\psit\!M\psi+\eta^{\rm\sss T}\psi)
= (\det M)^{1/2}\exp(\half\eta^{\rm\sss T}M^{-1}\eta) \;.
\label{intrs} 
\end{equation}
(In verifying this, remember that $M$ and its inverse are
both antisymmetric.)
Similarly, by making the replacements 
$\chi\to\chi-M^{-1}\eta$ and
$\chi^\dagger\to\chi^\dagger-\eta^\dagger M^{-1}$ 
in \eq{intgchibar}, we get
\begin{equation}
\int d^n\!\chi\,d^n\!\chibar\,
\exp(\chi^\dagger M\chi+\eta^\dagger\chi+\chi^\dagger\eta)
= (\det M)\exp(-\eta^\dagger M^{-1}\eta)\;.
\label{intcs}
\end{equation}

We can now see that \eq{zpsid043} is simply a particular case
of \eq{intcs}, with the index on the complex Grassmann variable
generalized to include both the ordinary spin index $\alpha$
and the continuous spacetime argument $x$ of the field 
$\Psi_\alpha(x)$.   Similarly, \eq{zmajd042} for the path integral
for a free Majorana field is simply a particular case of \eq{intrs}.
In both cases, the determinant factors are constants (that is,
independent of the fields and sources) that we simply absorb
into the overall normalization of the path integral.
We will meet determinants that cannot be so neatly absorbed
in sections 53 and 70.

\vfill\eject

\noindent Quantum Field Theory  \hfill   Mark Srednicki

\vskip0.5in

\begin{center}
\large{45: The Feynman Rules for Dirac Fields and Yukawa Theory}
\end{center}
\begin{center}
Prerequisite: 10, 13, 41, 43
\end{center}

\vskip0.5in

In this section we will derive the Feynman rules for {\it Yukawa theory},
a theory with a Dirac field $\Psi$ (with mass $m$) 
and a real scalar field $\ph$ (with mass $M$), interacting via
\begin{equation}
\L_1= g\ph\Psibar\Psi \;, 
\label{l1yuk}
\end{equation}
where $g$ is a coupling constant.
In this section, we will be concerned with tree-level processes only,
and so we omit renormalizing $Z$ factors.

In four spacetime dimensions,
$\ph$ has mass dimension $[\ph]=1$ and $\Psi$ has mass dimension
$[\Psi]={3\over2}$; thus the coupling constant $g$ is dimensionless: $[g]=0$.
As discussed in section 13, this is generally the most interesting situation.

Note that $\L_1$ is invariant under the U(1) transformation 
$\Psi\to e^{-i\alpha}\Psi$, as is the free Dirac lagrangian.
Thus, the corresponding Noether current $\Psibar\g^\mu\Psi$
is still conserved, and the associated charge $Q$ (which counts the number
of $b$-type particles minus the number of $d$-type particles) is constant
in time.

We can think of $Q$ as electric charge, and identify the $b$-type particle
as the electron $e^-$, and the $d$-type particle as the positron $e^+$.
The scalar particle is electrically neutral (and could, for example,
be thought of as the Higgs boson; see Part III).

We now use the general result of sections 9 and 43 to write
\begin{equation}
Z(\etabar,\eta,J) \propto
\exp\!\left[\,ig\int\dfx\left({1\over i}{\delta\over\delta J(x)}\right)\!\!
       \left(i{\delta\over\delta\eta_\alpha(x)}\right)\!\!
       \left({1\over i}{\delta\over\delta\etabar_\alpha(x)}\right)\right]\!
Z_0(\etabar,\eta)Z_0(J)\;,
\label{znbarnj}
\end{equation}
where
\begin{eqnarray}
Z_0(\etabar,\eta) &=&
\exp\!\left[i\int \dfx\,\dfy\;\etabar(x)S(x-y)\eta(y)\right],
\label{zpsid044} \\
\noalign{\medskip}
Z_0(J) &=&
\exp\!\left[{i\over 2}\int \dfx\,\dfy\;J(x)\Delta(x-y)J(y)\right],
\label{zj044}
\end{eqnarray}
and
\begin{eqnarray}
S(x-y) &=& \int{\dfp\over(2\pi)^4}\;{(-\psl+m)e^{ip(x-y)}
                 \over p^2+m^2-i\epsilon}\;,
\label{feyns44} \\
\noalign{\medskip}
\Delta(x-y) &=& \int{\dfk\over(2\pi)^4}\;{e^{ik(x-y)}
                 \over k^2+M^2-i\epsilon}
\label{feynd44}
\end{eqnarray}
are the appropriate Feynman propagators for the corresponding free fields.
We impose the normalization $Z(0,0,0)=1$, and write
\begin{equation}
Z(\etabar,\eta,J) = \exp[ iW(\etabar,\eta,J)]\;.
\label{wnbarnj}
\end{equation}
Then $iW(\etabar,\eta,J)$ can be expressed as a series of connected 
Feynman diagrams with sources.

\begin{figure}
\begin{center}
\epsfig{file=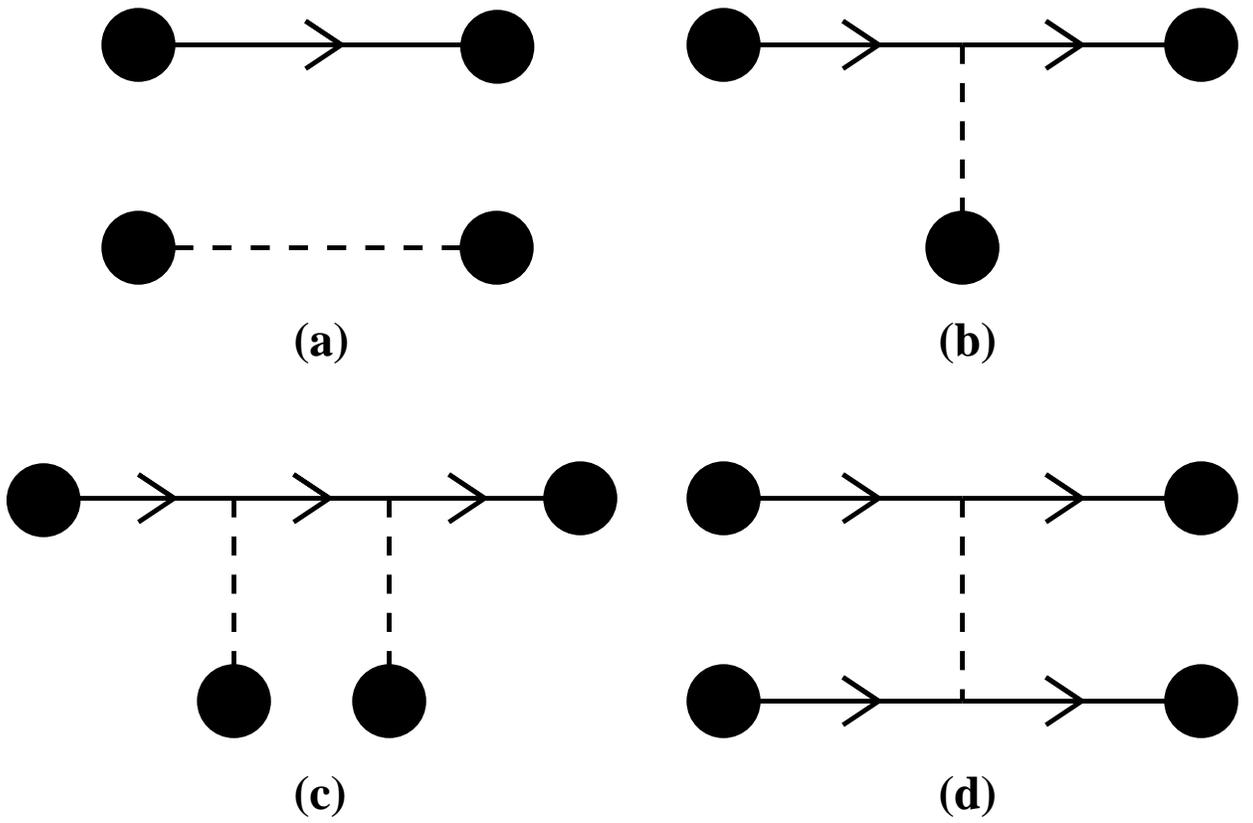}
\end{center}
\caption{Tree contributions to $iW(\etabar,\eta,J)$ with four or fewer sources.}
\label{yuktree}
\end{figure}

We use a dashed line to stand for the scalar propagator 
${1\over i}\Delta(x-y)$,
and a solid line to stand for the fermion propagator ${1\over i}S(x-y)$.
The only allowed vertex joins two solid lines and one dashed line;
the associated vertex factor is $ig$.
The blob at the end of a dashed line stands for the $\ph$ source
$i\int\dfx\,J(x)$, and the blob at the end of a solid line for
either the $\Psi$ source $i\int\dfx\,\etabar(x)$, or the $\Psibar$
source $i\int\dfx\,\eta(x)$.  To tell which is which,
we adopt the ``arrow rule'' of problem 9.3:
the blob stands for $i\int\dfx\,\eta(x)$
if the arrow on the attached line points {\it away\/} from the blob,
and the blob stands for $i\int\dfx\,\etabar(x)$ if the arrow on the attached
line points {\it towards\/} the blob.  Because
$\L_1$ involves one $\Psi$ and one $\Psibar$,
we also have the rule that, at each vertex, 
{\it one arrow must point towards the vertex, and one away}.
The first few tree diagrams that contribute to 
$iW(\etabar,\eta,J)$ are shown in \fig{yuktree}.
We omit tadpole diagrams; as in $\ph^3$ theory, these can 
be cancelled by shifting the $\ph$ field, or, equivalently,
adding a term linear in $\ph$ to $\L$.  The LSZ formula is 
valid only after all tadpole diagrams have been cancelled in
this way.

The spin indices on the fermionic sources and propagators 
are all contracted in the obvious way.  
For example, the complete expression corresponding to \fig{yuktree}(b) is
\begin{eqnarray}
\hbox{\Fig{yuktree}(b)} &=&
i^3\Bigl({\ts{1\over i}}\Bigr)^{\!\!3}\,(ig)\int\dfx\,\dfy\,\dfz\,\dfw\,
\nonumber \\
&& \qquad\qquad {} \times
\Bigl[\etabar(x)S(x-y)S(y-z)\eta(z)\Bigr] 
\nonumber \\
&& \qquad\qquad {} \times
\Delta(y-w)J(w) \;.
\label{3j}
\end{eqnarray}

Our main purpose in this section is to compute the tree-level amplitudes 
for various two-body elastic scattering processes, such as
$e^-\ph\to e^-\ph$ and $e^+ e^-\to\ph\ph$;
for these, we will need to evaluate the tree-level contributions to
connected correlation functions of the form 
$\la0|{\rm T}\Psi\Psibar\ph\ph|0\ra_{\rm C}$.
Other processes of interest include 
$e^- e^- \to e^- e^-$ and $e^+ e^- \to e^+ e^-$;
for these, we will need to evaluate the tree-level contributions to
connected correlation functions of the form 
$\la0|{\rm T}\Psi\Psibar\Psi\Psibar|0\ra_{\rm C}$.

\begin{figure}
\begin{center}
\epsfig{file=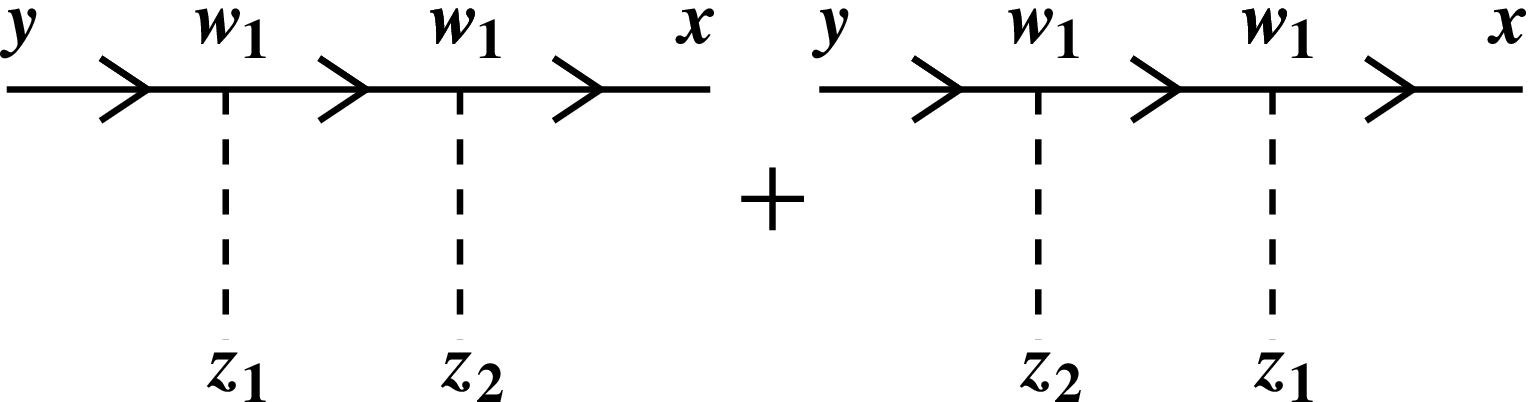}
\end{center}
\caption{Diagrams corresponding to \eq{cc2}.}
\label{2psi2ph}
\end{figure}

For $\la0|{\rm T}\Psi\Psibar\ph\ph|0\ra_{\rm C}$,
the relevant tree-level contribution to $iW(\etabar,\eta,J)$ 
is given by \fig{yuktree}(d).  We have
\begin{eqnarray}
&& \la0|{\rm T}\Psi_\alpha(x)\Psibar_\beta(y)\ph(z_1)\ph(z_2)|0\ra_{\rm C}
\nonumber \\
\noalign{\medskip}
&& \quad {} =
{1\over i}{\delta\over\delta\etabar_\alpha(x)}\,
i{\delta\over\delta\eta_\beta(y)}\,
{1\over i}{\delta\over\delta J(z_1)}\,
{1\over i}{\delta\over\delta J(z_2)} iW(\etabar,\eta,J)
          \Big|_{\etabar=\eta=J=0}
\nonumber \\
\noalign{\medskip}
&& \quad {} = \left({\ts{1\over i}}\right)^{\!\!5}(ig)^2\int d^4w_1\,d^4w_2\,
\nonumber \\
&& \qquad\qquad\qquad {} \times
                   [S(x{-}w_1)
                    S(w_1{-}w_2)
                    S(w_2{-}y)]_{\alpha\beta}
\nonumber \\
&& \qquad\qquad\qquad {} \times
                    \Delta(z_1{-}w_1) \Delta(z_2{-}w_2)
\nonumber \\
\noalign{\medskip}
&& \quad\qquad {} + \Bigl(z_1 \leftrightarrow z_2\Bigr) + O(g^4)\;.
\label{cc2}
\end{eqnarray}
The corresponding diagrams, with sources removed, are shown in 
\fig{2psi2ph}.

\begin{figure}
\begin{center}
\epsfig{file=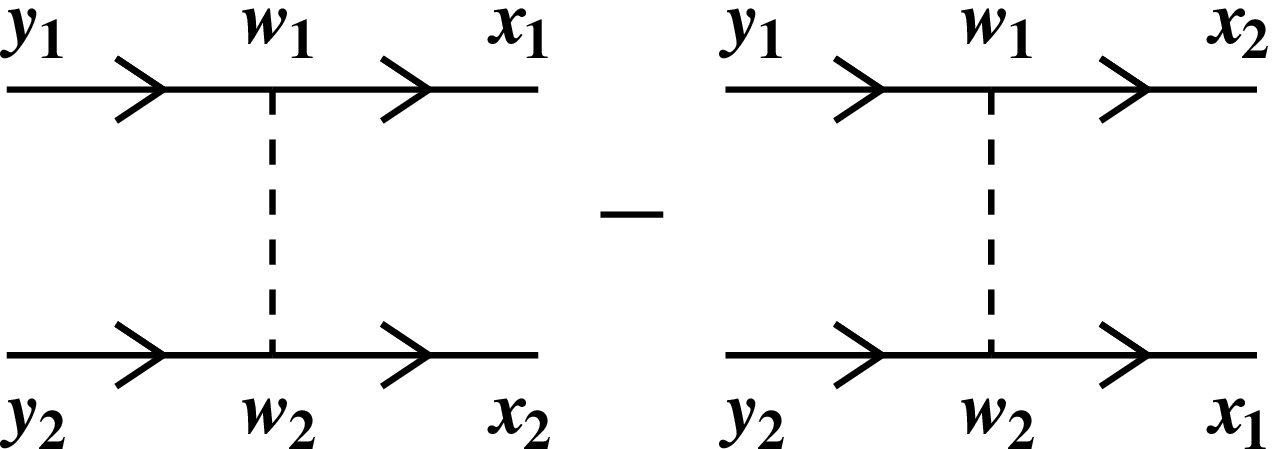}
\end{center}
\caption{Diagrams corresponding to \eq{cc3}.}
\label{4psi}
\end{figure}

For $\la0|{\rm T}\Psi\Psibar\Psi\Psibar|0\ra_{\rm C}$,
the relevant tree-level contribution to $iW(\etabar,\eta,J)$ is 
given by \fig{yuktree}(c),
which has a symmetry factor $S=2$.  We have
\begin{eqnarray}
&& \la0|{\rm T}\Psi_{\alpha_1}(x_1)\Psibar_{\beta_1}(y_1)
               \Psi_{\alpha_2}(x_2)\Psibar_{\beta_2}(y_2) |0\ra_{\rm C}
\nonumber \\
\noalign{\medskip}
&& \quad {} =
{1\over i}{\delta\over\delta\etabar_{\alpha_1}(x_1)}\,
i{\delta\over\delta\eta_{\beta_1}(y_1)}\,
{1\over i}{\delta\over\delta\etabar_{\alpha_2}(x_2)}\,
i{\delta\over\delta\eta_{\beta_2}(y_2)} iW(\etabar,\eta,J)
          \Big|_{\etabar=\eta=J=0}\;.
\label{cc30}
\end{eqnarray}
The two $\eta$ derivatives can act on the two $\eta$'s in the diagram
in two different ways; ditto for the two $\etabar$ derivatives.
This results in four different terms, but two of them are
algebraic duplicates of the other two; this duplication cancels
the symmetry factor (which is a general result).  We get
\begin{eqnarray}
&& \la0|{\rm T}\Psi_{\alpha_1}(x_1)\Psibar_{\beta_1}(y_1)
               \Psi_{\alpha_2}(x_2)\Psibar_{\beta_2}(y_2) |0\ra_{\rm C}
\nonumber \\
\noalign{\medskip}
&& \quad {} = \left({\ts{1\over i}}\right)^{\!\!5}(ig)^2\int d^4w_1\,d^4w_2\,
\nonumber \\
&& \qquad\qquad\qquad {} \times
              [S(x_1{-}w_1)
               S(w_1{-}y_1)]_{\alpha_1\beta_1}
\nonumber \\
&& \qquad\qquad\qquad {} \times
                   \Delta(w_1{-}w_2)
\nonumber \\
&& \qquad\qquad\qquad {} \times
              [S(x_2{-}w_2)
               S(w_2{-}y_2)]_{\alpha_2\beta_2}
\nonumber \\
\noalign{\medskip}
&& \quad\qquad {} - \Bigl((y_1,\beta_1) \leftrightarrow (y_2,\beta_2)\Bigr)
  + O(g^4)\;.
\label{cc3}
\end{eqnarray}
The corresponding diagrams, with sources removed, are shown in \fig{4psi}.
Note, however, that we now have a {\it relative minus sign\/} between the
two diagrams, due to the anticommutation of the derivatives with respect
to $\etabar$.  The general rule is this: there is a relative minus sign between
any two diagrams that are identical {\it except for a swap of the position and
spin labels between two external fermion lines.} 

\begin{figure}
\begin{center}
\epsfig{file=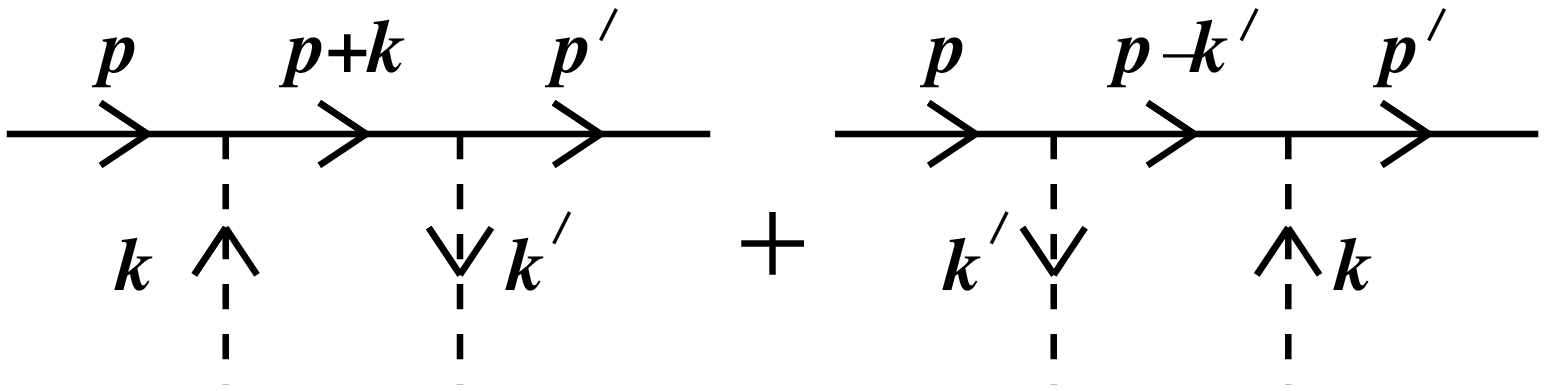}
\end{center}
\caption{Diagrams for $e^-\ph\to e^-\ph$, corresponding to 
\eq{emph-emph}.}
\label{emph}
\end{figure}

Let us now consider a particular scattering process: $e^-\ph\to e^-\ph$.
The scattering amplitude is
\begin{equation}
\la f|i\ra = \la0|\,{\rm T}\, a(\k')_{\rm out} b_{s'}(\p')_{\rm out}
                  \bd_s(\p)_{\rm in}\ad(\k)_{\rm in}\,|0\ra \;.
\label{fit44}
\end{equation}
Next we use the replacements
\begin{eqnarray}
\bd_s(\p)_{\rm in} &\to&  i\int\dfy\;
\Psibar(y)(+i{\buildrel\leftarrow\over\dsl} + m)
 u_s(\p)\,e^{+ipy}\;,
\label{bdsub44} \\
\noalign{\smallskip}
b_{s'}(\p')_{\rm out} &\to&  i\int\dfx\;
e^{-ipx}\;\ubar_{s'}(\p')(-i\dsl+m)\Psi(x)\;,
\label{bsub44} \\
\noalign{\smallskip}
\ad(\k)_{\rm in} &\to&  i\int\dfz_1\;
e^{+ikz_1}(-\d^2+m^2)\ph(z_1)\;,
\label{adsub44} \\
\noalign{\smallskip}
a(\k')_{\rm out} &\to&  i\int\dfz_2\;
e^{-ik'z_2}(-\d^2+m^2)\ph(z_2)\;.
\label{asub44}
\end{eqnarray}
We substitute these into \eq{fit44}, and then use \eq{cc2}.
The wave operators (either Klein-Gordon or Dirac) act on the
external propagators, and convert them to delta functions.
After using \eqs{feyns44} and (\ref{feynd44}) for the internal
propagators, all dependence on the various spacetime coordinates
is in the form of plane-wave factors, as in section 10.
Integrating over the internal coordinates then generates
delta functions that conserve four-momentum at each vertex.
The only new feature arises from the spinor factors $u_s(\p)$
and $\ubar_{s'}(\p')$.  We find that $u_s(\p)$ is associated with 
the external fermion line whose arrow points {\it towards\/} the vertex,
and that $\ubar_{s'}(\p')$
is associated with the external fermion line whose arrow points {\it away\/}
from the vertex.
We can therefore draw the momentum-space diagrams of \fig{emph},
and write down the associated tree-level expression for the $e^-\ph\to e^-\ph$
scattering amplitude,
\begin{equation}
i\T_{e^-\ph\to e^-\ph}
= {\ts{1\over i}} (ig)^2 \,
\ubar_{s'}(\p')\!\left[
{-\psl-\ksl + m\over -s+m^2}+
{-\psl+\ksl' + m\over -u+m^2}\right]\!
u_s(\p)\;,
\label{emph-emph}
\end{equation}
where $s=-(p+k)^2$ and $u=-(p-k')^2$.
(We can safely ignore the $i\eps$'s in the propagators,
because their denominators cannot vanish for any physically allowed
values of $s$ and $u$.)
We will see how to turn this into a more useful expression
in section 46.

\begin{figure}
\begin{center}
\epsfig{file=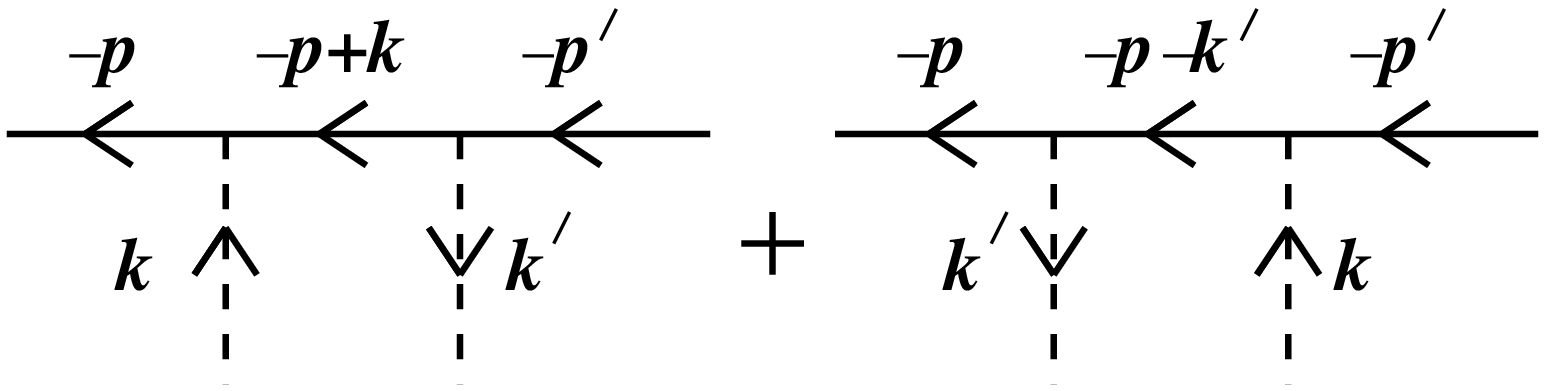}
\end{center}
\caption{Diagrams for $e^+\ph\to e^+\ph$, corresponding to 
\eq{epph-epph}.}
\label{epph}
\end{figure}

Next consider the process $e^+\ph\to e^+\ph$.  We now have
\begin{equation}
\la f|i\ra = \la0|\,{\rm T}\, a(\k')_{\rm out} d_{s'}(\p')_{\rm out}
                  \dd_s(\p)_{\rm in}\ad(\k)_{\rm in}\,|0\ra \;.
\label{fit442}
\end{equation}
The relevant replacements are
\begin{eqnarray}
\dd_s(\p)_{\rm in} &\to&  -i\int\dfx\;
e^{+ipx}\;\vbar_s(\p)(-i\dsl+m)\Psi(x)\;,
\label{ddsub44} \\
\noalign{\smallskip}
d_{s'}(\p')_{\rm out} &\to& -i\int\dfy\;
\Psibar(y)(+i{\buildrel\leftarrow\over\dsl} + m)v_{s'}(\p')\,e^{-ipy} \;,
\label{dsub44} \\
\noalign{\smallskip}
\ad(\k)_{\rm in} &\to&  i\int\dfz_1\;
e^{+ikz_1}(-\d^2+m^2)\ph(z_1)\;,
\label{adsub442} \\
\noalign{\smallskip}
a(\k')_{\rm out} &\to&  i\int\dfz_2\;
e^{-ikz_2}(-\d^2+m^2)\ph(z_2)\;.
\label{asub442}
\end{eqnarray}
We substitute these into \eq{fit442}, and then use \eq{cc2}.
This ultimately leads to the momentum-space Feynman diagrams of \fig{epph}.
Note that we must now label the external fermion lines with {\it minus\/} their
four-momenta; this is characteristic of $d$-type particles.  (The same
phenomenon occurs for a complex scalar field; see problem 10.1.)
Regarding the spinor factors, we find that $-\vbar_s(\p)$ is associated with
the external fermion line whose arrow points {\it away\/} from the vertex,
and $-v_{s'}(\p')$ with the external fermion line whose arrow points 
{\it towards\/} the vertex.
The minus signs attached to each $v$ and $\vbar$ can be consistently dropped,
however, as they only affect the overall sign of the amplitude (and not the
relative signs among contributing diagrams).
The tree-level expression for the $e^+\ph\to e^+\ph$ amplitude is then
\begin{equation}
i\T_{e^+\ph\to e^+\ph}
= {\ts{1\over i}} (ig)^2 \,
\vbar_{s}(\p)\!\left[
{\psl-\ksl' + m\over -u+m^2}+
{\psl+\ksl + m\over -s+m^2}\right]\!
v_{s'}(\p')\;,
\label{epph-epph}
\end{equation}
where again $s=-(p+k)^2$ and $u=-(p-k')^2$.

After working out a few more of these (you might try your hand
at some of them before reading ahead), we can abstract the following
set of Feynman rules.

1) For each {\it incoming electron}, draw a solid line with an arrow pointed
{\it towards\/} the vertex, 
and label it with the electron's four-momentum, $p_i$.

2) For each {\it outgoing electron}, draw a solid line with an arrow pointed
{\it away\/} from the vertex, 
and label it with the electron's four-momentum, $p'_i$.

3) For each {\it incoming positron}, draw a solid line with an arrow pointed
{\it away\/} from the vertex, 
and label it with {\it minus} the positron's four-momentum, $-p_i$.

4) For each {\it outgoing positron}, draw a solid line with an arrow pointed
{\it towards\/} the vertex, 
and label it with {\it minus} the positron's four-momentum, $-p'_i$.

5) For each {\it incoming scalar}, draw a dashed line with an arrow pointed
{\it towards\/} the vertex, 
and label it with the scalar's four-momentum, $k_i$.

6) For each {\it outgoing scalar}, draw a dashed line with an arrow pointed
{\it away\/} from the vertex, 
and label it with the scalar's four-momentum, $k'_i$.

7) The only allowed vertex joins two solid lines, one with an arrow
pointing towards it and one with an arrow pointing away from it,
and one dashed line (whose arrow can point in either direction).
Using this vertex, join up all the external lines, including extra
internal lines as needed.
In this way, draw all possible diagrams that are
{\it topologically inequivalent}.

8) Assign each internal line its own four-momentum. 
Think of the four-momenta as flowing along the arrows,
and conserve four-momentum at each vertex.
For a tree diagram, this fixes the momenta on all the internal lines.

9) The value of a diagram consists of the following factors:
{\obeylines
for each incoming or outgoing scalar, 1;
for each incoming electron, $u_{s_i}(\p_i)$;
for each outgoing electron, $\ubar_{s'_i}(\p'_i)$;
for each incoming positron, $\vbar_{s_i}(\p_i)$;
for each outgoing positron, $v_{s'_i}(\p'_i)$;
for each vertex, $ig$;
for each internal scalar line, $-i/(k^2+M^2-i\eps)$, 
where $k$ is the four-momentum of that line;
for each internal fermion line, $-i(-\psl+m)/(p^2+m^2-i\eps)$, 
where $p$ is the four-momentum of that line.}

10) Spinor indices are contracted by starting at one end of a fermion
line: specifically, the end that has the arrow pointing away from the vertex.
The factor associated with the external line is either 
$\ubar$ or $\vbar$.  Go along the complete fermion line, following the arrows
backwards, and write down (in order from left to right) 
the factors associated with the
vertices and propagators that you encounter.  The last factor is either
a $u$ or $v$.  Repeat this procedure for the other fermion lines, if any.

11) Two diagrams that are identical 
{\it except for the momentum and spin labels on two external fermion lines\/}
that {\it have their arrows pointing in the same direction\/} (either both
towards or both away from the vertex) have a relative minus sign.  

12) The value of $i\T$ (at tree level)
is given by a sum over the values of the contributing diagrams.

There are additional rules for counterterms and loops, but we will postpone
those to section 51.  

\begin{figure}
\begin{center}
\epsfig{file=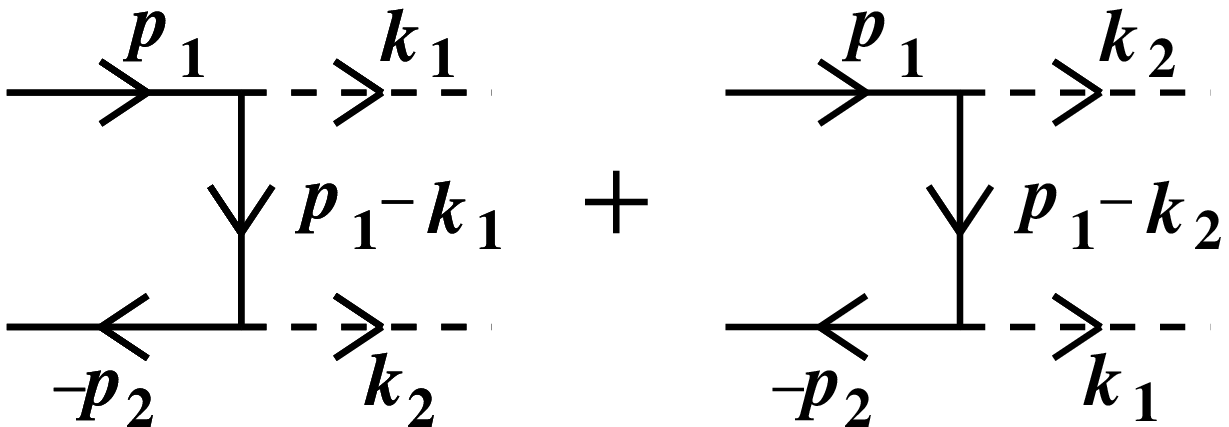}
\end{center}
\caption{Diagrams for $e^+ e^-\to \ph\ph$, corresponding to 
\eq{epem-phph}.}
\label{epem}
\end{figure}

Let us apply these rules to $e^+ e^-\to\ph\ph$.  Let the initial electron
and positron have four-momenta $p_1$ and $p_2$, respectively, 
and the two final scalars have four-momenta $k'_1$ and $k'_2$.
The relevant diagrams are shown in \fig{epem}, and the result is
\begin{equation}
i\T_{e^+ e^-\to\ph\ph}
= {\ts{1\over i}} (ig)^2 \,
\vbar_{s_s}(\p_2)\!\left[
{-\psl_1+\ksl'_1 + m\over -t+m^2}
+
{-\psl_1+\ksl'_2 + m\over -u+m^2}
\right]\!
u_{s_1}(\p_1)\;,
\label{epem-phph}
\end{equation}
where $t=-(p_1-k'_1)^2$ and $u=-(p_1-k'_2)^2$.

\begin{figure}
\begin{center}
\epsfig{file=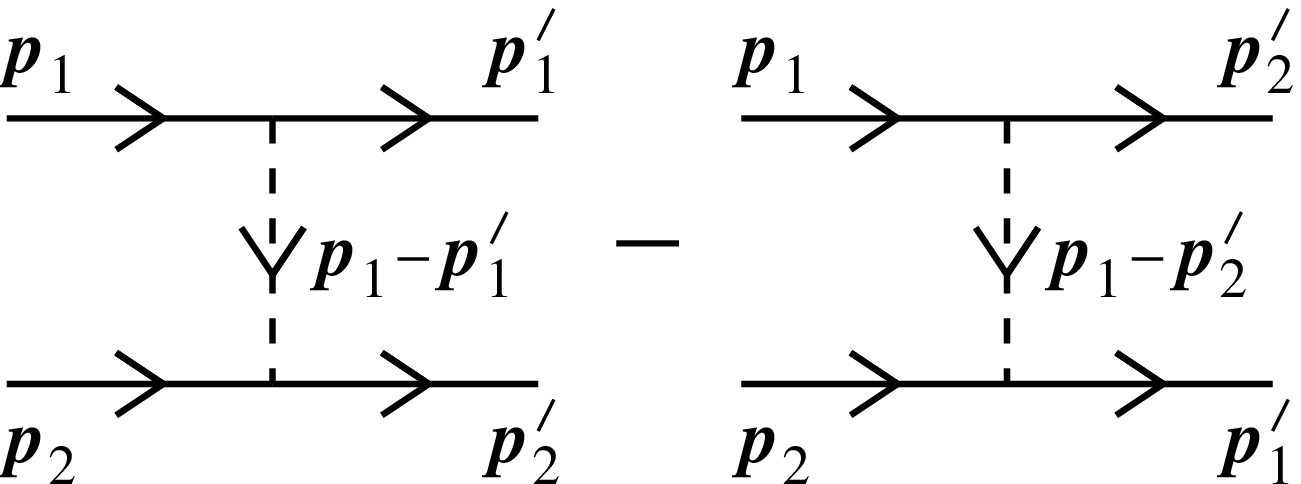}
\end{center}
\caption{Diagrams for $e^- e^-\to e^- e^-$, corresponding to 
\eq{emem-emem}.}
\label{emem}
\end{figure}

Next, consider $e^- e^- \to e^- e^-$.  
Let the initial electrons 
have four-momenta $p_1$ and $p_2$, 
and the final electrons
have four-momenta $p'_1$ and $p'_2$.
The relevant diagrams are shown in \fig{emem}.  
It is clear that they
are identical except for the labels on the two external fermion
lines that have arrows pointing away from their vertices.  Thus, 
according to rule \#11, these diagrams have a relative minus sign.
(Which diagram gets the extra minus sign is a matter of convention, 
and is physically irrelevant.)  Thus the result is
\begin{equation}
i\T_{e^- e^-\to e^- e^-} = {\ts{1\over i}} (ig)^2 \left[
  {(\ubarp_{\!1} u_1)(\ubarp_{\!2} u_2) \over -t+M^2}
- {(\ubarp_{\!2} u_1)(\ubarp_{\!1} u_2) \over -u+M^2} \right],
\label{emem-emem}
\end{equation}
where $u_1$ is short for $u_{s_1}(\p_1)$, etc.,
and $t=-(p_1-p'_1)^2$, $u=-(p_1-p'_2)^2$.

\begin{figure}
\begin{center}
\epsfig{file=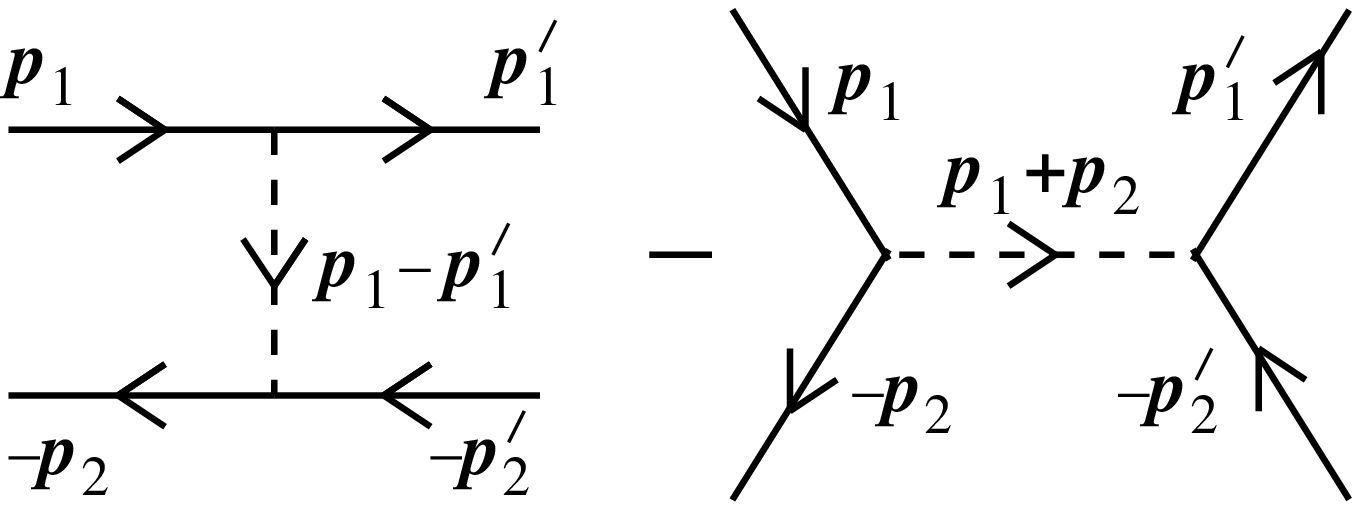}
\end{center}
\caption{Diagrams for $e^+ e^-\to e^+ e^-$, corresponding to 
\eq{epem-epem}.}
\label{epemepem}
\end{figure}

\begin{figure}
\begin{center}
\epsfig{file=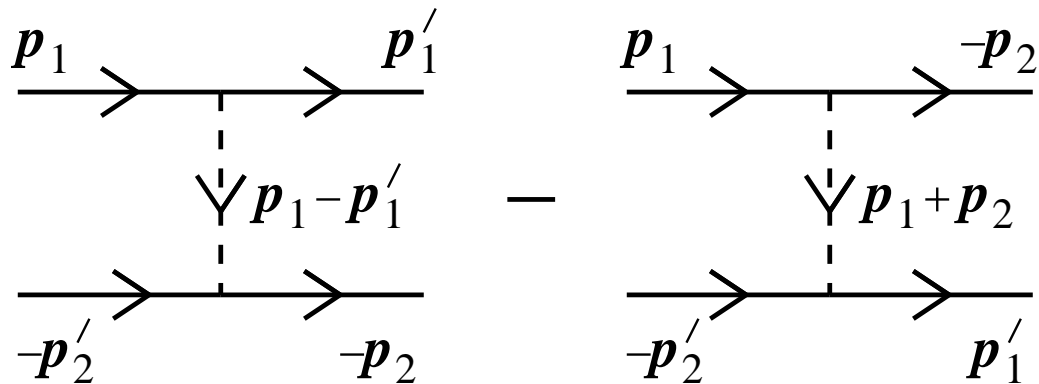}
\end{center}
\caption{Same as \fig{epemepem}, but with the diagrams redrawn
to show more clearly that, according to rule \#11, they have
a relative minus sign.}
\label{epemepem2}
\end{figure}

One more:  $e^+ e^- \to e^+ e^-$.  
Let the initial electron and positron 
have four-momenta $p_1$ and $p_2$, respectively, 
and the final electron and positron 
have four-momenta $p'_1$ and $p'_2$, respectively. 
The relevant diagrams are shown in \fig{epemepem}.
If we redraw them in the topologically equivalent manner shown in 
\fig{epemepem2}, then it becomes clear that they
are identical except for the labels on the two external fermion
lines that have arrows pointing away from their vertices.  Thus, 
according to rule \#11, these diagrams have a relative minus sign.
(Which diagram gets the extra minus sign is a matter of convention, 
and is physically irrelevant.)  Thus the result is
\begin{equation}
i\T_{e^+ e^-\to e^+ e^-} = {\ts{1\over i}} (ig)^2 \left[
  {(\ubarp_{\!1} u_1)(\vbar_2 v'_2) \over -t+M^2}
- {(\vbar_2 u_1)(\ubarp_{\!1} v'_2) \over -u+M^2} \right],
\label{epem-epem}
\end{equation}
where $s=-(p_1+p_2)^2$ and $t=-(p_1-p'_1)^2$.

\vskip0.5in

\begin{center}
Problems
\end{center}

\vskip0.25in

45.1a) Determine how $\ph(x)$ must transform under parity, time reversal,
and charge conjugation in order for these to all be symmetries of the theory.
(Prerequisite: 39)

b) Same question, but with the interaction given by
$\L_1=ig\ph\Psibar\g_5\Psi$ instead of \eq{l1yuk}.

45.2) Use the Feynman rules to write down (at tree level)
$i\T$ for the processes $e^+ e^+ \to e^+ e^+$ and
$\ph\ph \to e^+ e^-$.

\vfill\eject

\noindent Quantum Field Theory  \hfill   Mark Srednicki

\vskip0.5in

\begin{center}
\large{46: Spin Sums}
\end{center}
\begin{center}
Prerequisite: 45
\end{center}

\vskip0.5in

In the last section, we calculated various tree-level scattering amplitudes
in Yukawa theory.  For example, for $e^-\ph\to e^-\ph$ we found
\begin{equation}
\T = g^2 \,
\ubar_{s'}(\p')\!\left[
{-\psl-\ksl + m\over -s+m^2}+
{-\psl+\ksl' + m\over -u+m^2}\right]\!
u_s(\p)\;,
\label{emph2}
\end{equation}
where $s=-(p+k)^2$ and $u=-(p-k')^2$.
In order to compute the corresponding cross section, we must 
evaluate $|\T|^2=\T\T^*$.  
We begin by simplifying \eq{emph2} a little;
we use $(\psl+m)u_s(\p)=0$ to replace the $-\psl$
in each numerator with $m$.  We then abbreviate \eq{emph2} as
\begin{equation}
\T = \ubarp \! A u \;,
\label{tuau}
\end{equation}
where
\begin{equation}
A \equiv g^2 \left[
{-\ksl + 2m\over m^2-s}+
{\ksl' + 2m\over m^2-u}\right].
\label{biga}
\end{equation}
Then we have
\begin{equation}
\T^* = \overline{\T} = \overline{\ubarp\! A u} = \ubar\overline{A}u' \;,
\label{tstar}
\end{equation}
where in general $\overline{A} \equiv \beta A^\dagger\beta$, and, for the particular
$A$ of \eq{biga}, $\overline{A} =A$.  Thus we have 
\begin{eqnarray}
|\T|^2 &=& (\ubarp\!A u)(\ubar A u') 
\nonumber \\
\noalign{\medskip}
&=& \sum_{\alpha\beta\gamma\delta}
         \ubarp_\alpha A_{\alpha\beta}u_\beta
         \ubar_\gamma A_{\gamma\delta} u'_\delta
\nonumber \\
&=&  \sum_{\alpha\beta\gamma\delta}
         u'_\delta\ubarp_\alpha A_{\alpha\beta}u_\beta
         \ubar_\gamma A_{\gamma\delta}
\nonumber \\
&=& \tr\Bigl[(u'\ubarp)A(u\ubar)A\Bigr]\;.
\label{tsq}
\end{eqnarray}
Next, we use a result from section 38:
\begin{equation}
u_s(\p)\ubar_s(\p) = \half(1{-}s\g_5\zsl)(-\psl+m) \;,
\label{uuss}
\end{equation}
where $s=\pm$ tells us whether the spin is up or down along
the spin quantization axis $z$.  We then have
\begin{equation}
|\T|^2 =  {\ts{1\over4}}\tr\Bigl[ 
(1{-}s'\g_5\zsl')(-\pslp+m)A(1{-}s\g_5\zsl)(-\psl+m) A\Bigr]\;.
\label{tsq2}
\end{equation}
We now simply need to take traces of products of gamma matrices;
we will work out the technology for this in the next section.

However, in practice, we are often not interested in (or are unable
to easily measure or prepare) the spin states of the scattering particles.
Thus, if we know that an electron with momentum $p'$ landed in
our detector, but know nothing about its spin,
we should {\it sum\/} $|\T|^2$ over the two possible spin states
of this outgoing electron.  Similarly, if the spin state of the initial electron
is not specially prepared for each scattering event, then we should
{\it average\/} $|\T|^2$ over the two possible spin states of this 
initial electron.  Then we can use
\begin{equation}
\sum_{s=\pm}u_s(\p)\ubar_s(\p) = -\psl+m 
\label{uusss}
\end{equation}
in place of \eq{uuss}.  

Let us, then, take $|\T|^2$, sum over all final spins, and average over all
initial spins, and call the result $\la |\T|^2\ra$.
In the present case, we have
\begin{eqnarray}
\la |\T|^2\ra &\equiv& \half\sum_{s,s'}|\T|^2
\nonumber \\
&=& {\ts{1\over2}}\tr\Bigl[(-\pslp+m)A(-\psl+m)A\Bigr]\;,
\label{tsq3}
\end{eqnarray}
which is much less cumbersome than \eq{tsq2}.

Next let's try something a little harder, namely $e^+ e^-\to e^+ e^-$.
We found in section 45 that
\begin{equation}
\T = g^2
\left[ { (\ubarp_{\!1} u_1) (\vbar_2  v'_2) \over M^2-t } 
      -{ (\vbar_2  u_1) (\ubarp_{\!1} v'_2) \over M^2-s } \right].
\label{epem2}
\end{equation}
We then have
\begin{equation}
\overline\T = g^2
\Biggl[
{ (\ubar_1 u'_1)(\vbarp_{\!2} v_2) \over M^2-t } -
{ (\ubar_1  v_2)(\vbarp_{\!2} u'_1) \over M^2-s }\Biggr].
\label{epemst}
\end{equation}
When we multiply $\T$ by $\overline\T$, we will get four terms.
We want to arrange the factors in each of them so that every $u$ and every $v$
stands just to the left of the corresponding $\ubar$ and $\vbar$.
In this way, we get
\begin{eqnarray}
|\T|^2 &=& {} + {g^4\over(M^2{-}t)^2}
\,\tr\Bigl[u_1\ubar_1 u'_1\ubarp_{\!1}\Bigr]
\,\tr\Bigl[v'_2\vbarp_{\!2} v_2\vbar_2\Bigr]
\nonumber \\
\noalign{\medskip}
&& {} + {g^4\over(M^2{-}s)^2}
\,\tr\Bigl[u_1\ubar_1 v_2\vbar_2\Bigr]
\,\tr\Bigl[v'_2\vbarp_{\!2} u'_1\ubarp_{\!1}\Bigr]
\nonumber \\
\noalign{\medskip}
&& {} - {g^4\over(M^2{-}t)(M^2{-}s)}
\,\tr\Bigl[u_1\ubar_1 v_2\vbar_2 v'_2\vbarp_{\!2}u'_1\ubarp_{\!1}\Bigr]
\nonumber \\
\noalign{\medskip}
&& {} - {g^4\over(M^2{-}s)(M^2{-}t)}
\,\tr\Bigl[u_1\ubar_1 u'_1\ubarp_{\!1} v'_2\vbarp_{\!2}v_2\vbar_2\Bigr]\;.
\label{epemsq}
\end{eqnarray}
Then we average over initial spins and sum over final spins, and use
\eq{uusss} and
\begin{equation}
\sum_{s=\pm}v_s(\p)\vbar_s(\p) = -\psl-m \;. 
\label{vvsss}
\end{equation}
We then must evaluate traces of products of up to four gamma matrices.





\vfill\eject

\noindent Quantum Field Theory  \hfill   Mark Srednicki

\vskip0.5in

\begin{center}
\large{47: Gamma Matrix Technology}
\end{center}
\begin{center}
Prerequisite: 36
\end{center}

\vskip0.5in

In this section, we will learn some tricks for handling gamma
matrices.  We need the following information as a starting point:
\begin{eqnarray}
\{\g^\mu,\gamma^\nu\} &=& -2g^{\mu\nu}\;,
\label{gmgn} \\
\noalign{\smallskip}
\gamma_5^2 &=& 1\;,
\label{g5g5} \\
\noalign{\smallskip}
\{\g^\mu,\gamma_5\} &=& 0\;,
\label{g5g} \\
\noalign{\smallskip}
\tr\,1 &=& 4 \;.
\label{tr1}
\end{eqnarray}
Now consider the trace of the product of $n$ gamma matrices.  We have
\begin{eqnarray}
\tr[\g^{\mu_1}\ldots\g^{\mu_n}]
&=& \tr[\g_5^2\g^{\mu_1}\g_5^2\ldots \g_5^2\g^{\mu_n}]
\nonumber \\
\noalign{\smallskip}
&=& \tr[(\g_5\g^{\mu_1}\g_5)\ldots (\g_5\g^{\mu_n}\g_5)]
\nonumber \\
\noalign{\smallskip}
&=& \tr[(-\g_5^2\g^{\mu_1})\ldots (-\g_5^2\g^{\mu_n})]
\nonumber \\
\noalign{\smallskip}&=& (-1)^n\,\tr[\g^{\mu_1}\ldots \g^{\mu_n}] \;.
\label{trn} 
\end{eqnarray}
We used \eq{g5g5} to get the first equality, the cyclic property of the
trace for the second, \eq{g5g} for the third, and \eq{g5g5}
again for the fourth. 
If $n$ is odd, \eq{trn} tells us that this trace is equal to minus itself, 
and must therefore be zero:
\begin{equation}
\tr[\,\hbox{odd \# of $\g^\mu$'s}\,] = 0\;.
\label{trodd} 
\end{equation}
Similarly,
\begin{equation}
\tr[\,\g_5\,(\,\hbox{odd \# of $\g^\mu$'s}\,)\,] = 0\;.
\label{trodd5} 
\end{equation}

Next, consider $\tr[\g^\mu\g^\nu]$.  We have
\begin{eqnarray}
\tr[\g^\mu\g^\nu]
&=& \tr[\g^\nu\g^\mu]
\nonumber \\
\noalign{\smallskip}
&=& \half\tr[\g^\mu\g^\nu+\g^\nu\g^\mu]
\nonumber \\
\noalign{\smallskip}
&=& -g^{\mu\nu}\,\tr\,1
\nonumber \\
\noalign{\smallskip}
&=& -4g^{\mu\nu}  \;.
\label{trmunu} 
\end{eqnarray}
The first equality follows from the cyclic property of the trace,
the second averages the left- and right-hand sides of the first,
the third uses \eq{gmgn}, and the fourth uses \eq{tr1}.

A slightly nicer way of expressing \eq{trmunu} is to introduce two
arbitrary four-vectors $a^\mu$ and $b^\mu$, and write
\begin{equation}
\tr[\asl\bsl] = -4(ab) \;,
\label{trab}
\end{equation}
where $\asl=a_\mu\g^\mu$, $\bsl=b_\mu\g^\mu$, and $(ab)=a^\mu b_\mu$.

Next consider $\tr[\asl\bsl\csl\ddsl]$.  We evaluate this by moving
$\asl$ to the right, using \eq{gmgn}, which is now more usefully written as
\begin{equation}
\asl\bsl = -\bsl\asl - 2(ab) \;.
\label{abba}
\end{equation}
Using this repeatedly, we have
\begin{eqnarray}
\tr[\asl\bsl\csl\ddsl] 
&=& -\tr[\bsl\asl\csl\ddsl] - 2(ab)\tr[\csl\ddsl]
\nonumber \\
\noalign{\smallskip}
&=& +\tr[\bsl\csl\asl\ddsl] 
+ 2(ac)\tr[\bsl\ddsl]
- 2(ab)\tr[\csl\ddsl]
\nonumber \\
\noalign{\smallskip}
&=& -\tr[\bsl\csl\ddsl\asl] 
- 2(ad)\tr[\bsl\csl]
+ 2(ac)\tr[\bsl\ddsl]
- 2(ab)\tr[\csl\ddsl] \;. \qquad
\label{trabcd1}
\end{eqnarray}
Now we note that the first term on the right-hand side of the last line
is, by the cyclic property of the trace, actually equal to minus
the left-hand side.  We can then move this term to the left-hand side to get
\begin{equation}
2\,\tr[\asl\bsl\csl\ddsl] 
= {} - 2(ad)\tr[\bsl\csl]
+ 2(ac)\tr[\bsl\ddsl]
- 2(ab)\tr[\csl\ddsl] \;.
\label{trabcd2}
\end{equation}
Finally, we evaluate each $\tr[\asl\bsl]$ with \eq{trab}, and divide by two:
\begin{equation}
\tr[\asl\bsl\csl\ddsl] 
= 4\Bigl[(ad)(bc) - (ac)(bd) + (ab)(cd)\Bigr] \;.
\label{trabcd3}
\end{equation}
This is our final result for this trace.

Clearly, we can use the same technique to evaluate the trace of the
product of any even number of gamma matrices. 

Next, let's consider traces that involve $\g_5$'s and $\g^\mu$'s.
Since $\{\g_5,\g^\mu\}=0$, we can always bring all
the $\g_5$'s together by moving them through the $\g^\mu$'s 
(generating minus signs as we go).
Then, since $\g_5^2=1$, we end up with either one $\g_5$ or none.
So we need only consider $\tr[\g_5\g^{\mu_1}\ldots\g^{\mu_n}]$.
And, according to \eq{trodd5}, we need only 
be concerned with even $n$. 

Recall that an explicit formula for $\g_5$ is
\begin{equation}
\gamma_5 = i\g^0 \g^1 \g^2 \g^3 \;.
\label{g5}
\end{equation}
\Eq{trabcd3} then implies
\begin{equation}
\tr\,\gamma_5 = 0 \;.
\label{tr5}
\end{equation}
Similarly, the six-matrix generalization of \eq{trabcd3} yields
\begin{equation}
\tr[\gamma_5\g^\mu\g^\nu] = 0 \;.
\label{tr5mn}
\end{equation}
Finally, consider $\tr[\g_5\g^\mu\g^\nu\g^\rho\g^\sigma]$.  
The only way to get a nonzero result is to have the four vector indices
take on four different values.  If we consider the special case    
$\tr[\g_5\g^3\g^2\g^1\g^0]$, plug in \eq{g5}, and then use
$(\g^i)^2=-1$ and $(\g^0)^2=1$, we get $i(-1)^3\,\tr\,1=-4i$, or equivalently
\begin{equation}
\tr[\g_5\g^\mu\g^\nu\g^\rho\g^\sigma] = -4i\e^{\mu\nu\rho\sigma}\;,
\label{tr5mnps}
\end{equation}
where $\e^{0123}=\e^{3210}=+1$.

Another category of gamma matrix combinations that we will eventually encounter
is $\g^\mu\asl\ldots\g_\mu$.  The simplest of these is
\begin{eqnarray}
\g^\mu\g_\mu
&=& g_{\mu\nu}\g^\mu\g^\nu
\nonumber \\
\noalign{\smallskip}
&=& \half g_{\mu\nu}\{\g^\mu,\g^\nu\}
\nonumber \\
\noalign{\smallskip}
&=& -g_{\mu\nu}g^{\mu\nu}
\nonumber \\
\noalign{\smallskip}
&=& -d \;.
\label{gg}
\end{eqnarray}
To get the second equality, we used the fact that $g_{\mu\nu}$ is symmetric,
and so only the symmetric part of $\g^\mu\g^\nu$ contributes.
In the last line, $d$ is the number of spacetime dimensions.
Of course, our entire spinor formalism has been built around $d=4$,
but we will need formal results for $d=4{-}\e$ when we dimensionally regulate
loop diagrams involving fermions.  

We move on to evaluate
\begin{eqnarray}
\g^\mu\asl\g_\mu
&=& \g^\mu(-\g_\mu\asl-2a_\mu)
\nonumber \\
\noalign{\smallskip}
&=& -\g^\mu\g_\mu\asl-2\asl
\nonumber \\
\noalign{\smallskip}
&=& (d{-}2)\asl\;.
\label{gag}
\end{eqnarray}
We continue with
\begin{eqnarray}
\g^\mu\asl\bsl\g_\mu
&=& (-\asl\g^\mu-2a^\mu)(-\g_\mu\bsl-2b_\mu)
\nonumber \\
\noalign{\smallskip}
&=& \asl\g^\mu\g_\mu\bsl+2\asl\bsl+2\asl\bsl+4(ab)
\nonumber \\
\noalign{\smallskip}
&=& 4(ab)-(d{-}4)\asl\bsl\;.
\label{gabg}
\end{eqnarray}
And finally,
\begin{eqnarray}
\g^\mu\asl\bsl\csl\g_\mu
&=& (-\asl\g^\mu-2a^\mu)\bsl(-\g_\mu\csl-2c_\mu)
\nonumber \\
\noalign{\smallskip}
&=& \asl\g^\mu\bsl\g_\mu\csl+2\bsl\asl\csl+2\asl\csl\bsl+4(ac)\bsl
\nonumber \\
\noalign{\smallskip}
&=& (d{-}2)\asl\bsl\csl+2\bsl\asl\csl+2[\asl\csl+2(ac)]\bsl
\nonumber \\
\noalign{\smallskip}
&=& (d{-}2)\asl\bsl\csl+2\bsl\asl\csl-2\csl\asl\bsl
\nonumber \\
\noalign{\smallskip}
&=& (d{-}2)\asl\bsl\csl+2[-\asl\bsl-2(ab)]\csl-2\csl\asl\bsl
\nonumber \\
\noalign{\smallskip}
&=& (d{-}4)\asl\bsl\csl-4(ab)\csl-2\csl\asl\bsl
\nonumber \\
\noalign{\smallskip}
&=& (d{-}4)\asl\bsl\csl + 2\csl\,[-2(ab)-\asl\bsl\,] 
\label{gabcg}
\nonumber \\
\noalign{\smallskip}
&=& 2\csl\bsl\asl + (d{-}4)\asl\bsl\csl  \;.
\end{eqnarray}

\vfill\eject

\noindent Quantum Field Theory  \hfill   Mark Srednicki

\vskip0.5in

\begin{center}
\large{48: Spin-Averaged Cross Sections in Yukawa Theory}
\end{center}
\begin{center}
Prerequisite: 46, 47
\end{center}

\vskip0.5in

In section 46, we computed $|\T|^2$ for (among other processes)
$e^+ e^-\to e^+ e^-$.  We take the incoming and outgoing electrons 
to have momenta $p_1$ and $p'_1$, respectively, and the incoming and outgoing
positrons to have momenta $p_2$ and $p'_2$, respectively.
We have $p_i^2=p_i^{\prime 2}=-m^2$, 
where $m$ is the electron (and positron) mass.
The Mandelstam variables are
\begin{eqnarray}
s &=& -(p_1+p_2)^2 = -(p'_1+p'_2)^2 \;,
\nonumber \\
\noalign{\smallskip}
t &=& -(p_1-p'_1)^2 = -(p_2-p'_2)^2 \;,
\nonumber \\
\noalign{\smallskip}
u &=& -(p_1-p'_2)^2 = -(p_2-p'_1)^2 \;,
\label{man}
\end{eqnarray}
and they obey $s+t+u=4m^2$.  Our result was
\begin{equation}
|\T|^2 = g^4\!\left[\,{\Phi_{ss}\over(M^2-s)^2}
- {\Phi_{st}+\Phi_{ts}\over(M^2-s)(M^2-t)}
+{\Phi_{tt}\over(M^2-t)^2}\,\right],
\label{tsq47}
\end{equation}
where $M$ is the scalar mass, and
\begin{eqnarray}
\Phi_{ss} &=& 
\tr\Bigl[u_1\ubar_1 v_2\vbar_2\Bigr]
\,\tr\Bigl[v'_2\vbarp_{\!2} u'_1\ubarp_{\!1}\Bigr]\,,
\nonumber \\
\noalign{\medskip}
\Phi_{tt} &=& 
\tr\Bigl[u_1\ubar_1 u'_1\ubarp_{\!1}\Bigr]
\,\tr\Bigl[v'_2\vbarp_{\!2} v_2\vbar_2\Bigr]\,,
\nonumber \\
\noalign{\medskip}
\Phi_{st} &=& 
\tr\Bigl[u_1\ubar_1 u'_1\ubarp_{\!1} v'_2\vbarp_{\!2}v_2\vbar_2\Bigr]\,,
\nonumber \\
\noalign{\medskip}
\Phi_{ts} &=& 
\tr\Bigl[u_1\ubar_1 v_2\vbar_2 v'_2\vbarp_{\!2}u'_1\ubarp_{\!1}\Bigr]\,.
\label{ast}
\end{eqnarray}

Next, we average over the two initial spins and sum over the two final spins
to get
\begin{equation}
\la|\T|^2\ra = {\ts{1\over4}}\sum\nolimits_{s_1,s_2,s'_1,s'_2}|\T|^2\;.
\label{lat2ra}
\end{equation}
Then we use
\begin{eqnarray}
\sum_{s=\pm}u_s(\p)\ubar_s(\p) &=& -\psl+m \;,
\nonumber \\
\sum_{s=\pm}v_s(\p)\vbar_s(\p) &=& -\psl-m \;,
\label{uvsss}
\end{eqnarray}
to get
\begin{eqnarray}
\la \Phi_{ss}\ra &=& {\ts{1\over4}}
\tr\Bigl[(-\psl_1{+}m)(-\psl_2{-}m)\Bigr]
\,\tr\Bigl[(-\pslp_{\!2}{-}m)(-\pslp_{\!1}{+}m)\Bigr]\,,
\label{assav} \\
\noalign{\medskip}
\la \Phi_{tt}\ra &=& {\ts{1\over4}}
\tr\Bigl[(-\psl_1{+}m)(-\pslp_{\!1}{+}m)\Bigr]
\,\tr\Bigl[(-\pslp_{\!2}{-}m)(-\psl_2{-}m)\Bigr]\,,
\label{attav} \\
\noalign{\medskip}
\la \Phi_{st}\ra &=& {\ts{1\over4}}
\tr\Bigl[(-\psl_1{+}m)(-\pslp_{\!1}{+}m)(-\pslp_{\!2}{-}m)(-\psl_2{-}m)\Bigr]\,,
\label{astav} \\
\noalign{\medskip}
\la \Phi_{ts}\ra &=& {\ts{1\over4}}
\tr\Bigl[(-\psl_1{+}m)(-\psl_2{-}m)(-\pslp_{\!2}{-}m)(-\pslp_{\!1}{+}m)\Bigr]\,.
\label{atsav}
\end{eqnarray}
It is now merely tedious to evaluate these traces with the technology
of section 47.

For example,
\begin{eqnarray}
\tr\Bigl[(-\psl_1{+}m)(-\psl_2{-}m)\Bigr]
&=& \tr[\psl_1\psl_2] - m^2\,\tr\,1 
\nonumber \\
\noalign{\smallskip}
&=& -4(p_1 p_2) - 4m^2 \;,
\label{t1}
\end{eqnarray}
It is convenient to write four-vector products in terms of the
Mandelstam variables.  We have
\begin{eqnarray}
p_1 p_2 = p'_1 p'_2 &=& -\half(s-2m^2) \;,
\nonumber \\
\noalign{\smallskip}
p_1 p'_1 = p_2 p'_2 &=& +\half(t-2m^2) \;,
\nonumber \\
\noalign{\smallskip}
p_1 p'_2 = p'_1 p_2 &=& +\half(u-2m^2) \;,
\label{man2}
\end{eqnarray}
and so
\begin{equation}
\tr\Bigl[(-\psl_1{+}m)(-\psl_2{-}m)\Bigr] = 2s-8m^2\;.
\label{t2}
\end{equation}
Thus, we can easily work out \eqs{assav} and (\ref{attav}):
\begin{eqnarray}
\la \Phi_{ss}\ra &=& (s-4m^2)^2 \;,
\label{assav2} \\
\noalign{\smallskip}
\la \Phi_{tt}\ra &=& (t-4m^2)^2 \;.
\label{attav2}
\end{eqnarray}
Obviously, if we start with $\la \Phi_{ss}\ra$ and make the swap $s\lra t$, we get
$\la \Phi_{tt}\ra$.  We could have anticipated this from \eqs{assav} and
(\ref{attav}): if we start with the right-hand side of \eq{assav}
and make the swap $p_2\lra -p'_1$, we get the right-hand side of \eq{attav}.
But from \eq{man2}, we see that this momentum swap is equivalent to $s\lra t$.

Let's move on to $\la \Phi_{st}\ra$ and $\la \Phi_{ts}\ra$.  These two are also
related by $p_2\lra -p'_1$, and so we only need to compute one of them.  We have
\begin{eqnarray}
\la \Phi_{st}\ra 
&=& {\ts{1\over4}}\tr[\psl_1\pslp_{\!1}\pslp_{\!2}\psl_2]
+{\ts{1\over4}}m^2\,\tr[\psl_1\pslp_{\!1} 
                      - \psl_1\pslp_{\!2}
                      - \psl_1\psl_2
                      - \pslp_{\!1}\pslp_{\!2}
                      - \pslp_{\!1}\psl_2
                      + \pslp_{\!2}\psl_2]
\nonumber \\
&& {} + {\ts{1\over4}} m^4\,\tr\,1 
\nonumber \\
\noalign{\smallskip}
&=& (p_1 p'_1)(p_2 p'_2) - (p_1 p'_2)(p_2 p'_1) + (p_1 p_2)(p'_1 p'_2)
\nonumber \\
&& {} - m^2[p_1 p'_1 - p_1 p'_2 - p_1 p_2 - p'_1 p'_2 + p'_1 p_2 + p_2 p'_2] + m^4
\nonumber \\
\noalign{\smallskip}
&=& -\half s t + 2m^2 u \;.
\label{astav2}
\end{eqnarray}
To get the last line, we used \eq{man2}, and then simplified it as much
as possible via $s+t+u=4m^2$.
Since our result is symmetric on $s\lra t$, we have
$\la \Phi_{ts}\ra = \la \Phi_{st}\ra$.  

Putting all of this together, we get
\begin{equation}
\la|\T|^2\ra = g^4\!\left[\,{(s-4m^2)^2\over(M^2-s)^2}
+ {st - 4m^2 u \over(M^2-s)(M^2-t)}
+{(t-4m^2)^2\over(M^2-t)^2}\,\right].
\label{tsq472}
\end{equation}
This can then be converted to a differential cross section (in
any frame) via the formulae of section 11.

Let's do one more: $e^-\ph\to e^-\ph$.  We take the incoming
and outgoing electrons to have momenta $p$ and $p'$, respectively, 
and the incoming and outgoing scalars to have momenta $k$ and $k'$,
respectively.  We then have $p^2=p^{\prime 2}=-m^2$ and 
$k^2=k^{\prime 2}=-M^2$.  The Mandelstam variables are
\begin{eqnarray}
s &=& -(p+k)^2 = -(p'+k')^2 \;,
\nonumber \\
\noalign{\smallskip}
t &=& -(p-p')^2 = -(k-k')^2 \;,
\nonumber \\
\noalign{\smallskip}
u &=& -(p-k')^2 = -(k-p')^2 \;,
\label{man3}
\end{eqnarray}
and they obey $s+t+u=2m^2+2M^2$.  Our result in section 46 was
\begin{equation}
\la |\T|^2\ra 
= {\ts{1\over2}}\tr\Bigl[A
(-\psl+m)A(-\pslp+m)\Bigr]\;,
\label{tsq473}
\end{equation}
where
\begin{equation}
A = g^2 \left[
{-\ksl + 2m\over m^2-s}+
{\ksl' + 2m\over m^2-u}\right].
\label{biga2}
\end{equation}
Thus we have
\begin{equation}
\la|\T|^2\ra = g^4\!\left[\,{\la\Phi_{ss}\ra\over(m^2-s)^2}
+ {\la\Phi_{su}\ra+\la\Phi_{us}\ra\over(m^2-s)(m^2-u)}
+{\la\Phi_{uu}\ra\over(m^2-u)^2}\,\right],
\label{tsq474}
\end{equation}
where now
\begin{eqnarray}
\la \Phi_{ss}\ra &=& {\ts{1\over2}}
\tr\Bigl[(-\pslp{+}m)
(-\ksl{+}2m)
(-\psl{+}m)
(-\ksl{+}2m)
\Bigr]\;,
\label{assav4} \\
\noalign{\medskip}
\la \Phi_{uu}\ra &=& {\ts{1\over2}}
\tr\Bigl[(-\pslp{+}m)
(+\ksl'{+}2m)
(-\psl{+}m)
(+\ksl'{+}2m)
\Bigr]\;,
\label{auuav4} \\
\noalign{\medskip}
\la \Phi_{su}\ra &=& {\ts{1\over2}}
\tr\Bigl[(-\pslp{+}m)
(-\ksl{+}2m)
(-\psl{+}m)
(+\ksl'{+}2m)
\Bigr]\;,
\label{asuav4} \\
\noalign{\medskip}
\la \Phi_{us}\ra &=& {\ts{1\over2}}
\tr\Bigl[(-\pslp{+}m)
(+\ksl'{+}2m)
(-\psl{+}m)
(-\ksl{+}2m)
\Bigr]\;.
\label{ausav4}
\end{eqnarray}
We can evaluate these in terms of the Mandelstam variables by using our trace 
technology, along with
\begin{eqnarray}
p k = p' k' &=& -\half(s-m^2-M^2) \;,
\nonumber \\
\noalign{\smallskip}
p p' &=& +\half(t-2m^2) \;,
\nonumber \\
\noalign{\smallskip}
k k' &=& +\half(t-2M^2) \;,
\nonumber \\
\noalign{\smallskip}
p k' = p' k &=& +\half(u-m^2-M^2) \;.
\label{man4}
\end{eqnarray}
Examining \eqs{assav4} and (\ref{auuav4}), we see that
$\la \Phi_{ss}\ra$ and $\la \Phi_{uu}\ra$ are transformed into each other by
$k\lra -k'$.
Examining \eqs{asuav4} and (\ref{ausav4}), we see that
$\la \Phi_{su}\ra$ and $\la \Phi_{us}\ra$ 
are also transformed into each other by
$k\lra -k'$.  From \eq{man4}, we see that this is equivalent to $s\lra u$.
Thus we need only compute $\la \Phi_{ss}\ra$ and $\la \Phi_{su}\ra$, 
and then take
$s\lra u$ to get $\la \Phi_{uu}\ra$ and $\la \Phi_{us}\ra$.
This is, again, merely tedious, and the results are
\begin{eqnarray}
\la \Phi_{ss}\ra &=& 
-su+m^2(9s+u)+7m^4-8m^2 M^2+M^4 \;,
\label{assav5} \\
\noalign{\medskip}
\la \Phi_{uu}\ra &=& 
-su+m^2(9u+s)+7m^4-8m^2 M^2+M^4 \;, 
\label{auuav5} \\
\noalign{\medskip}
\la \Phi_{su}\ra &=&
+ s u + 3 m^2(s+u) + 9m^4 - 8m^2 M^2 - M^4  \;.
\label{asuav5} \\
\noalign{\medskip}\la \Phi_{us}\ra &=&
+ s u + 3 m^2(s+u) + 9m^4 - 8m^2 M^2 - M^4  \;.
\label{ausav5} 
\end{eqnarray}

\vfill\eject

\begin{center}
Problems
\end{center}

\vskip0.25in

48.1) The tedium of these calculations is greatly alleviated by making use
of a symbolic manipulation program like Mathematica or Maple.
One approach is brute force: 
compute $4\times 4$ matrices like $\psl$ in the CM frame,
and take their products and traces.  
If you are familiar with a symbolic-manipulation
program, write one that does this.
See if you can verify eqs.$\,$(\ref{assav5}--\ref{ausav5}).

48.2) Compute $\la|\T|^2\ra$ for $e^- e^- \to e^- e^-$.  You should find
that your result is the same as that for 
$e^+ e^- \to e^+ e^-$, but with $s\lra u$, and an extra overall minus sign.
This relationship is known as {\it crossing symmetry}. 

48.3) Compute $\la|\T|^2\ra$ for $e^+ e^- \to \ph\ph$.  You should find
that your result is the same as that for 
$e^- \ph \to e^- \ph$, but with $s\lra t$, and an extra overall minus sign.
This is another example of crossing symmetry.

48.4) Suppose that $M>2m$, so that the scalar can decay to an 
electron-positron pair.

a) Compute the decay rate, summed over final spins.

b) Compute $|\T|^2$
for decay into an electron with spin $s_1$ and a
positron with spin $s_2$.   Take the fermion three-momenta to be along
the $z$ axis, and let the $x$-axis be the spin-quantization axis.
You should find that $|\T|^2=0$ if $s_1=-s_2$, or if $M=2m$ (so that
the outgoing three-momentum of each fermion is zero).
Discuss this in light of conservation of angular momentum
and of parity.  (Prerequisite: 40.)

c) Compute 
the rate for decay into an electron with helicity $s_1$ and a
positron with helicity $s_2$.  (See section 38 for the definition
of helicity.)
You should find that the decay rate is zero if $s_1=-s_2$.
Discuss this in light of conservation of angular momentum
and of parity.  

d) Now consider changing the interaction to $\L_1 = ig\ph\Psibar\g_5\Psi$,
and compute the spin-summed decay rate.
Explain (in light of conservation of angular momentum
and of parity) why the decay rate is larger than it was without the $i\g_5$
in the interaction.

e) Repeat parts (b) and (c) for the new form of the interaction,
and explain any differences in the results.

\vfill\eject

\hyphenation{coun-ter-terms}

\noindent Quantum Field Theory  \hfill   Mark Srednicki

\vskip0.5in

\begin{center}
\large{49: The Feynman Rules for Majorana Fields}
\end{center}
\begin{center}
Prerequisite: 45
\end{center}

\vskip0.5in

In this section we will deduce the Feynman rules for Yukawa theory,
but with a Majorana field instead of a Dirac field.
We can think of the particles associated with the Majorana field
as massive neutrinos.

We have
\begin{eqnarray}
\L_1 &=& \half g\ph\Psibar\Psi 
\nonumber \\
\noalign{\medskip}
&=& \half g\ph\Psit \C\Psi \;, 
\label{l1maj}
\end{eqnarray}
where $\Psi$ be a Majorana field (with mass $m$) and
$\ph$ is a real scalar field (with mass $M$), and
$g$ is a coupling constant.   In this section, we will
be concerned with tree-level processes only, and so 
we omit renormalizing $Z$ factors.

From section 41, we have the LSZ rules appropriate for a Majorana field,
\begin{eqnarray}
\bd_s(\p)_{\rm in} &\to&  -i\int\dfx\;
e^{+ipx}\;\vbar_s(\p)(-i\dsl+m)\Psi(x)
\label{bdsub1} \\
\noalign{\medskip}
&& = + i\int\dfx\;
\Psit(x)\C(+i{\buildrel\leftarrow\over\dsl} + m) u_s(\p)e^{+ipx}\;,
\label{bdsub2} \\
\noalign{\medskip}
b_{s'}(\p')_{\rm out} &\to&  +i\int\dfx\;
e^{-ip'x}\;\ubar_{s'}(\p')(-i\dsl+m)\Psi(x)\;,
\label{bsub1} \\
\noalign{\medskip}
&& = - i\int\dfx\;
e^{-ip'x}\;\Psit(x)\C(+i{\buildrel\leftarrow\over\dsl} + m) 
   v_{s'}(\p')e^{-ip'x}\;.
\label{bsub2}
\end{eqnarray}
\Eq{bdsub2} follows from \eq{bdsub1} by taking the transpose of
the right-hand side, and using 
$\vbar_{s'}(\p')^{\rm\sss T}=-\C u_{s'}(\p')$ and
$(-i\dsl+m)^{\rm\sss T}=\C(+i\dsl+m)\C^{-1}$; similarly,
\eq{bsub2} follows from \eq{bsub1}.  Which form we use depends on
convenience, and is best chosen on a diagram-by-diagram basis,
as we will see shortly.

Eqs.$\,$(\ref{bdsub1}--\ref{bsub2}) lead us to compute correlation
functions containing $\Psi$'s, but not $\Psibar$'s. In position space,
this leads to Feynman rules where the fermion propagator is
${1\over i}S(x-y)\C^{-1}$, and the $\ph\Psi\Psi$ vertex is $ig\C$;
the factor of $\half$ in $\L_1$ is killed by a symmetry 
factor of $2!$ that arises from having two identical
$\Psi$ fields in $\L_1$.
In a particular diagram, as we move along a fermion line,
the $\C^{-1}$ in the propagator will cancel against the $\C$ in
the vertex, leaving over a final $\C^{-1}$ at one end.  This $\C^{-1}$
can be canceled by a $\C$ from \eq{bdsub2} (for an incoming
particle) or (\ref{bsub2}) (for an outgoing particle).
On the other hand, for the other end of the
same line, we should use either \eq{bdsub1} (for an incoming
particle) or \eq{bsub1} (for an outgoing particle) to avoid introducing
an extra $\C$ at {\it that\/} end.  In this way, we can avoid ever having
explicit factors of $\C$ in our Feynman rules.

Using this approach, the Feynman rules for this theory are as follows.

1) The total number of incoming and outgoing neutrinos is always even;
call this number $2n$.  Draw $n$ solid lines.  Connect them with internal
dashed lines, using a vertex that joins one dashed and two solid lines. 
Also, attach an external dashed line for each incoming or outgoing
scalar. In this way, draw all possible diagrams that are
{\it topologically inequivalent}.

2) Draw arrows on each segment of each solid line; keep the arrow direction
continuous along each line.

3) Label each external dashed line with the momentum of an
incoming or outgoing scalar.  If the particle is incoming, draw an arrow
on the dashed line that points {\it towards\/} the vertex;
If the particle is outgoing, draw an arrow
on the dashed line that points {\it away\/} from the vertex.

4) Label each external solid line with the momentum of an
incoming or outgoing neutrino, but include a minus sign with the momentum
if (a) the particle is incoming and the arrow points {\it away\/} from the vertex,
or (b) the particle is outgoing and the arrow points {\it towards\/} the vertex.

5) Do this labeling of external lines in all possible {\it inequivalent\/} ways.
Two diagrams are considered {\it equivalent\/} if they can be transformed into
each other by reversing all the arrows on one or more fermion lines,
and correspondingly changing the signs of the external momenta
on each reversed-arrow line.  The process of arrow reversal contributes
a minus sign for each reversed-arrow line.

6) Assign each internal line its own four-momentum. 
Think of the four-momenta as flowing along the arrows,
and conserve four-momentum at each vertex.
For a tree diagram, this fixes the momenta on all the internal lines.

9) The value of a diagram consists of the following factors:
{\obeylines
for each incoming or outgoing scalar, 1;
for each incoming neutrino labeled with $+p_i$, $u_{s_i}(\p_i)$;
for each incoming neutrino labeled with $-p_i$, $\vbar_{s_i}(\p_i)$;
for each outgoing neutrino labeled with $+p'_i$, $\ubar_{s'_i}(\p'_i)$; 
for each outgoing neutrino labeled with $-p'_i$, $v_{s'_i}(\p'_i)$;
for each vertex, $ig$;
for each internal scalar line, $-i/(k^2+M^2-i\eps)$, 
where $k$ is the four-momentum of that line;
for each internal fermion line, $-i(-\psl+m)/(p^2+m^2-i\eps)$, 
where $p$ is the four-momentum of that line.}

10) Spinor indices are contracted by starting at one end of a fermion
line: specifically, the end that has the arrow pointing away from the vertex.
The factor associated with the external line is either 
$\ubar$ or $\vbar$.  Go along the complete fermion line, following the arrows
backwards, and writing down (in order from left to right) 
the factors associated with the
vertices and propagators that you encounter.  The last factor is either
a $u$ or $v$.  Repeat this procedure for the other fermion lines, if any.

11) Two diagrams that are identical
{\it except for the momentum and spin labels on two external fermion lines\/}
that {\it have their arrows pointing in the same direction\/} (either both
towards or both away from the vertex) have a relative minus sign.

12) The value of $i\T$ is given by a sum over the values of all these diagrams.

There are additional rules for counterterms and loops, but we will postpone
those to section 51.  

\begin{figure}
\begin{center}
\epsfig{file=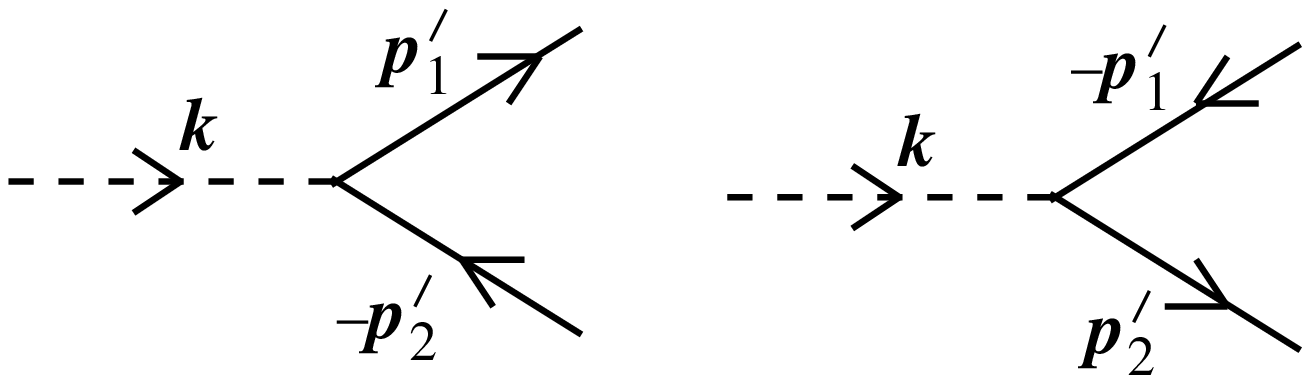}
\end{center}
\caption{Two equivalent diagrams for $\ph\to\nu\nu$,
corresponding to \eqs{phnunu1} and (\ref{phnunu2}), respectively.}
\label{phnunu}
\end{figure}

Let's look at the simplest process, $\ph\to\nu\nu$.  There are two possible
diagrams for this, shown in \fig{phnunu}.  However, according to rule \#5, 
these two diagrams are equivalent.  The first one evaluates to
\begin{equation}
i\T = ig\,\vbarp_{\!2}u'_1 \;,
\label{phnunu1}
\end{equation}
while the second gives
\begin{equation}
i\T = -ig\,\vbarp_{\!1}u'_2 \;.
\label{phnunu2}
\end{equation}
The minus sign comes from the last part of rule \#5: reversing the arrows
on one fermion line gives an extra minus sign.
These two versions of $\T$ should of course yield the same result;
to check this, note that
\begin{eqnarray}
\vbar_1 u_2 &=&[\vbar_1 u_2]^{\rm\sss T}
\nonumber \\
\noalign{\smallskip}
&=& u_2^{\rm\sss T}\,\vbar_1^{\rm\sss T}
\nonumber \\
\noalign{\smallskip}
&=& \vbar_2 \C^{-1}\C^{-1}u_1
\nonumber \\
\noalign{\smallskip}
&=& -\vbar_2 u_1 \;,
\label{phnunu3}
\end{eqnarray}
as required.

In general, for processes with a total of just two incoming 
and outgoing neutrinos, such as 
$\nu\ph\to\nu\ph$ or $\nu\nu\to\ph\ph$, these rules give (up to an 
irrelevant overall sign) 
the same result for $i\T$ as we would get for the corresponding process
in the Dirac case,  $e^-\ph\to e^-\ph$ or $e^+ e^-\to\ph\ph$.
(Note, however, that in the Dirac case, we have $\L_1=g\ph\Psibar\Psi$,
as compared with $\L_1=\half g\ph\Psibar\Psi$ in the Majorana case.)

\begin{figure}
\begin{center}
\epsfig{file=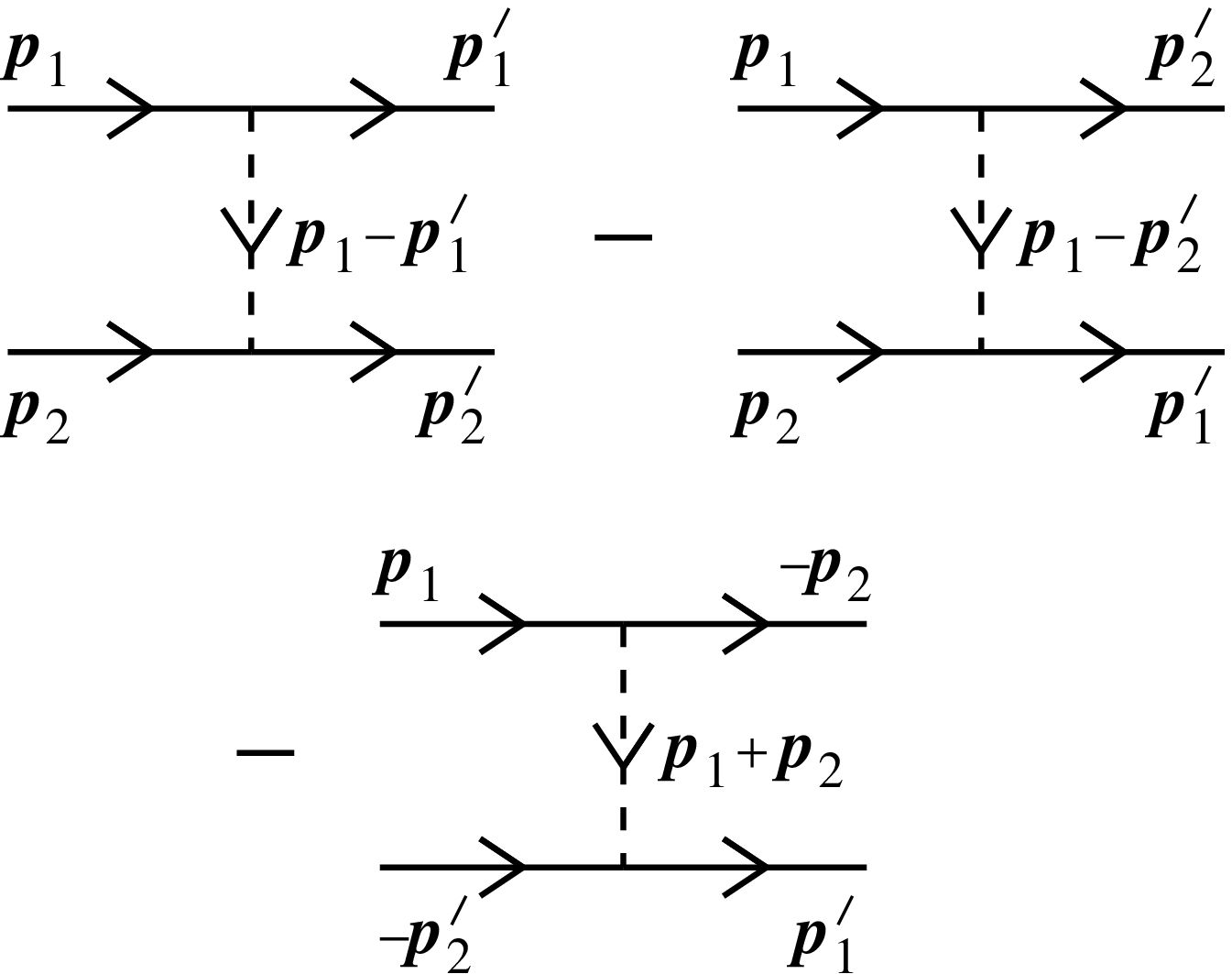}
\end{center}
\caption{Diagrams for $\nu\nu\to\nu\nu$, corresponding to \eq{4nu2}.}
\label{4nu}
\end{figure}

The differences between Dirac and Majorana fermions become
more pronounced for $\nu\nu\to\nu\nu$.  Now there are {\it three\/}
inequivalent contributing diagrams, shown in \fig{4nu}.  
The corresponding amplitude can be written as
\begin{equation}
i\T = {\ts{1\over i}} (ig)^2 
\left[
  { (\ubarp_{\!1}u_1)(\ubarp_{\!2}u_2) \over -t+M^2 } 
- { (\ubarp_{\!2}u_1)(\ubarp_{\!1}u_2) \over -u+M^2 }
+ { (\vbar_2 u_1)(\ubarp_{\!1}v'_2)    \over -s+M^2 }\right],
\label{4nu2}
\end{equation}
where $s=-(p_1+p_2)^2$, $t=-(p_1-p'_1)^2$ and $u=-(p_1-p'_2)^2$.
Note the relative signs.
After taking the absolute square of this expression,
we can use relations like \eq{phnunu3} on a term-by-term basis
to put everything into a form that allows the spin sums to be performed
in the standard way.  

In fact, we have already done all the necessary work in the Dirac case.
The $s$-$s$, $s$-$t$, and $t$-$t$ terms in $\la|\T|^2\ra$ 
for $\nu\nu\to\nu\nu$ are the same
as those for $e^+ e^-\to e^+ e^-$, while the
$t$-$t$, $t$-$u$, and $u$-$u$ terms are the same as those for 
the crossing-related process $e^- e^- \to e^- e^-$.  
Finally, the $s$-$u$ terms can be obtained from the
$s$-$t$ terms via $t\leftrightarrow u$, or equivalently from the
$t$-$u$ terms via $t\leftrightarrow s$.  Thus the result is
\begin{eqnarray}
\la|\T|^2\ra &=&
 g^4 \Biggl[\,
{(s-4m^2)^2\over(M^2-s)^2} + {st - 4m^2 u \over(M^2-s)(M^2-t)}
\nonumber \\
\noalign{\medskip}
&& \; {} + 
{(t-4m^2)^2\over(M^2-t)^2} + {tu - 4m^2 s \over(M^2-t)(M^2-u)}
\nonumber \\
\noalign{\medskip}
&& \; {} + 
{(u-4m^2)^2\over(M^2-u)^2} + {us - 4m^2 t \over(M^2-u)(M^2-s)}
\Biggr],
\label{nunununu}
\end{eqnarray}
which is neatly symmetric on permutations of $s$, $t$, and $u$.




\vfill\eject

\noindent Quantum Field Theory  \hfill   Mark Srednicki

\vskip0.5in

\begin{center}
\large{50: Massless Particles and Spinor Helicity}
\end{center}
\begin{center}
Prerequisite: 48
\end{center}

\vskip0.5in

Scattering amplitudes often simplify greatly if the particles
are massless (or can be approximated as massless because the
Mandelstam variables all have magnitudes much larger than
the particle masses squared).  In this section we will explore
this phenomenon for spin-one-half (and spin-zero) particles.  
We will begin developing
the technology of {\it spinor helicity}, which will prove to be of
indispensible utility in Part III.

Recall from section 38 that the $u$ spinors for a massless spin-one-half
particle obey
\begin{equation}
u_s(\p)\ubar_s(\p) = \half(1+s\g_5)(-\psl) \;,
\label{usp}
\end{equation}
where $s=\pm$ specifies the {\it helicity}, the component of the particle's
spin measured along the axis specified by its three-momentum; in this
notation the helicity is $\half s$.  The $v$ spinors obey a similar relation,
\begin{equation}
v_s(\p)\vbar_s(\p) = \half(1-s\g_5)(-\psl) \;.
\label{vsp}
\end{equation}
In fact, in the massless case, with the phase conventions of section 38,
we have $v_s(\p)=u_{-s}(\p)$.  Thus we can confine our discussion to $u$-type spinors only, since we need merely
change the sign of $s$ to accomodate $v$-type spinors. 

Let us consider a $u$ spinor for a particle of negative helicity.  We have
\begin{equation}
u_-(\p)\ubar_-(\p) = \half(1-\g_5)(-\psl) \;.
\label{uspm}
\end{equation}
Let us define
\begin{equation}
p_{a\adot} \equiv p_\mu\sigma^\mu_{a\adot} \;.
\label{paadot} 
\end{equation}
Then we also have
\begin{equation}
p^{\adot a} = \e^{ac}\e^{\adot\ccdot}p_{c\ccdot} =
p_\mu\bar\sigma^{\mu \adot a} \;.
\label{padota}
\end{equation}
Then, using
\begin{equation}
\gamma^\mu =
\pmatrix{0 & \sigma^\mu \cr
\noalign{\medskip}
\bar\sigma^\mu & 0 \cr} ,
\qquad
\half(1-\g_5) = \pmatrix{1 & 0 \cr
\noalign{\medskip}
           0 & 0 \cr}
\label{gammamu}
\end{equation}
in \eq{uspm}, we find
\begin{equation}
u_-(\p)\ubar_-(\p) = \pmatrix{0 & -p_{a\adot} \cr
                     \noalign{\medskip}
                              0 & 0 \cr}.
\label{uspm2}
\end{equation}
On the other hand, we know that the lower two components of $u_-(\p)$ vanish,
and so we can write
\begin{equation}
u_-(\p) = \pmatrix{\phi_a \cr 
\noalign{\medskip}
                              0 \cr}.
\label{upphi}
\end{equation}
Here $\phi_a$ is a two-component numerical spinor; 
it is not an anticommuting object.  
Such a commuting spinor is sometimes called a {\it twistor}.
An explicit numerical formula for it (verified in problem 50.1) is 
\begin{equation}
\phi_a = \sqrt{2\omega} \pmatrix{-\sin(\half\theta)e^{-i\phi} \cr
\noalign{\medskip}
                                 +\cos(\half\theta)\phantom{e^{-i\phi}}  \cr},
\label{phia}
\end{equation}
where $\theta$ and $\phi$ are the polar and azimuthal angles
that specify the direction of the three-momentum $\p$, and $\w=|\p|$.
Barring \eq{upphi} yields
\begin{equation}
\ubar_-(\p) = \pmatrix{0, & \phi^*_\adot \cr},
\label{upbarphi}
\end{equation}
where $\phi_\adot^*=(\phi_a)^*$.
Now, combining \eqs{upphi} and (\ref{upbarphi}), we get
\begin{equation}
u_-(\p)\ubar_-(\p) = \pmatrix{0 & \phi_a\phi^*_\adot \cr
                     \noalign{\medskip}
                              0 & 0 \cr}.
\label{uspm3}
\end{equation}
Comparing with \eq{uspm2}, we see that
\begin{equation}
p_{a\adot}=-\phi_a\phi^*_\adot\;.
\label{paaphi}
\end{equation}
This expresses the four-momentum of the particle neatly in terms of the twistor
that describes its spin state.  The essence of the spinor helicity method is to 
treat $\phi_a$ as the fundamental object, and to express the particle's 
four-momentum in terms of it, via \eq{paaphi}.  

Given \eq{upphi}, and the phase conventions of section 38, 
the positive-helicity spinor is
\begin{equation}
u_+(\p) = \pmatrix{0 \cr
\noalign{\medskip} 
                                \phi^{*\adot} \cr},
\label{uppphi}
\end{equation}
where $\phi^{*\adot}=\e^{\adot\ccdot}\phi^*_\ccdot$.  
Barring \eq{uppphi} yields
\begin{equation}
\ubar_+(\p) = \pmatrix{\phi^a, & 0 \cr}.
\label{uppbarphi}
\end{equation}
Computation of $u_+(\p)\ubar_+(\p)$ via \eqs{uppphi} and 
(\ref{uppbarphi}), followed by comparison with \eq{usp} with $s={+}$,
then reproduces \eq{paaphi}, but with the indices raised.

In fact, the decomposition of $p_{a\adot}$ into the direct product
of a twistor and its complex conjugate is unique
(up to an overall phase for the twistor).  To see this, use
$\sigma^\mu=(I,\vec\sigma)$ to write
\begin{equation}
p_{a\adot}= \pmatrix{-p^0+p^3 & p^1-ip^2 \cr
                     \noalign{\medskip}
                              p^1+ip^2 & -p^0-p^3 \cr}.
\label{paamat}
\end{equation}
The determinant of this matrix is $-(p^0)^2+\p^2$, and this
vanishes because the particle is (by assumption) massless.
Thus $p_{a\adot}$ has a zero eigenvalue.  Therefore, it
can be written as a projection onto the eigenvector corresponding
to the nonzero eigenvalue.  That is what \eq{paaphi} represents,
with the nonzero eigenvalue absorbed into the normalization of the 
eigenvector $\phi_a$.

Let us now introduce some useful notation.  Let $p$ and $k$ be two 
four-momenta, and $\phi_a$ and $\kappa_a$ the corresponding twistors.  
We define the twistor product
\begin{equation}
[p\,k] \equiv \phi^a\kappa_a\;.
\label{pk}
\end{equation}
Because $\phi^a\kappa_a=\e^{ac}\phi_c\kappa_a$, 
and the twistors commute, we have
\begin{equation}
[k\,p] = -[p\,k] \;.
\label{kp}
\end{equation}
From \eqs{upphi} and (\ref{uppbarphi}), we can see that 
\begin{equation}
\ubar_+(\p)u_-(\k) =  [p\,k] \;.
\label{upum}
\end{equation}
Similarly, let us define
\begin{equation}
\la p\,k\ra \equiv \phi_\adot^*\kappa^{*\adot}\;.
\label{pk2}
\end{equation}
Comparing with \eq{pk} we see that
\begin{equation}
\la p\,k\ra = [k\,p]^*\;,
\label{pkkp*}
\end{equation}
which implies that this product is also antisymmetric,
\begin{equation}
\la k\,p\ra = -\la p\,k\ra\;.
\label{kp2}
\end{equation}
Also, from \eqs{upbarphi} and (\ref{uppphi}), we have
\begin{equation}
\ubar_-(\p)u_+(\k) =  \la p\,k\ra \;.
\label{umup}
\end{equation}
Note that the other two possible spinor products vanish:
\begin{equation}
\ubar_+(\p)u_+(\k) = \ubar_-(\p)u_-(\k) =  0\;.
\label{upup}
\end{equation}
The twistor products $\la p\,k\ra$ and $[p\,k]$ 
satisfy another important relation,
\begin{eqnarray}
\la p\,k\ra [k\,p] &=& (\phi^*_\adot\kappa^{*\adot}) (\kappa^a\phi_a)
\nonumber \\
\noalign{\medskip}
&=&  (\phi^*_\adot\phi_a)(\kappa^a\kappa^{*\adot})
\nonumber \\
\noalign{\medskip}
&=&  p_{\adot a}k^{a\adot}
\nonumber \\
\noalign{\medskip}
&=&  -2p^\mu k_\mu \;,
\label{pkkp}
\end{eqnarray}
where the last line follows from 
$\bar\sigma^{\mu\adot a}\sigma^\nu_{a\adot}=-2g^{\mu\nu}$.

Let us apply this notation to the tree-level scattering amplitude for 
$e^-\ph\to e^-\ph$ in Yukawa theory, 
which we first computed in Section 44, and which reads
\begin{equation}
\T_{s's}= g^2 \,
\ubar_{s'}(\p')\Bigl[\tilde S(p{+}k)+\tilde S(p{-}k')\Bigr]u_s(\p)\;.
\label{epep}
\end{equation}
For a massless fermion, $\tilde S(p)=-\psl/p^2$.  
If the scalar is also massless,
then $(p+k)^2=2p\cdot k$ and $(p-k')^2=-2p\cdot k'$.  Also, we can remove
the $\psl$'s in the propagator numerators in \eq{epep},
because $\psl u_s(\p)=0$.  Thus we have
\begin{equation}
\T_{s's}= g^2 \,
\ubar_{s'}(\p')\!\left[{-\ksl\over 2p\cd k} + {-\ksl'\over 2p\cd k'}\right]\!u_s(\p)\;.
\label{epep2}
\end{equation}
Now consider the case $s'=s=+$.  From \eqs{uppphi}, (\ref{uppbarphi}), and
\begin{equation}
-\ksl = \pmatrix{0 & \kappa_a\kappa^*_\adot \cr
                     \noalign{\medskip}
                           \kappa^{*\adot}\kappa^a & 0 \cr},
\label{kslkk}
\end{equation}
we get
\begin{eqnarray}
\ubar_+(\p')(-\ksl)u_+(\p) 
&=& \phi^{\prime a}\kappa_a\kappa^*_\adot\phi^{*\adot}
\nonumber \\
\noalign{\medskip}
&=& [ p'\,k]\,\la k\,p\ra\;.
\label{ubpkup}
\end{eqnarray}
Similarly, for $s'=s=-$, we find
\begin{eqnarray}
\ubar_-(\p')(-\ksl)u_-(\p) &=& 
\phi^{\prime *}_\adot\kappa^{*\adot}\kappa^a\phi_a
\nonumber \\
\noalign{\medskip}
&=& \la p'\,k\ra\,[k\,p]\;,
\label{ubmkum}
\end{eqnarray}
while for $s'\ne s$, the amplitude vanishes:
\begin{equation}
\ubar_-(\p')(-\ksl)u_+(\p) = \ubar_+(\p')(-\ksl)u_-(\p) = 0 \;.
\label{ubarpmuzero}
\end{equation}
Then, using \eq{pkkp} on the denominators in \eq{epep2}, we find
\begin{eqnarray}
\T_{++}&=& 
-g^2 \left(  {[p'\,k]\over [p\,k]} 
           + {[p'\,k']\over [p\,k']} \right),
\nonumber \\
\noalign{\medskip}
\T_{--}&=& 
-g^2 \left(  {\la p'\,k\ra\over\la p\,k\ra} 
           + {\la p'\,k'\ra\over\la p\,k'\ra} \right),
\label{tpptmm}
\end{eqnarray}
while
\begin{equation}
\T_{+-}=\T_{-+}=0\;.
\label{tpmtmp}
\end{equation}
Thus we have rather simple expressions for the fixed-helicity scattering 
amplitudes in terms of twistor products. 

We can simplify the derivation of these results 
by setting up a bra-ket notation.  Let
\begin{eqnarray}
|p] &=& u_-(\p)=v_+(\p)\;,
\nonumber \\
\noalign{\medskip}
|p\ra &=& u_+(\p)=v_-(\p)\;,
\nonumber \\
\noalign{\medskip}
[p | &=& \ubar_+(\p)=\vbar_-(\p)\;,
\nonumber \\
\noalign{\medskip}
\la p | &=& \ubar_-(\p)=\vbar_+(\p)\;.
\label{bra-ket}
\end{eqnarray}
We then have
\begin{eqnarray}
\la k|\;|p\ra &=& \la k\,p\ra\;,
\nonumber \\
\noalign{\medskip}
[ k|\;|p] &=& [ k\,p]\;,
\nonumber \\
\noalign{\medskip}
\la k|\;|p]  &=& 0\;,
\nonumber \\
\noalign{\medskip}
[ k|\;|p\ra &=& 0\;.
\label{braket1}
\end{eqnarray}
We also can write
\begin{equation}
-\psl = |p\ra[p| + |p]\la p|\;,
\label{pslbraket}
\end{equation}
where $p$ is any massless four-momentum.
With this notation, we can easily reproduce the results of 
eqs.$\;$(\ref{ubpkup}--\ref{ubarpmuzero}).

\vskip0.5in

\begin{center}
Problems
\end{center}

\vskip0.25in

50.1a)
Use \eqs{phia} and (\ref{paamat}) to verify \eq{paadot}.

b) 
Show that $\psl u_-(\p) = p^{\adot a}\phi_a$.
Then use \eq{paadot} to show that that $p^{\adot a}\phi_a=0$.

c) 
Let the three-momentum $\p$ be in the $\bf +\hat z$ direction.
Use \eq{uvp} in section 38 to compute $u_\pm(\p)$ explicitly in the
massless limit (corresponding to the limit $\eta\to\infty$, where
$\sinh\eta = |\p|/m$.
Verify that, when $\theta=0$, your results agree with 
\eqs{upphi}, (\ref{phia}), and (\ref{uppphi}).
Hint: if a matrix $M$ has eigenvalues $\pm1$ only, then
$\exp(aM)=\cosh(a)+\sinh(a)M$.

50.2)
Prove the Schouten identity, \begin{equation}
 \la p\,q\ra\,\la r\,s\ra 
+\la p\,r\ra\,\la s\,q\ra 
+\la p\,s\ra\,\la q\,r\ra = 0\;.
\label{schouten}
\end{equation}
Hint: note that the left-hand side is completely antisymmetric in
the three labels $q$, $r$, and $s$, 
and that each corresponding twistor has only two components.

50.3)
Show that 
\begin{equation}
 \la p\,q\ra\,[q\,r]\,\la r\,s\ra\,[s\,p]
= \tr\,\half(1{-}\g_5)\psl\qsl\rsl\ssl \;, 
\label{4prod}
\end{equation}
and evaluate the right-hand side.

50.4a) 
Prove the useful identities
\begin{eqnarray}
\la p|\g^\mu|k] &=& [k|\g^\mu|p\ra\;,
\label{fierz0} \\
\noalign{\medskip}
\la p|\g^\mu|k]^* &=& \la k|\g^\mu|p]\;,
\label{fierz1} \\
\noalign{\medskip}
\la p|\g^\mu|p] &=& 2p^\mu \;,
\label{fierz2} \\
\noalign{\medskip}
\la p|\g^\mu|k\ra &=& 0 \;,
\label{fierz3} \\ 
\noalign{\medskip}
[ p|\g^\mu|k ] &=& 0\;.
\label{fierz4} 
\end{eqnarray}

b)
Extend the last two identies of part (a): show that the product of
an odd number of gamma matrices sandwiched 
between either $\la p|$ and $|k\ra$ or $[p|$ and $|k]$ vanishes.  
Also show that the product of an even number of gamma matrices
between either $\la p|$ and $|k]$ or $[p|$ and $|k\ra$ vanishes.

c)
Prove the Fierz identities,
\begin{eqnarray}
-\half \la p|\g_\mu|q]\gamma^\mu &=& |q]\la p| + |p\ra[q| \;, 
\label{fierzza} \\
\noalign{\medskip}
-\half [p|\g_\mu|q\ra\gamma^\mu &=& |q\ra[p| + |p]\la q| \;.
\label{fierzzb}
\end{eqnarray}
Now take the matrix element of \eq{fierzzb} between 
$\la r|$ and $|s]$ to get yet another form of the Fierz identity,
\begin{equation}
[p|\g^\mu|q\ra\,\la r|\g_\mu|s] = 2\,[p\,s]\,\la q\,r\ra\;.
\label{fierzc}
\end{equation}

\vfill\eject

\noindent Quantum Field Theory  \hfill   Mark Srednicki

\vskip0.5in

\begin{center}
\large{51: Loop Corrections in Yukawa Theory}
\end{center}
\begin{center}
Prerequisite: 19, 40, 48
\end{center}

\vskip0.5in

In this section we will compute the one-loop corrections in Yukawa
theory with a Dirac field.  The basic concepts are all the same as for
a scalar, and so we will mainly be concerned with the extra technicalities
arising from spin indices and anticommutation.

First let us note that the general discussion of sections 18 and 29 leads us
to expect that we will need to add to the lagrangian all possible
terms whose coefficients have positive or zero mass dimension,
and that respect the symmetries of the original lagrangian.
These include Lorentz symmetry, the U(1) phase symmetry of the
Dirac field, and the discrete symmetries of parity, time reversal,
and charge conjugation.

The mass dimensions of the fields (in four spacetime dimensions)
are $[\ph]=1$ and $[\Psi]={3\over2}$.  Thus any power of $\ph$
up to $\ph^4$ is allowed.  But there are no additional required
terms involving $\Psi$: the only candidates contain either 
$\g_5$ (e.g., $i\Psibar\g_5\Psi$) and are forbidden by parity,
or $\C$ (e.g, $\Psi^{\rm\sss T}\C\Psi$) and are forbidden by the
U(1) symmetry.

Nevertheless, having to deal with the addition of three new terms
($\ph$, $\ph^3$, $\ph^4$)
is annoying enough to prompt us to
look for a simpler example.  Consider, then, 
a modified form of the Yukawa interaction,
\begin{equation}
\L_1 = ig\ph\Psibar\g_5\Psi \;.
\label{l1g5}
\end{equation}
This interaction will conserve parity 
if and only if $\ph$ is a pseudoscalar: 
\begin{equation}
P^{-1}\ph(\x,t)P = -\ph(-\x,t) \;.
\label{piphp49}
\end{equation}
Then, $\ph$ and $\ph^3$ are odd under parity, and so 
we will {\it not\/} need to add them to $\L$.
The one term we will need to add is $\ph^4$.

Therefore, the theory we will consider is
\begin{eqnarray}
\L &=& \L_0 + \L_1 \;,
\label{ll0l1} \\
\noalign{\medskip}
\L_0 &=& i\Psibar\dsl\Psi-m\Psibar\Psi 
-\half\d^\mu\ph\d_\mu\ph-\half M^2\ph^2 \;,
\label{l0} \\
\noalign{\medskip}
\L_1 &=& iZ_g g\ph\Psibar\g_5\Psi - {\ts{1\over24}}Z_\lam \lam\ph^4
+ \L_{\rm ct}\;,
\label{l1} \\
\noalign{\medskip}
\L_{\rm ct} &=& 
i(Z_\Psi{-}1)\Psibar\dsl\Psi-(Z_m{-}1)m\Psibar\Psi 
\nonumber \\
&& {} - \half(Z_\ph{-}1)\d^\mu\ph\d_\mu\ph-\half(Z_M{-}1)M^2\ph^2
\label{lct}
\end{eqnarray}
where $\lam$ is a new coupling constant.
We will use an on-shell renormalization scheme.
The lagrangian parameter $m$ is then the actual mass
of the electron.  We will define the couplings $g$ and $\lam$
as the values of appropriate vertex functions when the
external four-momenta vanish. 
Finally, the fields are normalized according to the requirements
of the LSZ formula.  In practice, this means that the
scalar and fermion propagators must have appropriate poles with unit residue.

We will assume that $M<2m$, 
so that the scalar is stable against decay into an 
electron-positron pair.  The exact scalar propagator
(in momentum space) can be then written in \LK form as
\begin{equation}
\tbfd(k^2)=
{1\over k^2+M^2-i\eps}
+  \int_{M_{\rm th}^2}^\infty ds\,\rho_\ph(s)\,{1\over k^2+s-i\eps} \;,
\label{tbfd49}
\end{equation}
where the spectral density $\rho_\ph(s)$ is real and nonnegative.
The threshold mass $M_{\rm th}$ is either $2m$
(corresponding to the contribution of
an electron-positron pair) or
$3M$ (corresponding to the contribution of three scalars;
by parity, there is no contribution from two scalars), whichever is less.

We see that $\tbfd(k^2)$ has a pole at $k^2=-M^2$ with residue one.
This residue corresponds to the field normalization that is needed
for the validity of the LSZ formula.

We can also write the exact scalar propagator in the form
\begin{equation}
\tbfd(k^2)^{-1}=k^2 + M^2 - i\eps - \Pi(k^2) \;,
\label{tbfdi49}
\end{equation}
where $i\Pi(k^2)$ is given by the sum of 1PI diagrams with two
external scalar lines, and the external propagators removed. 
The fact that $\tbfd(k^2)$ has a pole at $k^2=-M^2$ with residue one
implies that $\Pi(-M^2)=0$ and $\Pi'(-M^2)=0$; this fixes the 
coefficients $Z_\ph$ and $Z_M$.

All of this is mimicked for the Dirac field.  When parity is conserved,
the exact propagator (in momentum space) can be written in \LK form as
\begin{equation}
\tbfs(\psl)=
{-\psl + m\over p^2+m^2-i\eps} +  \int_{m_{\rm th}^2}^\infty ds\,
\rho_\Psi(s)\,{-\psl+\sqrt{s}\over p^2+s-i\eps}\;,
\label{tbfs}
\end{equation}
real and nonnegative.
The threshold mass $m_{\rm th}$ is $m+M$ 
(corresponding to the contribution of a fermion and a scalar),
which, by assumption, is less than $3m$
(corresponding to the contribution of three fermions;
by Lorentz invariance, there is no contribution from two fermions).

Since $p^2=-\psl\psl$, we can rewrite \eq{tbfs} as
\begin{equation}
\tbfs(\psl)=
{1\over \psl +m-i\eps} + \int_{m_{\rm th}^2}^\infty ds\,
\rho_\Psi(s)\,{1\over \psl+\sqrt{s}-i\eps}\;,
\label{tbfs2}
\end{equation}
with the understanding that $1/(\ldots)$ refers to the matrix inverse.
However, since $\psl$ is the only matrix involved, we can think of 
$\tbfs(\psl)$ as an analytic function of the single variable $\psl$.
With this idea in mind, we
see that $\tbfs(\psl)$ has a pole at $\psl=-m$ with residue one.
This residue corresponds to the field normalization that is needed
for the validity of the LSZ formula.

We can also write the exact fermion propagator in the form
\begin{equation}
\tbfs(\psl)^{-1} = \psl + m - i\eps - \Sigma(\psl) \;,
\label{tbfsi}
\end{equation}
where $i\Sigma(\psl)$ is given by the sum of 1PI diagrams with two
external fermion lines, and the external propagators removed. 
The fact that $\tbfs(\psl)$ has a pole at $\psl=-m$ with residue one
implies that $\Sigma(-m)=0$ and $\Sigma'(-m)=0$; this fixes the 
coefficients $Z_\Psi$ and $Z_m$.

We proceed to the diagrams.  The Yukawa vertex carries a factor
of $i(iZ_g g)\gamma_5=-Z_g g\gamma_5$.  Since $Z_g=1+O(g^2)$, 
we can set $Z_g=1$ in the one-loop diagrams.

\begin{figure}
\begin{center}
\epsfig{file=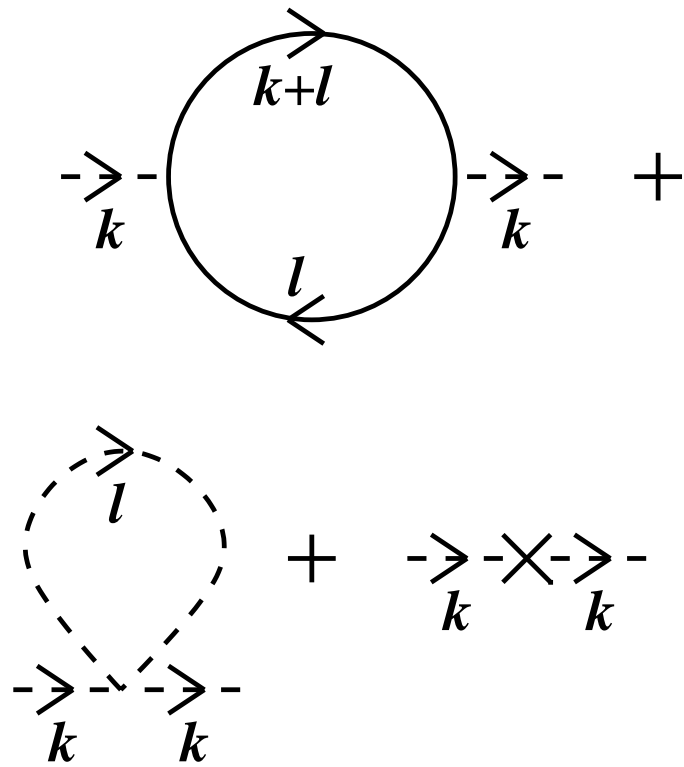}
\end{center}
\caption{The one-loop and counterterm corrections to the scalar propagator
in Yukawa theory.}
\label{piyuk}
\end{figure}

Consider first $\Pi(k^2)$, which receives
the one-loop (and counterterm) corrections shown in \fig{piyuk}.  
The first diagram has a closed fermion loop.  
As we will see in problem~51.1 (and section 53), 
anticommutation of the fermion
fields results in an extra factor of minus one for
each closed fermion loop. 
The spin indices on the propagators and vertices 
are contracted in the usual way, following the arrows backwards.
Since the loop closes on itself, we end up with a trace over the
spin indices.  Thus we have
\begin{equation}
i\Pi_{\Psi\,\rm loop}(k^2) = 
(-1)(-g)^2\!\left({\ts{1\over i}}\right)^{\!2}
\int{\dfl\over(2\pi)^4}\,
\tr\Bigl[\st(\lsl{+}\ksl)\g_5\st(\lsl)\g_5\Bigr] \;, 
\label{piloop}
\end{equation}
where 
\begin{equation}
\st(\psl) = {-\psl + m\over p^2 + m^2 -i\eps}
\label{st}
\end{equation}
is the free fermion propagator in momentum space.

We now proceed to evaluate \eq{piloop}.  We have
\begin{eqnarray}
\tr[(-\lsl-\ksl+m)\g_5(-\lsl+m)\g_5] 
&=& \tr[(-\lsl-\ksl+m)(+\lsl+m)] 
\nonumber \\
\noalign{\medskip}
&=& 4[(\ell+k)\ell + m^2] 
\nonumber \\
\noalign{\medskip}
&\equiv& 4N \;. 
\label{numtr}
\end{eqnarray}
The first equality follows from $\g_5^2=1$ and $\g_5\psl\g_5 = -\psl$.

Next we combine the denominators with Feynman's formula.
Suppressing the $i\eps$'s, we have
\begin{equation}
{1\over (\ell{+}k)^2 + m^2}\, {1\over \ell^2 + m^2}
= \int_0^1 dx\,{1\over(q^2 + D)^2} \;,
\label{ellqs}
\end{equation}
where $q=\ell+xk$ and $D = x(1{-}x)k^2 + m^2$.

We then change the integration variable in \eq{piloop} from $\ell$ to $q$;
the result is
\begin{equation}
i\Pi_{\Psi\,\rm loop}(k^2) = 
4g^2 \int_0^1 dx
\int{\dfq\over(2\pi)^4}\,{N\over (q^2 + D)^2 } \;,
\label{piloop2}
\end{equation}
where now $N=(q + (1{-}x)k)(q-xk) + m^2$.
The integral diverges, and so we analytically continue it to 
$d=4-\e$ spacetime dimensions.  
(Here we ignore a subtlety with the definition of $\g_5$ in $d$ dimensions,
and assume that $\g_5^2=1$ and $\g_5\psl\g_5=-\psl$ continue to hold.)
We also make the replacement
$g\to g\mut^{\e/2}$, where $\mut$ has dimensions of mass, 
so that $g$ remains dimensionless.

Expanding out the numerator, we have
\begin{equation}
N = q^2 - x(1{-}x)k^2 + m^2 + (1{-}2x)kq\;.
\label{pinum}
\end{equation}
The term linear in $q$ integrates to zero.   For the rest,
we use the general result of section 14 to get 
\begin{eqnarray}
\mut^\e\int{\ddq\over(2\pi)^d}\,{1\over(q^2 + D)^2} 
&=& {i\over16\pi^2}\left[\,{2\over\e} - \ln(D/\mu^2)\right] \;,
\label{intyuk0} \\
\noalign{\medskip}
\mut^\e\int{\ddq\over(2\pi)^d}\,{q^2\over(q^2 + D)^2} 
&=& {i\over16\pi^2}\left[\,{2\over\e} + \half -\ln(D/\mu^2)\right](-2D) \;,
\label{intyuk2} 
\end{eqnarray}
where $\mu^2=4\pi e^{-\gamma}\mut^2$, and we have dropped terms
of order $\e$.
Plugging \eqs{intyuk0} and (\ref{intyuk2}) into \eq{piloop2} yields
\begin{eqnarray}
\Pi_{\Psi\,\rm loop}(k^2) &=& 
-{g^2\over4\pi^2}\Biggl[
      {1\over\e}(k^2+2m^2)
     +{\ts{1\over6}}k^2+m^2
\nonumber \\
&& \qquad {}
- \int_0^1 dx \left(3x(1{-}x)k^2 + m^2\right)\!\ln(D/\mu^2)\Biggr].
\label{piloop3}
\end{eqnarray}
We see that the divergent term has (as expected) 
a form that permits cancellation by the counterterms.

We evaluated the second
diagram of \fig{piyuk} in section 30, with the result
\begin{equation}
\Pi_{\ph\,\rm loop}(k^2) 
= {\lam\over(4\pi)^2}\left[\,{1\over\e}+\half
            -\half\ln(M^2/\mu^2)\right]\!M^2\;.
\label{phloop}
\end{equation}
The third diagram gives the contribution of the counterterms,
\begin{equation}
\Pi_{\rm ct}(k^2) 
= -(Z_\ph{-}1)k^2 - (Z_M{-}1)M^2 \;.
\label{pi_ct}
\end{equation}
Adding up eqs.$\,$(\ref{piloop3}--\ref{pi_ct}), we see that 
finiteness of $\Pi(k^2)$ requires
\begin{eqnarray}
Z_\ph &=& 1 - {g^2\over 4\pi^2}\!\left({1\over\e} + \hbox{finite}\right),
\label{zph49} \\
\noalign{\medskip}
Z_M &=& 1 + \left({\lam\over 16\pi^2}
                     -{g^2\over 2\pi^2}{m^2\over M^2}\right)\!\!
\left({1\over\e} + \hbox{finite}\right),
\label{zM49}
\end{eqnarray}
plus higher-order (in $g$ and/or $\lam$) corrections.  
Note that, although there is an $O(\lam)$
correction to $Z_M$, there is not an $O(\lam)$ correction to $Z_\ph$. 

We can impose $\Pi(-M^2)=0$ by writing
\begin{equation}
\Pi(k^2) = 
{g^2\over4\pi^2}
\!\left[\,\int_0^1 dx \left(3x(1{-}x)k^2 + m^2\right)\!\ln(D/D_0)
+ \kappa_\Pi(k^2+M^2)\right],
\label{pi49}
\end{equation}
where $D_0 = -x(1{-}x)M^2 + m^2$, and $\kappa_\Pi$ 
is a constant to be determined.
We fix $\kappa_\Pi$ by imposing $\Pi'(-M^2)=0$, which yields
\begin{equation}
\kappa_\Pi = \int_0^1 dx\, x(1{-}x)[3x(1{-}x)M^2 - m^2]/\!D_0 \;.
\label{k49}
\end{equation}
Note that, in this on-shell renormalization 
scheme, there is no $O(\lam)$ correction to $\Pi(k^2)$.

\begin{figure}
\begin{center}
\epsfig{file=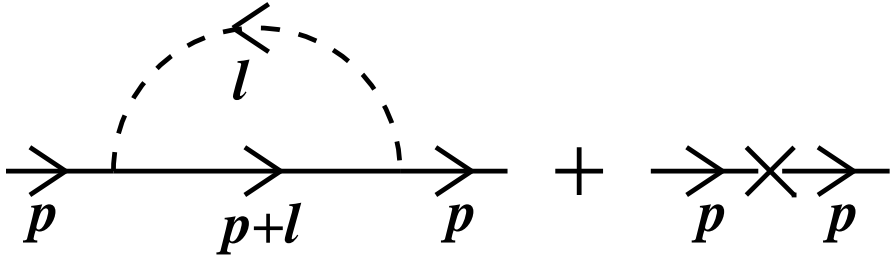}
\end{center}
\caption{The one-loop and counterterm corrections to the fermion
propagator in Yukawa theory.}
\label{psiyuk}
\end{figure}

Next we turn to the $\Psi$ propagator, which receives
the one-loop (and counterterm) corrections shown in \fig{psiyuk}.  
The spin indices are contracted in the usual way, 
following the arrows backwards.  We have
\begin{equation}
i\Sigma_{\rm 1\,loop}(\psl) = 
(-g)^2\!\left({\ts{1\over i}}\right)^{\!2}
\int{\dfl\over(2\pi)^4}\,
\Bigl[\g_5\st(\psl+\lsl)\g_5\Bigr]\td(\ell^2) \;,
\label{sigloop}
\end{equation}
where $\st(\psl)$ is given by \eq{st}, and
\begin{equation}
\td(\ell^2) = {1\over \ell^2 + M^2 -i\eps}
\label{td49}
\end{equation}
is the free scalar propagator in momentum space.

We evaluate \eq{sigloop} with the usual bag of tricks.  The result is
\begin{equation}
i\Sigma_{\rm 1\,loop}(\psl) = -g^2 \int_0^1 dx
\int{\dfq\over(2\pi)^4}\,{ N \over (q^2 + D)^2 } \;,
\label{sigloop2}
\end{equation}
where $q=\ell+xp$ and 
\begin{eqnarray}
N &=& \qsl + (1{-}x)\psl + m \;,
\label{n49} \\
\noalign{\medskip}
D &=& x(1{-}x)p^2 + xm^2 + (1{-}x)M^2 \;.
\label{d49}
\end{eqnarray}
The integral diverges, and so we analytically continue it to 
$d=4-\e$ spacetime dimensions, make the replacement $g\to g\mut^{\e/2}$,
and take the limit as $\e\to 0$.  The term linear in $q$ in \eq{n49}
integrates to zero.  Using \eq{intyuk0}, we get
\begin{equation}
\Sigma_{\rm 1\,loop}(\psl)
= -{g^2\over 16\pi^2} \Biggl[ {1\over\e}(\psl + 2m)
- \int_0^1 dx \,\Bigl((1{-}x)\psl + m\Bigr)\!\ln(D/\mu^2)\Biggr].
\label{sigloop3}
\end{equation}
We see that the divergent term has (as expected) 
a form that permits cancellation by the counterterms, which give
\begin{equation}
\Sigma_{\rm ct}(\psl) 
= -(Z_\Psi{-}1)\psl - (Z_m{-}1)m \;.
\label{sigma_ct}
\end{equation}
Adding up \eqs{sigloop3} and (\ref{sigma_ct}), we see that
finiteness of $\Sigma(\psl)$ requires 
\begin{eqnarray}
Z_\Psi &=& 1 - {g^2\over 16\pi^2}\!\left({1\over\e} + \hbox{finite}\right),
\label{zpsi49} \\
\noalign{\medskip}
Z_m &=& 1 - {g^2\over 8\pi^2}\left({1\over\e} + \hbox{finite}\right),
\label{zm49}
\end{eqnarray}
plus higher-order corrections.

We can impose $\Sigma(-m)=0$ by writing
\begin{equation}
\Sigma(\psl) =  
{g^2\over 16\pi^2}\left[\int_0^1 dx \,\Bigl((1{-}x)\psl + m\Bigr)\!\ln(D/D_0)
+ \kappa_\Sigma(\psl + m)\right],
\label{sig5}
\end{equation}
where $D_0$ is $D$ evaluated at $p^2=-m^2$, and $\kappa_\Sigma$ 
is a constant to be determined.
We fix $\kappa_\Sigma$ by imposing $\Sigma'(-m)=0$.  In differentiating
with respect to $\psl$, we take the $p^2$ in $D$, \eq{d49},
to be $-\psl{\kern1pt}^2$; we find
\begin{equation}
\kappa_\Sigma = -2\int_0^1 dx\, x^2(1{-}x)m^2\!/\!D_0 \;.
\label{kt49}
\end{equation}

\begin{figure}
\begin{center}
\epsfig{file=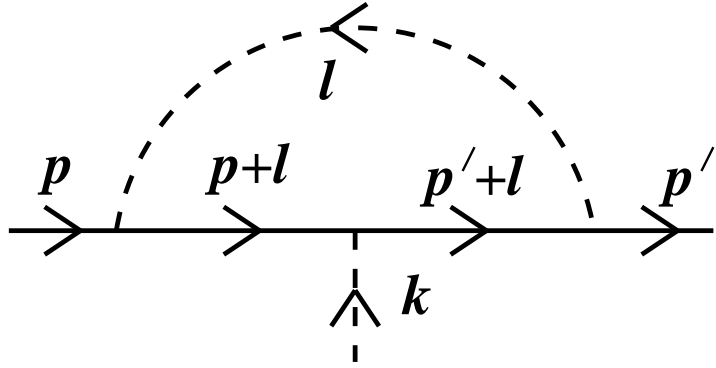}
\end{center}
\caption{The one-loop correction to the scalar-fermion-fermion vertex
in Yukawa theory.}
\label{vyuk}
\end{figure}

Next we turn to the correction to the Yukawa vertex.
We define the vertex function $i\V_Y(p',p)$
as the sum of one-particle irreducible diagrams with one 
incoming fermion with momentum $p$, one outgoing fermion
with momentum $p'$, and one incoming scalar with momentum $k=p'-p$.
The original vertex $-Z_g g\gamma_5$ is the first term
in this sum, and the diagram of \fig{vyuk} is the second.  Thus we have
\begin{equation}
i\V_Y(p',p) = -Z_g g\gamma_5 + i\V_{Y,\,1\,\rm loop}(p',p) + O(g^5)\;,
\label{vy}
\end{equation}
where
\begin{equation}
i\V_{Y,\,1\,\rm loop}(p',p) =
(-g)^3\!\left({\ts{1\over i}}\right)^{\!3} \int{\ddl\over(2\pi)^d}\,
\Bigl[\g_5\st(\pslp{+}\lsl)\g_5\st(\psl{+}\lsl)\g_5\Bigr]\td(\ell^2) \;.
\label{vyloop}
\end{equation}
The numerator can be written as 
\begin{equation}
N=(\pslp+\lsl+m)(-\psl-\lsl+m)\g_5\;, 
\label{n3a}
\end{equation}
and the denominators combined in the usual way.  We then get
\begin{equation}
i\V_{Y,\,1\,\rm loop}(p',p)/g =
-ig^2\int dF_3\int{\dfq\over(2\pi)^4}\, {N \over (q^2 + D)^3 }\;,
\label{vyloop2}
\end{equation}
where the integral over Feynman parameters was defined in section 16, and now
\begin{eqnarray}
q &=& \ell + x_1 p + x_2 p' \;,
\label{q32} \\
\noalign{\medskip}
N &=& [\qsl - x_1\psl + (1{-}x_2)\pslp + m]
      [-\qsl-(1{-}x_1)\psl+x_2\pslp + m]\g_5 \;,
\label{n32} \\
\noalign{\medskip}
D &=& x_1(1{-}x_1)p^2 + x_2(1{-}x_2)p^{\prime 2} 
- 2x_1x_2 p\cd p' +(x_1{+}x_2)m^2 + x_3 M^2 \;. \qquad
\label{d32}
\end{eqnarray}
Using $\qsl\qsl = -q^2$, we can write $N$ as
\begin{equation}
N = q^2\g_5 + \widetilde{N} + \hbox{(linear in $q$)} \;,
\label{n332}
\end{equation}
where
\begin{equation}
\widetilde{N} = 
[-x_1\psl + (1{-}x_2)\pslp + m]
[-(1{-}x_1)\psl + x_2\pslp + m] \g_5 \;.
\label{nt3}
\end{equation}
The terms linear in $q$ in \eq{n332} integrate to zero, 
and only the first term is divergent.  Performing the usual
manipulations, we find
\begin{equation}
i\V_{Y,\,1\,\rm loop}(p',p)/g =
-{g^2\over8\pi^2}\left[\left({1\over\e}
-{\ts{1\over4}}-{\ts{1\over2}}\int dF_3\,\ln(D/\mu^2)\right)\!\g_5
+{\ts{1\over4}}\int dF_3\,{\widetilde{N}\over D}\right].
\label{vyloop3}
\end{equation}
From \eq{vy}, we see that finiteness of $\V_Y(p',p)$ requires
\begin{equation}
Z_g = 1 + {g^2\over 8\pi^2}\!\left({1\over\e} + \hbox{finite}\right),
\label{zg32}
\end{equation}
plus higher-order corrections.

To fix the finite part of $Z_g$, we need a condition to impose
on $\V_Y(p',p)$.  One possibility is to mimic what we did in $\ph^3$ theory in
section 16: require $\V_Y(0,0)$ to have the tree-level value $ig\g_5$.
As in $\ph^3$ theory, this is not well motivated physically, but has
the virtue of simplicity, and this is a good enough reason for us
to adopt it.  We leave the details to problem 51.2.

\begin{figure}
\begin{center}
\epsfig{file=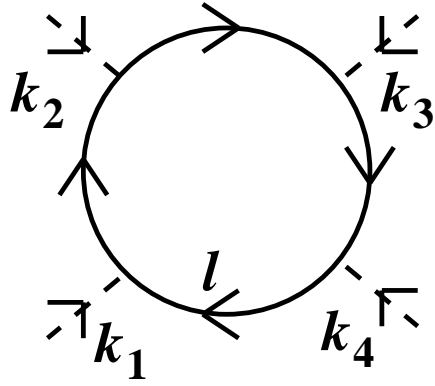}
\end{center}
\caption{One of six diagrams with a closed fermion loop and four external
scalar lines; the other five are obtained by permuting the external momenta
in all possible inequivalent ways.}
\label{4yuk}
\end{figure}

Next we turn to the corrections to the $\ph^4$ vertex $i\V_4(k_1,k_2,k_3,k_4)$;
the tree-level contribution is $-iZ_\lam\lam$.
There are diagrams with a closed fermion loop, as shown in \fig{4yuk},
plus one-loop diagrams with $\ph$ particles only that we evaluated 
in section 30.  We have
\begin{eqnarray}
i\V_{4,\,\Psi\,\rm loop} &=&
(-1)(-g)^4\!\left({\ts{1\over i}}\right)^{\!4} 
\int{\dfl\over(2\pi)^4}\,
\tr\Bigl[\st(\lsl)\g_5\st(\lsl{-}\ksl_1)\g_5
\nonumber \\
&& \qquad\qquad\qquad\quad {} \times
   \st(\lsl{+}\ksl_2{+}\ksl_3)\g_5\st(\lsl{+}\ksl_2)\g_5\Bigr] 
\nonumber \\
&& {} + \hbox{5 permutations of $(k_2,k_3,k_4)$}\;.
\label{ph4loop}
\end{eqnarray}
Again we can employ the standard methods; there are no unfamiliar
aspects.  This being the case, let us concentrate on obtaining
the divergent part; this will give us enough information to calculate
the one-loop contributions to the beta functions for $g$ and $\lam$.

To obtain the divergent part of \eq{ph4loop}, it is sufficient to set $k_i=0$.  
Then the numerator in \eq{ph4loop} becomes simply
$\tr\,(\lsl\g_5)^4 = 4(\ell^2)^2$, and the denominator is $(\ell^2+m^2)^4$.
Then we find, after including the identical contributions from the other five
permutations of the external momenta,
\begin{equation}
\V_{4,\,\Psi\,\rm loop} = 
-{3g^4\over\pi^2}\!\left({1\over\e} + \hbox{finite}\right).
\label{v4psidiv}
\end{equation}
From section 30, we have
\begin{equation}
\V_{4,\,\ph\,\rm loop} = 
{3\lam\over16\pi^2}\!\left({1\over\e} + \hbox{finite}\right).
\label{v4phdiv}
\end{equation}
Then, using
\begin{equation}
\V_4 = -Z_\lam \lam +
\V_{4,\,\Psi\,\rm loop} +
\V_{4,\,\ph\,\rm loop} + \ldots\;,
\label{v4total}
\end{equation}
we see that finiteness of $\V_4$ requires
\begin{equation}
Z_\lam = 1 + \left({3\lam\over 16\pi^2}
                     -{3g^4\over \pi^2\lam}\right)\!\!
             \left({1\over\e} + \hbox{finite}\right),
\label{zlam}
\end{equation}
plus higher-order corrections.

\vskip0.5in

\begin{center}
Problems
\end{center}

\vskip0.25in

51.1)
Derive the one-loop correction to the scalar propagator by working through
\eq{znbarnj}, and show that it has an extra minus sign (corresponding
to the closed fermion loop).

51.2)
Prove \eq{tbfs}.  
Hints:  Given a multiparticle state $|p,s,q,n\ra$ with four momentum $p^\mu$
and mass $M^2 = -p^2$, $J_z = \half s$, charge $q$, and other attributes
specified by $n$, show that $\la 0|\Psi(x)|p,s,q,n\ra$ vanishes unless
$s=\pm1$ and $q=+1$.  Argue that this is enough information to fix
$\la 0|\Psi(x)|p,s,q,n\ra \propto u_s(p)$, a spinor of mass $M$.

51.3)
Finish the computation of $\V_Y(p',p)$, imposing the condition
\begin{equation}
\V_Y(0,0)=ig\g_5 \;.
\label{VY00}
\end{equation}

\vfill\eject

\noindent Quantum Field Theory  \hfill   Mark Srednicki

\vskip0.5in

\begin{center}
\large{52: Beta Functions in Yukawa Theory}
\end{center}
\begin{center}
Prerequisite: 27, 51
\end{center}

\vskip0.5in

In this section we will compute the beta functions for the
Yukawa coupling $g$ and the $\ph^4$ coupling $\lam$ in Yukawa
theory, using the methods of section 27.  

The relations between
the bare and renormalized couplings are
\begin{eqnarray}
g_0 &=& Z_\ph^{-1/2}Z_\Psi^{-1}Z_g\mut^{\e/2}g \;,
\label{g50} \\
\noalign{\medskip}
\lam_0 &=& Z_\ph^{-2}Z_\lam\mut^{\e}\lam \;.
\label{lam50} 
\end{eqnarray}
Let us define
\begin{eqnarray}
\ln\Bigl(Z_\ph^{-1/2}Z_\Psi^{-1}Z_g\Bigr) &=&
\sum_{n=1}^\infty {G_n(g,\lam)\over\e^n} \;,
\label{gsum} \\
\noalign{\medskip}
\ln\Bigl(Z_\ph^{-2}Z_\lam\Bigr) &=&
\sum_{n=1}^\infty {L_n(g,\lam)\over\e^n} \;.
\label{lamsum}
\end{eqnarray}
From our results in section 51, we have
\begin{eqnarray}
G_1(g,\lam) &=& {5g^2\over 16\pi^2} + \ldots\;,
\label{g1} \\
\noalign{\medskip}
L_1(g,\lam) &=& {3\lam\over 16\pi^2} + {g^2\over 2\pi^2} 
                 - {3g^4\over \pi^2\lam} + \ldots\;,
\label{lam1}
\end{eqnarray}
where the ellipses stand for higher-order (in $g^2$ and/or $\lam$) 
corrections.

Taking the logarithm of \eqs{g50} and (\ref{lam50}), and using
\eqs{gsum} and (\ref{lamsum}), we get
\begin{eqnarray}
\ln g_0 &=& \sum_{n=1}^\infty {G_n(g,\lam)\over\e^n}
+ \ln g + \half \e\ln\mut \;,
\label{lng50} \\
\noalign{\medskip}
\ln \lam_0 &=& \sum_{n=1}^\infty {L_n(g,\lam)\over\e^n}
+ \ln\lam + \e\ln\mut \;.
\label{lnlam50} 
\end{eqnarray}
We now use the fact that $g_0$ and $\lam_0$ must be independent of $\mu$.
We differentiate \eqs{lng50} and (\ref{lnlam50})
with respect to $\ln\mu$; the left-hand sides vanish, and we
multiply the right-hand sides by $g$ and $\lam$, respectively.  The result is
\begin{eqnarray}
0 &=& \sum_{n=1}^\infty\left(g{\d G_n\over\d g}{dg\over d\ln\mu}
+g{\d G_n\over\d\lam}{d\lam\over d\ln\mu}\right)\!{1\over\e^n}
+{dg\over d\ln\mu} + \half\e g \;,
\label{zerog} \\
\noalign{\medskip}
0 &=& \sum_{n=1}^\infty\left(\lam{\d L_n\over\d g}{dg\over d\ln\mu}
+\lam{\d L_n\over\d\lam}{d\lam\over d\ln\mu}\right)\!{1\over\e^n}
+{d\lam\over d\ln\mu} + \e\lam \;.
\label{zerolam}
\end{eqnarray}
In a renormalizable theory, 
$dg/d\ln\mu$ and $d\lam/d\ln\mu$ must be finite in the
$\e\to 0$ limit.  Thus we can write
\begin{eqnarray}
{dg\over d\ln\mu} &=& -\half\e g + \beta_g(g,\lam)\;,
\label{betag} \\
\noalign{\medskip}
{d\lam\over d\ln\mu} &=& -\e\lam + \beta_\lam(g,\lam)\;.
\label{betalam} 
\end{eqnarray}
Substituting these into \eqs{zerog} and (\ref{zerolam}), and matching
powers of $\e$, we find
\begin{eqnarray}
\beta_g(g,\lam) 
&=& g\!\left(\half g{\d \over\d g} + \lam{\d \over\d\lam}\right)\!G_1\;,
\label{betag2} \\
\noalign{\medskip}
\beta_\lam(g,\lam) 
&=& \lam\!\left(\half g{\d \over\d g} + \lam{\d \over\d\lam}\right)\!L_1\;.
\label{betalam2} 
\end{eqnarray}
The coefficients of all higher powers of $1/\e$ must also vanish,
but this gives us no more information about the beta functions.

Using \eqs{g1} and (\ref{lam1}) in \eqs{betag2} and (\ref{betalam2}), we get
\begin{eqnarray}
\beta_g(g,\lam) 
&=& {5g^3\over 16\pi^2} + \ldots \;,
\label{betag3} \\
\noalign{\medskip}
\beta_\lam(g,\lam) 
&=& {1\over 16\pi^2}\Bigl(3\lam^2 + 8\lam g^2 - 48g^4\Bigr) + \ldots \;.
\label{betalam3} 
\end{eqnarray}
The higher-order corrections have extra factors of $g^2$ and/or $\lam$.

\vfill\eject

\noindent Quantum Field Theory  \hfill   Mark Srednicki

\vskip0.5in

\begin{center}
\large{53: Functional Determinants}
\end{center}
\begin{center}
Prerequisite: 44, 45
\end{center}

\vskip0.5in

In the section we will explore the meaning of the 
{\it functional determinants\/} that arise
when doing gaussian path integrals, either bosonic or fermionic.  
We will be interested in situations where the path integral over 
one particular field is gaussian, but generates a functional determinant
that depends on some other field.  We will see how to relate this
functional determinant to a certain infinite set of Feynman diagrams. 
We will need the technology we develop here to compute the path
integral for nonabelian gauge theory in section 70.

We begin by considering a theory of a complex scalar field $\chi$ with
\begin{equation}
{\cal L} = 
           -\d^\mu\chid\d_\mu\chi-m^2\chid\chi
           +g\ph\chid\chi \;,
\label{ellchi}
\end{equation}
where $\ph$ is a real scalar {\it background field}.  That is, $\ph(x)$
is treated as a fixed function of spacetime.
Next we define the path integral
\begin{equation}
Z(\ph) = \int {\cal D}\chid\,{\cal D}\chi\;e^{i\int\dfx\,\L}\;,
\label{zph}
\end{equation}
where we use the $\eps$ trick of section 6 to impose vacuum boundary 
conditions, and the normalization $Z(0)=1$ is fixed by hand.

Recall from section 44 that if we have $n$ complex variables $z_i$,
then we can evaluate gaussian integrals by the general formula
\begin{equation}
\int d^n\!z\,d^n\bar z\,\exp\left(-i\bar z_i M_{ij}z_j\right)
\propto (\det M)^{-1} \;.
\label{intchi}
\end{equation}
In the case of the functional integral in \eq{zph}, 
the index $i$ on the integration variable
is replaced by the continuous spacetime label $x$, and
the ``matrix'' $M$ becomes
\begin{equation}
M(x,y)=[-\d_x^2 + m^2 - g\ph(x)]\delta^4(x-y) \;.
\label{mxy}
\end{equation}
In order to apply \eq{intchi}, we have to understand what it
means to compute the determinant of this expression.

To this end, let us first note that we can write $M=M_0 \Mt$,
which is shorthand for
\begin{equation}
M(x,z)=\int\dfy\,M_0(x,y)\Mt(y,z) \;,
\label{mmm}
\end{equation}
where
\begin{eqnarray}
M_0(x,y) &=& (-\d_x^2 + m^2)\delta^4(x-y) \;,
\label{mxy0} \\
\noalign{\medskip}
\Mt(y,z) &=& \delta^4(y-z) - g\Delta(y-z)\ph(z) \;.
\label{tildem} 
\end{eqnarray}
Here $\Delta(y-z)$ is the Feynman propagator, which obeys
\begin{equation}
(-\d_y^2 + m^2)\Delta(y-z) = \delta^4(y-z) \;.
\label{dDd}
\end{equation}
After various integrations by parts, it is easy to see that
eqs.$\,$(\ref{mmm}--\ref{tildem}) reproduce \eq{mxy}.

Now we can use the general matrix relation 
\begin{equation}
\det AB =\det A\,\det B 
\label{detab}
\end{equation}
to conclude that
\begin{equation}
\det M=\det M_0\,\det\Mt \;.
\label{detmmm}
\end{equation}
The advantage of this decomposition is that $M_0$ 
is independent of the background
field $\ph$, and so the resulting factor of $(\det M_0)^{-1}$
in $Z(\ph)$ can simply be absorbed into the overall 
normalization.  Furthermore, we have $\Mt=I-G$, where 
\begin{equation}
I(x,y) = \delta^4(x-y)
\label{ixy}
\end{equation}
is the identity matrix, and 
\begin{equation}
G(x,y) = g\Delta(x-y)\ph(y) \;.
\label{gxy}
\end{equation}
Thus, for $\ph(x)=0$, we have $\Mt=I$ and so $\det\Mt=1$.
Then, using \eq{intchi} and the normalization condition $Z(0)=1$,
we see that for nonzero $\ph(x)$ we must have simply
\begin{equation}
Z(\ph) = (\det\Mt)^{-1} \;.
\label{zphmt}
\end{equation}

Next, we need the general matrix relation 
\begin{equation}
\det A = \exp\mathop{\rm Tr}\ln A \;,
\label{daexptrlna}
\end{equation}
which is most easily proven by 
remembering that the determinant and trace are both basis independent,
and then working in a basis where $A$ is in Jordan form (that is, all entries
below the main diagonal are zero).  Thus we can write
\begin{eqnarray}
\det\Mt &=& \exp\mathop{\rm Tr}\ln \Mt 
\nonumber \\
\noalign{\medskip}
&=& \exp\mathop{\rm Tr}\ln(I-G)
\nonumber \\
\noalign{\medskip}
&=& \exp\mathop{\rm Tr}\Biggl[-\sum_{n=1}^\infty {1\over n}G^n\Biggr].
\label{dettm}
\end{eqnarray}
Combining \eqs{zphmt} and (\ref{dettm}) we get
\begin{equation}
Z(\ph) = \exp\sum_{n=1}^\infty {1\over n}\mathop{\rm Tr}G^n \;,
\label{zphf}
\end{equation}
where
\begin{equation}
\mathop{\rm Tr}G^n = g^n\int\dfx_1\ldots\dfx_n\,\Delta(x_1{-}x_2)\ph(x_2)
                     \ldots \Delta(x_n{-}x_1)\ph(x_1)\;.
\label{trgn}
\end{equation}
This is our final result for $Z(\ph)$.

\begin{figure}
\begin{center}
\epsfig{file=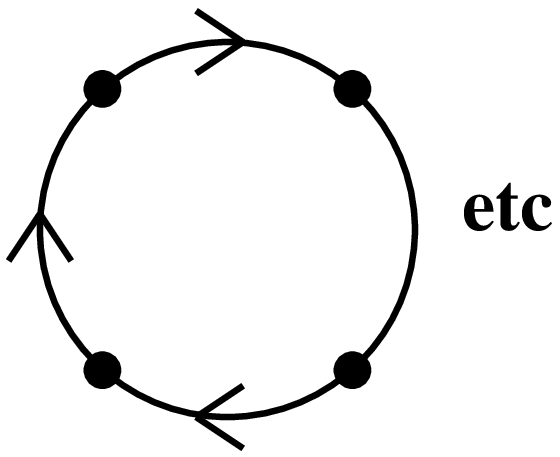}
\end{center}
\caption{All connected diagrams with $\ph(x)$ treated as an external field.
Each of the $n$ dots represents a factor of $ig\ph(x)$, 
and each solid line is a $\chi$ or $\Psi$ propagator.}  
\label{detfig}
\end{figure}

To better understand what it means, we will rederive it in a different way.
Consider treating the $g\ph\chid\chi$ term in $\L$ as an interaction.
This leads to a vertex that connects two $\chi$ propagators; the associated
vertex factor is $ig\ph(x)$.  According to the general analysis
of section 9, we have $Z(\ph)=\exp i\Gamma(\ph)$, where 
$i\Gamma(\ph)$ is given by a sum of connected diagrams.  
(We have called the exponent $\Gamma$ rather than $W$ because it
is naturally interpreted as a quantum action for $\ph$ after
$\chi$ has been integrated out.)
The only diagrams we can draw with these
Feynman rules are those of \fig{detfig}, with $n$ insertions of the vertex,
where $n\ge 1$.  The diagram with $n$ vertices has an $n$-fold cyclic
symmetry, leading to a symmetry factor of $S=n$.  The factor of $i$ 
associated with each vertex is canceled by the factor of $1/i$ associated
with each propagator.  Thus the value of the $n$-vertex diagram is
\begin{equation}
{1\over n}\,g^n\int\dfx_1\ldots\dfx_n\,\Delta(x_1{-}x_2)\ph(x_2)
                     \ldots \Delta(x_n{-}x_1)\ph(x_1)\;.
\label{ntrgn}
\end{equation}
Summing up these diagrams, and using \eq{trgn}, we find
\begin{equation}
i\Gamma(\ph) = \sum_{n=1}^\infty {1\over n}\mathop{\rm Tr}G^n \;.
\label{wph}
\end{equation}
This neatly reproduces \eq{zphf}.  Thus we see that a functional
determinant can be represented as an infinite sum of Feynman diagrams.

Next we consider a theory of a Dirac fermion $\Psi$ with
\begin{equation}
{\cal L} = i\Psibar\dsl\Psi - m \Psibar\Psi +g\ph\Psibar\Psi \;,
\label{ellphpsi}
\end{equation}
where $\ph$ is again a real scalar background field. 
We define the path integral
\begin{equation}
Z(\ph) = \int {\cal D}\Psibar\,{\cal D}\Psi\;e^{i\int\dfx\,\L}\;,
\label{zph2}
\end{equation}
where we again use the $\eps$ trick to impose vacuum boundary 
conditions, and the normalization $Z(0)=1$ is fixed by hand.

Recall from section 44 that if we have $n$ complex Grassmann 
variables $\psi_i$,
then we can evaluate gaussian integrals by the general formula
\begin{equation}
\int d^n\!\bar\psi\,d^n\!\psi\,\exp\left(-i\bar\psi_i M_{ij}\psi_j\right)
\propto \det M \;.
\label{intpsi}
\end{equation}
In the case of the functional integral in \eq{zph2}, 
the index $i$ on the integration variable
is replaced by the continuous spacetime label $x$ plus the spinor
index $\alpha$, and the ``matrix'' $M$ becomes
\begin{equation}
M_{\alpha\beta}(x,y)=[-i\dsl_x + m - g\ph(x)]_{\alpha\beta}\delta^4(x-y) \;.
\label{mxy2}
\end{equation}
In order to apply \eq{intpsi}, we have to understand what it
means to compute the determinant of this expression.

To this end, let us first note that we can write $M=M_0 \Mt$,
which is shorthand for
\begin{equation}
M_{\alpha\gamma}(x,z)=\int\dfy\,M_{0\alpha\beta}(x,y)
                               \Mt_{\beta\gamma}(y,z) \;,
\label{mmm2}
\end{equation}
where
\begin{eqnarray}
M_{0\alpha\beta}(x,y) &=& (-i\dsl_x + m)_{\alpha\beta}\delta^4(x-y) \;,
\label{mxy02} \\
\noalign{\medskip}
\Mt_{\beta\gamma}(y,z) &=& \delta_{\beta\gamma} \delta^4(y-z)
                       - g S_{\beta\gamma}(y-z)\ph(z) \;.
\label{tildem2} 
\end{eqnarray}
Here $S_{\beta\gamma}(y-z)$ is the Feynman propagator, which obeys
\begin{equation}
(-i\dsl_y + m)_{\alpha\beta}S_{\beta\gamma}(y-z) 
= \delta_{\alpha\gamma}\delta^4(y-z) \;.
\label{dDd2}
\end{equation}
After various integrations by parts, it is easy to see that
eqs.$\,$(\ref{mmm2}--\ref{tildem2}) reproduce \eq{mxy2}.

Now we can use \eq{detmmm}.
The advantage of this decomposition is that $M_0$ 
is independent of the background
field $\ph$, and so the resulting factor of $\det M_0$
in $Z(\ph)$ can simply be absorbed into the overall 
normalization.  Furthermore, we have $\Mt=I-G$, where 
\begin{equation}
I_{\alpha\beta}(x,y) = \delta_{\alpha\beta}\delta^4(x-y)
\label{ixy2}
\end{equation}
is the identity matrix, and 
\begin{equation}
G_{\alpha\beta}(x,y) = gS_{\alpha\beta}(x-y)\ph(y) \;.
\label{gxy2}
\end{equation}
Thus, for $\ph(x)=0$, we have $\Mt=I$ and so $\det\Mt=1$.
Then, using \eq{intpsi} and the normalization condition $Z(0)=1$,
we see that for nonzero $\ph(x)$ we must have simply
\begin{equation}
Z(\ph) = \det\Mt \;.
\label{zphmt2}
\end{equation}

Next, we use \eqs{dettm} and (\ref{zphmt2}) to get
\begin{equation}
Z(\ph) = \exp\,-\!\sum_{n=1}^\infty {1\over n}\mathop{\rm Tr}G^n \;,
\label{zphf2}
\end{equation}
where now
\begin{equation}
\mathop{\rm Tr}G^n = g^n\int\dfx_1\ldots\dfx_n\,{\rm tr}\,S(x_1{-}x_2)\ph(x_2)
                     \ldots S(x_n{-}x_1)\ph(x_1)\;,
\label{trgn2}
\end{equation}
and ``tr'' denotes a trace over spinor indices.
This is our final result for $Z(\ph)$.

To better understand what it means, we will rederive it in a different way.
Consider treating the $g\ph\Psibar\Psi$ term in $\L$ as an interaction.
This leads to a vertex that connects two $\Psi$ propagators; the associated
vertex factor is $ig\ph(x)$.  According to the general analysis
of section 9, we have $Z(\ph)=\exp i\Gamma(\ph)$, where 
$i\Gamma(\ph)$ is given by a sum of connected diagrams.  
(We have called the exponent $\Gamma$ rather than $W$ because it
is naturally interpreted as a quantum action for $\ph$ after
$\Psi$ has been integrated out.)
The only diagrams we can draw with these
Feynman rules are those of \fig{detfig}, with $n$ insertions of the vertex,
where $n\ge 1$.  The diagram with $n$ vertices has an $n$-fold cyclic
symmetry, leading to a symmetry factor of $S=n$.  The factor of $i$ 
associated with each vertex is canceled by the factor of $1/i$ associated
with each propagator.  The closed fermion loop implies a trace over the
spinor indices.  Thus the value of the $n$-vertex diagram is 
\begin{equation}
{1\over n}\,g^n\int\dfx_1\ldots\dfx_n\,{\rm tr}\,S(x_1{-}x_2)\ph(x_2)
                     \ldots S(x_n{-}x_1)\ph(x_1)\;.
\label{ntrgn2}
\end{equation}
Summing up these diagrams, we find
that we are missing the overall minus sign in \eq{zphf2}.
The appropriate conclusion is that we must associate 
an extra minus sign with each closed fermion loop.  

\vfill\eject

\end{document}